\documentclass[useAMS,usenatbib,onecolumn]{mn2e}
\usepackage{graphicx}
\usepackage{bm}
\usepackage{ulem}

\newcommand{\bea}{\begin{eqnarray}}
\newcommand{\eea}{\end{eqnarray}}
\newcommand{\beq}{\begin{equation}}
\newcommand{\eeq}{\end{equation}}

\newcommand{\dm}{\boldsymbol {\Delta}}
\newcommand{\ceta}{\boldsymbol \eta}

\newcommand{\simless}[0]{\mathbin{\lower 3pt\hbox
   {$\rlap{\raise 5pt\hbox{$\char'074$}}\mathchar"7218$}}}
\newcommand{\simgreat}[0]{\mathbin{\lower 3pt\hbox
   {$\rlap{\raise 5pt\hbox{$\char'076$}}\mathchar"7218$}}}

\newcommand{\lta}[0]{\simless}

\newcommand{\figref}[1]{figure \ref{#1}}
\newcommand{\figrefs}[1]{figures \ref{#1}}
\newcommand{\figrefbare}[1]{\ref{#1}}
\newcommand{\capfigref}[1]{Figure \ref{#1}}

\newcommand{\eqnref}[1]{eq. (\ref{#1})}
\newcommand{\eqnrefs}[1]{eqs. (\ref{#1})}
\newcommand{\eqnrefbare}[1]{(\ref{#1})}

\newcommand{\myreal}[0]{\Re}
\newcommand{\myimag}[0]{\Im}

\title[Extending the domain of validity of the Lagrangian approximation]{Extending the domain of validity of the Lagrangian approximation}
\author[Sharvari Nadkarni-Ghosh and David F. Chernoff]{Sharvari Nadkarni-Ghosh$^{1}$\thanks{E-mail:
smn27@cornell.edu} and David F. Chernoff$^{2}$\thanks{E-mail:
chernoff@astro.cornell.edu} \\
$^{1}$Department of Physics, Cornell University, Ithaca, NY 14853 USA\\
$^{2}$Department of Astronomy, Cornell University, Ithaca, NY 14853 USA}
\begin{document}

\date{}
\pagerange{\pageref{firstpage}--\pageref{lastpage}} \pubyear{}

\maketitle
\label{firstpage}

\begin{abstract}

We investigate convergence of Lagrangian Perturbation Theory (LPT) by
analysing the model problem of a spherical homogeneous top-hat in an
Einstein-deSitter background cosmology.  We derive the formal
structure of the LPT series expansion, working to arbitrary order in
the initial perturbation amplitude.  The factors that regulate LPT
convergence are identified by studying the exact, analytic solution
expanded according to this formal structure. The key methodology is to
complexify the exact solution, demonstrate that it is analytic and
apply well-known convergence criteria for power series expansions of
analytic functions.  The ``radius of convergence'' and the ``time of
validity'' for the LPT expansion are of great practical interest.  The
former describes the range of initial perturbation amplitudes which
converge over some fixed, future time interval.  The latter describes the extent
in time for convergence of a given initial amplitude. We determine the
radius of convergence and time of validity for a full sampling of
initial density and velocity perturbations.

This analysis fully explains the previously reported observation that LPT
fails to predict the evolution of an underdense, open region beyond a
certain time. It also implies the existence of other examples,
including overdense, closed regions,
for which LPT predictions should also fail. We show that this is
indeed the case by numerically computing the LPT expansion in these
problematic cases.

The formal limitations to the validity of LPT expansion are
considerably more complicated than simply the first occurrence of
orbit crossings as is often assumed. Evolution to a future time
generically requires re-expanding the solution in overlapping domains
that ultimately link the initial and final times, each domain subject
to its own time of validity criterion. We demonstrate that it is possible
to handle all the problematic cases by taking multiple steps (LPT
re-expansion). 

A relatively small number ($\sim 10$) of re-expansion steps suffices
to satisfy the time of validity constraints for calculating the
evolution of a non-collapsed, recombination-era perturbation up to the
current epoch. If it were possible to work to infinite Lagrangian
order then the result would be exact.  Instead, a finite expansion has
finite errors.  We characterise how the leading order numerical error
for a solution generated by LPT re-expansion varies with the choice of
Lagrangian order and of time step size. Convergence occurs when the
Lagrangian order increases and/or the time step size decreases in a
simple, well-defined manner. We develop a recipe for time step control
for LPT re-expansion based on these results.

\end{abstract}


\section{Introduction}
\label{intro}
Understanding the non-linear growth of structure in an expanding
universe has been an active area of research for nearly four decades.
Simulations have been instrumental in illustrating exactly what
happens to an initial power spectrum of small fluctuations but
analytic methods remain essential for elucidating the physical basis
of the numerical results. Perturbation theory, in particular,
is an invaluable tool for achieving a sophisticated understanding.

The Eulerian and Lagrangian frameworks are the two principal modes of
description of a fluid. The fundamental dependent variables in the
Eulerian treatment are the density $\rho({\bf x},t)$ and velocity
${\bf v}({\bf x}, t)$ expressed as functions of the grid coordinates
${\bf x}$ and time $t$, the independent variables.  In perturbation
theory the dependent functions are expanded in powers of a small
parameter. For cosmology that parameter typically encodes 
a characteristic small
spatial variation of density and/or velocity with respect to a
homogeneous cosmology at the initial time.  As a practical matter, the
first-order perturbation theory becomes inaccurate when the
perturbation grows to order unity.  Subsequently one must work to
higher order to handle the development of non-linearity (see
\citealt{bernardeau_large-scale_2002} for a review) or adopt an
alternative method of expansion.

In the Lagrangian framework, the fundamental dependent variable is the
physical position of a fluid element or particle (terms used
interchangeably here).  The independent variables are a set of labels ${\bf
  X}$, each of which follows a fluid element, and the time. Usually
${\bf X}$ is taken as the position of the element at some initial time
but other choices are possible. In any case, the physical position and
velocity of a fluid element are ${\bf r} = {\bf r} ({\bf X},t)$ and
${\dot {\bf r}}({\bf X},t)$, respectively. Knowledge of the motion of
each fluid element permits the full reconstruction of the Eulerian
density and velocity fields.  In cosmological applications of
Lagrangian perturbation theory (LPT), just like Eulerian perturbation
theory, the dependent variables are expanded in terms of initial
deviations with respect to a homogeneous background.  The crucial
difference is that the basis for the expansion is the variation in the
initial position and position-derivative
not the variation in the initial fluid density and velocity.
The Eulerian density and velocity may be reconstructed
from knowledge of the Lagrangian position using exact non-perturbative
definitions.  A linear approximation to the displacement field results
in a non-linear expression for the density contrast. The Lagrangian
description is well-suited to smooth, well-ordered initial conditions;
a single fluid treatment breaks down once particle crossings begin, caustics
form and the density formally diverges.

First-order LPT was originally introduced by \cite{zeldovich_gravitational_1970} to study the formation of
non-linear structure in cosmology. In his treatment the initial density
field was taken to be linearly proportional to the initial
displacement field (the ``Zeldovich approximation''). These results
were extended by many authors (\citealt{moutarde_precollapse_1991};
\citealt{buchert_lagrangian_1992};
\citealt{bouchet_weakly_1992};
\citealt{buchert_lagrangian_1993};
\citealt{buchert_lagrangian_1994};
\citealt{munshi_non-linear_1994};
 \citealt{catelan_lagrangian_1994};
\citealt{buchert_lagrangian_1995};
\citealt{bouchet_perturbative_1995}; 
\citealt{bouchet_introductory_1996};
 \citealt{ehlers_newtonian_1997}).
 The work pioneered by Bouchet focused
on Zeldovich initial conditions and established the link between LPT
variables and statistical observables.  The work by Buchert as well as
the paper by \cite{ehlers_newtonian_1997}
formalised the structure of the Newtonian perturbative series for
arbitrary initial conditions.  A general relativistic version of the
Zeldovich approximation was developed by \cite{kasai_tetrad-based_1995} and other relativistic descriptions
of the fluid in its rest frame were investigated by \cite{matarrese_post-newtonian_1996} and \cite*{matarrese_general-relativistic_1993, matarrese_relativistic_1994}.
LPT has been used for many applications including, recently, the
construction of non-linear halo mass functions
by \cite{monaco_lagrangian_1997} and \cite{scoccimarro_pthalos:fast_2002}.

\begin{figure}
\includegraphics[width=7cm,height=7cm]{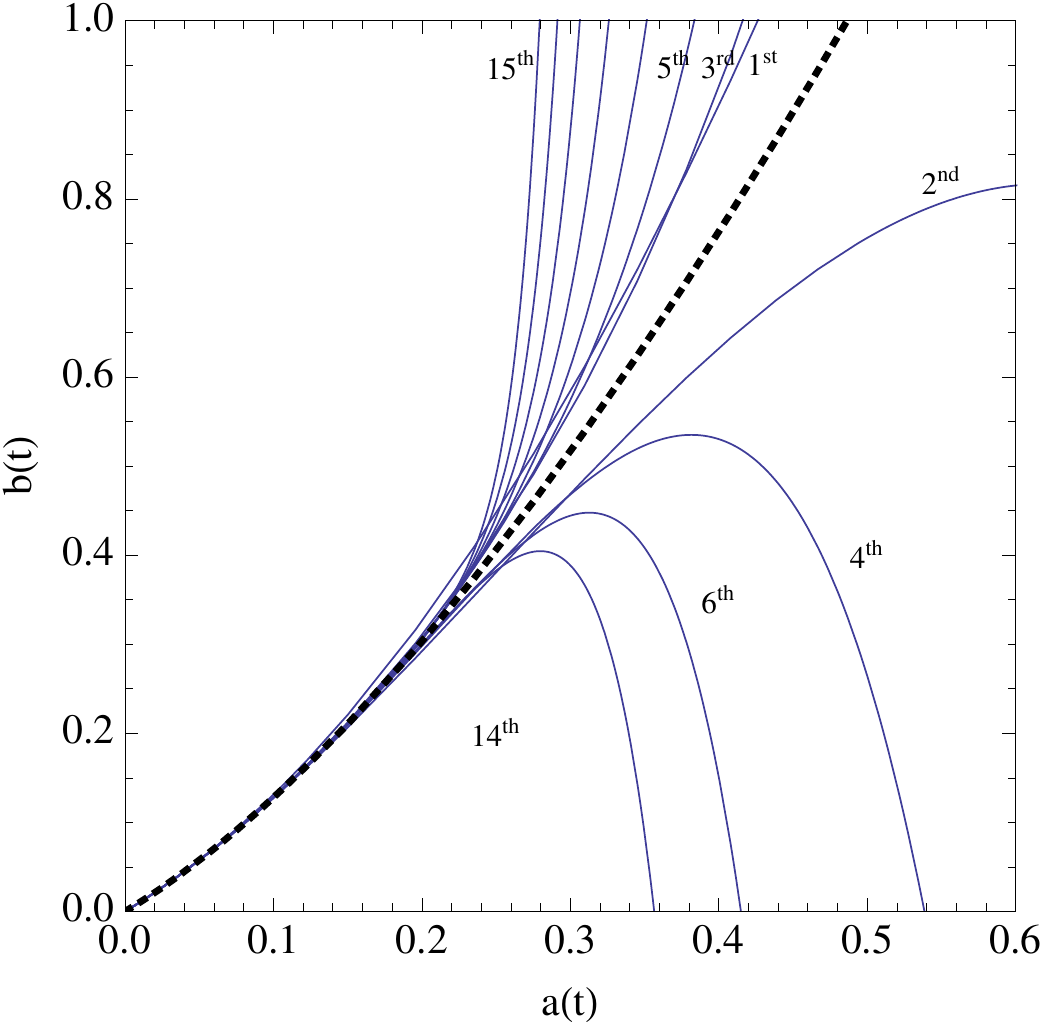}
\caption{The time-dependent scale factor $b$ of an initial spherical
  top-hat perturbation is plotted as a function of the background
  scale factor $a$. The perturbation is a pure growing mode, i.e. the
  density and velocity perturbations vanish at $t=0$.  The black dotted line
  is the exact solution.  The smooth blue lines are the LPT results obtained
by  working successively to higher and higher order. Series with even (odd) final order lie below
  (above) the exact solution.  Roughly speaking, LPT converges only
  for $a \lta 0.2$. Beyond that point {\it the higher order
    approximations deviate from the exact solution more than lower
    order ones}.}
\label{openplot}
\end{figure}

Not much has been written about the convergence of LPT although LPT
expansions are routinely employed. \cite{sahni_accuracy_1996} pointed
out that the formal series solution for the simplest problem, the
spherical top-hat, did not converge for the evolution of homogeneous
voids. \capfigref{openplot} illustrates the conundrum that the LPT
approximations diverge from the exact solution in a manner that
worsens as the order of the approximation increases. The details will
be described in the next section.

This paper explores LPT convergence for the spherical top-hat and
identifies the root cause for the lack of convergence. The analysis
naturally suggests a means of extending the range of validity of
LPT. This generalisation of LPT guarantees convergence to the exact
solution of the model problem at all times prior to the occurrence of
the first caustic.

\cite{tatekawa_improvinglagrangian_2007} attempted to treat the
divergence by applying the Shanks transformation to the LPT series.
Although non-linear transformations can sum a divergent series,
the correct answer is not guaranteed; comparison of
several different methods is usually necessary to yield trustworthy
results. Other approaches include the Shifted-Time-Approximation (STA) and Frozen-Time-Approximation (FTA) which have been investigated by \cite*{karakatsanis_temporal_1997}. These schemes modify lower order terms to mimic the behavior of higher order terms and/or extend the range of applicability in time. None of these techniques are considered here.

The organisation follows: 
\S\ref{LPTforSTH} sketches the
model problem, the evolution of a uniform sphere in a background homogeneous
Einstein-deSitter cosmology. The LPT equations, the structure of the
formal series and the term-by-term solution are outlined.  
\S\ref{convergencestudy} discusses the complexification of the
LPT solution and convergence of the series. 
This section introduces the ``radius of convergence"
and the ``time of validity" for LPT.  
\S\ref{exactsoln} outlines the real and complex forms of the
parametric solution and sets forth the equations that must be solved to
locate the poles which govern the convergence.
\S\ref{Results} presents
numerical results for the time of validity and radius of convergence
for a full range of possible initial conditions for
the top-hat. 
The notion of mirror model symmetry is introduced and used to explain
a connection in the convergence for open and closed models.
\S\ref{sec_extension} shows that the time of validity may be extended by
re-expanding the solution in overlapping domains that ultimately
link the initial and final times, each domain subject to an individual
time of validity criterion. The feasibility of this method
is demonstrated in some examples.
\S\ref{conclusion} summarises the work.

\section{The model problem and formal series solution}
\label{LPTforSTH}

This section describes the governing equations, the initial physical
conditions, the formal structure of the LPT series solution and
the order-by-order solution.

\subsection{Newtonian treatment}

Consider evolution on sub-horizon scales after recombination in a 
matter-dominated universe.  A Newtonian treatment of gravity based on solving
Poisson's equation for the scalar potential and on evaluating the force
in terms of the gradient of the potential gives an excellent
approximation for non-relativistic dynamics.  When there are no significant
additional forces on the fluid element (e.g. pressure forces) then it
is straightforward to eliminate the gradient of the potential in favour
of ${\ddot {\bf r}}$, the acceleration. The governing equations are
\bea 
\nabla_{x} \cdot {\ddot{\bf r}} &=& -4 \pi  G \rho ({\bf x},t) \label{maineq}\\
\nabla_{x} \times {\ddot {\bf r}}&=&0 \label{curleq} 
\eea
where $\rho({\bf x},t)$ is the background plus perturbation density,
$G$ is Newton's gravitational constant and $\nabla_{x}$ is the
Eulerian gradient operator. In the Lagrangian treatment, the
independent variables are transformed $({\bf x},t) \to ({\bf
  X},t)$ and the particle position ${\bf r} = {\bf r}({\bf X},t)$
adopted as the fundamental dependent quantity. For clarity note that ${\bf x}$ refers to a fixed Eulerian grid not a comoving coordinate.

\subsection{Spherical top-hat}

The starting physical configuration is a compensated spherical
perturbation in a homogeneous background cosmology. The perturbation
encompasses a constant density sphere about the centre of symmetry and
a compensating spherical shell. The shell that surrounds the sphere
may include vacuum regions plus regions of varying density.
Unperturbed background extends beyond the outer edge of the
shell. Physical distances are measured with respect to the centre of
symmetry. At initial time $t_0$ the background and the innermost
perturbed spherical region (hereafter, ``the sphere'') have Hubble
constants $H_0$ and $H_{p0}$, and densities $\rho_0$ and $\rho_{p0}$,
respectively. Let $r_{b,0}$ ($r_{p,0}$) be the physical
distance from the centre of symmetry to the inner edge of the
background (to the outer edge of the sphere)
at the initial time. Let $a_0$, $b_0$ be the initial scale factors for the background and the sphere respectively. 
Two sets of Lagrangian coordinates $Y=r_{b,0}/a_0$ and 
$X=r_{p,0}/b_0$ are defined.  A gauge choice sets $a_0=b_0$. Appendix \ref{formalsetup} provides a figure and gives a somewhat more detailed chain of reasoning that clarifies the construction of the physical and Lagrangian coordinate
systems. The initial perturbation is characterised by the independent
parameters
\bea
\delta & = & \frac{\rho_{p0}}{\rho_{0}} - 1 \nonumber \\
\delta_v & = & \frac{H_{p0}}{H_0} -1  .
\label{defdeltas}
\eea
Finally, assume that the background
cosmology is critical $\Omega_0=1$. The perturbed sphere has
\beq 
\Omega_{p0} = \frac{1+\delta}{(1+\delta_v)^2}. 
\eeq
The physical problem of interest here is the future evolution of an
arbitrary initial state unconstrained by the past history.  In general, the
background and the perturbation can have different big bang times.
Initial conditions with equal big bang times will be analysed as a
special case of interest and imply an additional relationship between
$\delta$ and $\delta_v$. 

While the previous paragraphs summarise the set up, they eschew the complications in modelling an inhomogeneous system in terms of
separate inner and outer homogeneous universes.  For example, matter
motions within the perturbed inner region may overtake the outer
homogeneous region so that there are problem-specific limits on how
long solutions for the scale factors $a(t)$ and $b(t)$ remain valid.  The appendix shows
that there exist inhomogeneous initial configurations for which the
limitations arising from the convergence of the LPT series are
completely independent of the limitations associated with collisions
or crossings of inner and outer matter-filled regions. A basic premise
of this paper is that it is useful to explore the limitations of
the LPT series independent of the additional complications
that inhomogeneity entails.

\subsection{Equation governing scale factors}

During the time that
the spherical perturbation evolves as an independent homogeneous
universe it may be fully described in terms of the motion of its
outer edge $r_p$. Write
\beq 
r_p(t) = b(t) X 
\label{defn}
\eeq 
where $b(t)$ is the scale factor and $X$ is the
Lagrangian coordinate of the edge. 
The initial matter density of the homogeneous sphere 
$\rho(X,t_0) = \rho_{p0} = \rho_{0} (1+\delta)$. 
The physical density of the
perturbation at time $t$ is
\beq 
\rho (X,t)=  \frac{\rho(X,t_0) J(X,t_0)}{J(X,t)} 
\eeq
where the Jacobian of
the transformation relating the Lagrangian and physical spaces is
\beq 
J(X,t) = \det \left( \frac{\partial {\vec r}}{\partial {\vec X}} \right ). 
\eeq
Since
\eqnref{defn} implies  $J(X,t) = b(t)^3$ and the choice $a_0=b_0$
implies $J(X,t_0) = a_0^3$ the perturbation matter density at later times is
\beq 
\label{eqn:perturbedrho}
\rho_p (t)  = \frac{\rho_{0} (1 + \delta) a_0^3}{b(t) ^3}. 
\eeq
Substituting for $\rho_p$ and $r_p$ in \eqnref{maineq} gives
\beq 
\frac{{\ddot b}}{b} = -\frac{1}{2}  \frac{H_0^2 a_0^3(1+\delta)}{b^3} 
\label{secondorder} 
\eeq
with initial
conditions $b(t_0) = a_0$ and ${\dot b}(t_0) = {\dot a}_0(1+
\delta_v) $. The curl of the acceleration (i.e. \eqnref{curleq}) vanishes
by spherical symmetry. The corresponding equation for the background
scale factor is
\beq 
\label{eqn:unperturbedscalefactor}
\frac{{\ddot a}}{a} = -\frac{1}{2} \frac{H_0^2 a_0^3}{a^3}
\eeq
with initial conditions $a(t_0)=a_0$ and ${\dot a}(t_0)= {\dot a}_0 = a_0 H_0$.
The solution for $b(t)$ will be expressed in terms of its deviations
from $a(t)$.

In summary, the physical setup is an $\Omega_0=1$ background model and
a compensated spherical top-hat (over- or underdense). The properties
of interest are the relative scale factors $a(t)/a_0$ and $b(t)/a_0$
(the choice of $a_0$ is arbitrary and $b_0=a_0$). The evolution of the
relative scale factors is fully specified by $H_0$, $H_{p0}$ and
$\Omega_{p0}$ at time $t_0$. The perturbed physical quantities,
$H_{p0}$ and $\Omega_{p0}$, may be equivalently specified by a choice
of $\delta$ and $\delta_v$. Appendix 
\ref{formalsetup} contains a systematic description and enumerates degrees
of freedom, parameters, constraints, etc.

\subsection{Perturbations in phase space }

\begin{figure}
\includegraphics[width=17cm]{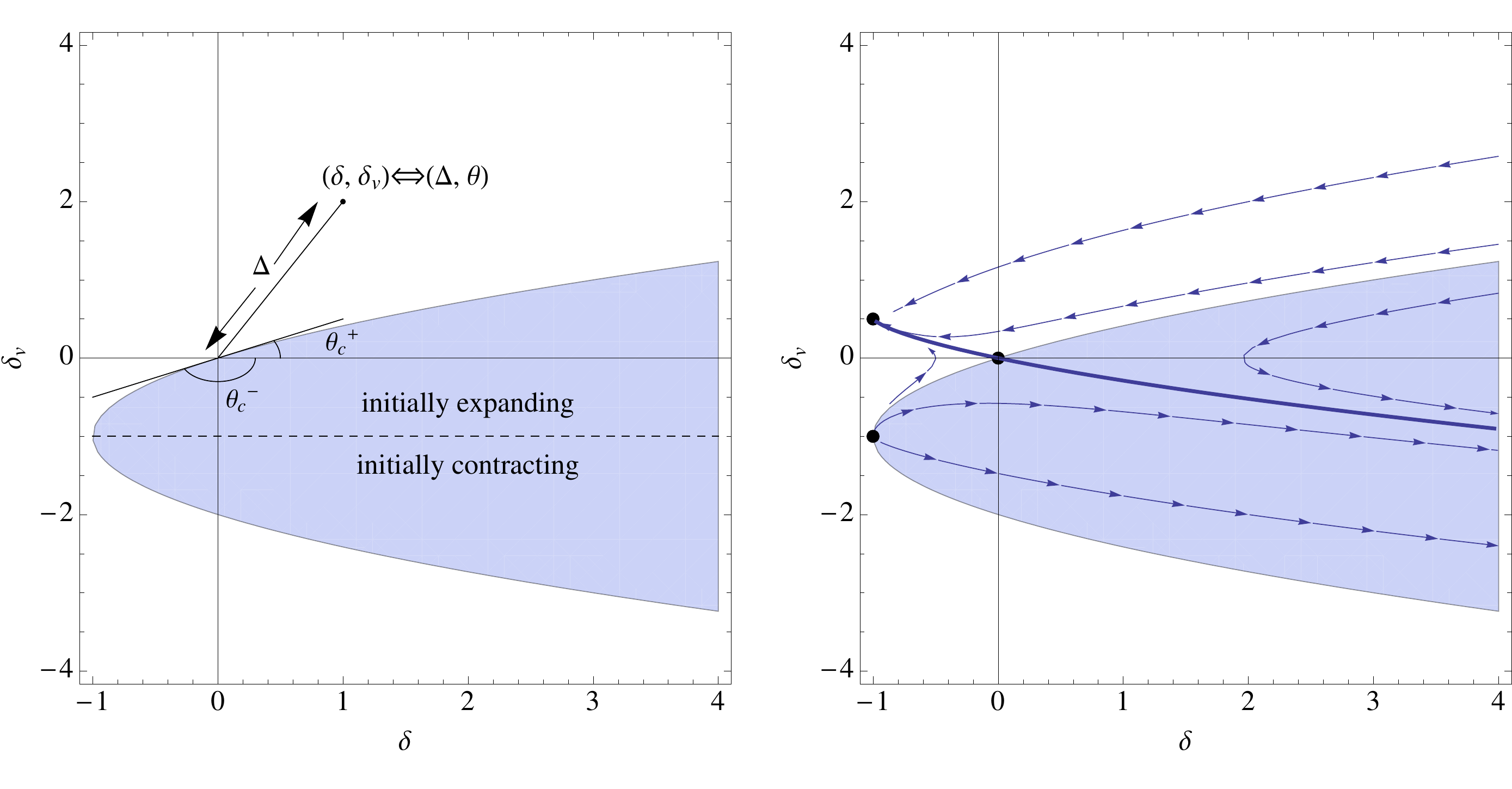}
\caption{Phase diagram of density and velocity perturbations $(\delta, \delta_v)$. Physical initial conditions require $-1 < \delta
  < \infty$ and $-\infty < \delta_v < \infty$. The left panel highlights the 
  qualitatively different initial conditions. The shaded (unshaded)
  region corresponds to closed (open) model with negative (positive)
  total energy. For small $\Delta$, models with $\theta_c^- < \theta <
  \theta_c^+$ are closed. Initially expanding and contracting models
  are separated by the dashed horizontal line ($\delta_v=-1$). 
 The right panel shows the evolution of $\delta$ and $\delta_v$. The solid blue line corresponds to the ``Zeldovich" condition i.e no perturbation at $t=0$. The points $(-1,-1),(-1,0.5)$ and $(0,0)$ are unstable, stable and saddle fixed points of the phase space flow. The flow lines (indicated by the blue vectors) converge along the Zeldovich curve either to the stable fixed point at $(-1,0.5)$ or move parallel to the Zeldovich curve to a future density singularity. Further discussion follows in \S\ref{discussion} and \S\ref{flow}. }

 \label{phaseplot} 
\end{figure}

The initial density and velocity perturbations are taken to be of the same order in
the formalism developed by \cite{buchert_lagrangian_1992, buchert_lagrangian_1994}, \cite{buchert_lagrangian_1993} and \cite{ehlers_newtonian_1997}. We assume the same ordering here.
Write the initial perturbation ($\delta$, $\delta_v$) in terms of
magnitude $\Delta$ and angle $\theta$
\beq 
\Delta = \sqrt{\delta^2 + \delta_v^2}
\eeq
so that
\bea 
\delta &=& \Delta \cos \theta \\ 
\delta_v &=& \Delta \sin\theta.
\eea
To map physical perturbations $(\delta,\delta_v)$ in
a unique manner to $(\Delta,\theta)$ adopt the ranges
$\Delta \ge 0$ and $-\pi<\theta \leq \pi$. 
\capfigref{phaseplot} (left panel) shows the phase space of initial
perturbations. Since density is non-negative the
regime of physical interest is $\delta \ge -1$.
Open (closed) models with positive (negative)
total energy are the regions that are unshaded (shaded).
Initially expanding models, $1+\delta_v>0$, lie above
the horizontal dashed line. The right panel of \figref{phaseplot} summarises the overall evolution of the system. The initial choice of $\delta$ and $\delta_v$ dictates the trajectory in the plane. Cosmologically relevant initial conditions generally assume there to be no perturbation at $t=0$. We adopt the name ``Zeldovich'' initial conditions for models that satisfy this condition. This establishes a specific relation between $\delta$ and $\delta_v$ which is indicated by the sold blue line. The exact mathematical relationship is given in \S\ref{discussion}. Starting from a general initial point ($\delta, \delta_v$), the system as it evolves traces out a curve in phase space indicated by the blue arrows. There are three fixed points visible. The origin $(\delta, \delta_v) \equiv (0,0) $, which corresponds to a unperturbed background model, is a saddle point. The vacuum  static model at point $(-1,-1)$ is a unstable node and the vacuum, expanding  model at $(-1,0.5)$ is a degenerate attracting node. Far to the right and below the dashed line the models collapse to a future singularity. The phase portrait illustrates that the trajectories either converge to the vacuum, expanding model or to the singular, collapsing model. The equations that govern the flow and further relevance of the Zeldovich solution is discussed in \S\ref{flow} and \S\ref{discussion}.

\subsection{Generating the Lagrangian series solution}

The scale factor is formally expanded
\beq
b(t) = \sum_{n=0}^{\infty}  b^{(n)}(t) \Delta^n 
\eeq
where $b^{(n)}$ denotes an $n$-th order term.
The initial conditions are
\bea 
b(t_0) &=& a(t_0) \\ 
{\dot b}(t_0) &=& \dot{a}_0 (1 + \delta_v) =  \dot{a}_0 (1 + \Delta \sin\theta) .
\eea 
Substitute the expansion for $b(t)$ into \eqnref{secondorder},
equate orders of $\Delta$ to give at {\it zeroth order }
\beq 
{\ddot b}^{(0)} + \frac{1}{2} \frac{H_0^2 a_0^3 }{{b^{(0)}}^2} =0 
\eeq
which is identical in form to \eqnref{eqn:unperturbedscalefactor}
for the unperturbed background scale factor.
The initial conditions at zeroth order:
\bea 
b^{(0)}(t_0)        &=& a_0 \\
{\dot b}^{(0)}(t_0) &=&  {\dot a}_0  .
\eea
The equation and initial conditions for $b^{(0)}(t)$
simply reproduce the background scale factor evolution
$b^{(0)} (t) = a(t)$. Without loss of generality
assume that the background model has
big bang time $t=0$ so that
\beq
a(t) = a_0 \left(\frac{t}{t_0}\right )^{2/3} = a_0 \left(\frac{3 H_0 t}{2} \right )^{2/3}.
\eeq
At {\it first order}
\beq 
{\ddot b}^{(1)} - \frac{H_0^2 a_0^3b^{(1)}}{a^3}  =  - \frac{1}{2} \frac{H_0^2 a_0^3  \cos \theta}{a^2}
\eeq
and, in general, 
\beq 
 {\ddot b}^{(n)} - \frac{H_0^2 a_0^3b^{(n)}}{a^3} =  S^{(n)} 
\eeq
where $S^{(n)}$ depends upon lower order approximations
($b^{(0)}, b^{(1)} \dots b^{(n-1)}$)
as well as $\theta$. The first few are:

\bea
S^{(2)} & = & -\frac{1}{2} \frac{H_0^2a_0^3  }{a^4}\left[ 
 b^{(1)} \left \{
3b^{(1)} - 2 a \cos \theta 
\right\}
\right]  \\
S^{(3)} & = & - \frac{1}{2} \frac{H_0^2a_0^3  }{a^5} \left[
b^{(1)} \left\{ -4 \left(b^{(1)}\right )^2 +
 6 a b^{(2)} +
  3 a b^{(1)}\cos \theta \right \} -
  2 a^2 b^{(2)}\cos \theta \right ] \\
S^{(4)} & = & - \frac{1}{2} \frac{H_0^2a_0^3  }{a^6} 
\left[ \left(b^{(1)}\right )^2 
\left\{ 
5 \left(b^{(1)}\right )^2   -
12 a  b^{(2)}   - 4 a b^{(1)}\cos \theta
\right\} +
\right. \nonumber\\
&& \phantom{- \frac{1}{2} \frac{H_0^2a_0^3  }{a^6}  \left[\right. }
\left.
6 a^2 b^{(1)}
\left\{ b^{(3)}  +  b^{(2)} \cos \theta  \right\}+  3 a^2 \left(b^{(2)}\right )^2 -2 a^3 b^{(3)}  \cos \theta  
 \right].
\eea
These terms can be easily generated by
symbolic manipulation software. The initial conditions are
\bea 
b^{(1)}(t_0) &=& 0  \\
{\dot b}^{(1)}(t_0) &=&  {\dot a}_0 \sin \theta
\eea
and for $n > 1$
\bea 
b^{(n)}(t_0) &=& 0  \\
{\dot b}^{(n)}(t_0) &=&  0 .
\eea
The ordinary differential equations for $b^{(n)}$ may
be solved order-by-order.

To summarise, the structure of the hierarchy and
the simplicity of the initial conditions allows the
evaluation of the solution at any given order in terms of the
solutions with lower order. This yields a formal expansion
for the scale factor of the sphere
\beq 
b = \sum_{n=0}^{\infty} b^{(n)}(t) \Delta^n  
\label{seriesexpand}
\eeq
which encapsulates the Lagrangian perturbation treatment.
The right hand size explicitly depends
upon the size of the perturbation and time and implicitly upon
$a_0$, $H_0$, and $\theta$.
This hierarchy of equations is identical to that generated
by the full formalism developed by Buchert and collaborators
when it is applied to the top-hat problem. The convergence
properties in time and in $\Delta$ are distinct; a simple
illustrative example of this phenomenon
is presented in Appendix \ref{example}.

\section{Convergence properties of the LPT series solution}
\label{convergencestudy}

The series solution outlined in the previous section does not converge
at all times. \capfigref{openplot} is a practical demonstration of this
non-convergence for the case of an expanding void.  An understanding
of the convergence of the LPT series is achieved by extending the
domain of the expansion variable $\Delta$ from the real positive axis
to the complex plane. 

\subsection{Complexification}
\label{singularity}

The differential \eqnref{secondorder} and initial conditions for
the physical system are 
\bea
\label{rdiffeq}
\ddot b(t) &=& -\frac{1}{2} \frac{H_0^2 a_0^3 (1 + \Delta \cos \theta)}{b(t)^2} \nonumber\\
b(t_0) &=& a_0 \nonumber\\
\dot{b}(t_0) &=& {\dot a}_0 (1 + \Delta  \sin \theta)
\eea
where $t$, $b(t)$, $\Delta$ and all zero-subscripted quantities
are real. This set may be extended by allowing $\Delta$ and $b$
to become complex quantities, denoted hereafter, $\dm$ and ${\bf b}$, while the
rest of the variables remain real. The complex set is
\bea
\label{cdiffeq}
\ddot {\bf b}(t) &=& -\frac{1}{2} \frac{H_0^2 a_0^3 (1+\dm \cos \theta)}{{\bf b}(t)^2} \nonumber\\
{\bf b}(t_0) &=& a_0 \nonumber\\
\dot{\bf b}(t_0) &=& {\dot a}_0 (1+\dm  \sin \theta).
\eea

The theory of differential equations (for example, \citealt{chicone})
guarantees that the solution to a
real initial value problem is unique and smooth in the initial
conditions and parameters of the equation and can be extended in time
as long as there are no singularities in the differential equation (hereafter,
the maximum extension of the solution).
First, note that each complex quantity in
\eqnref{cdiffeq} may be represented by a real pair, i.e.
${\bf b}= u + i v$ by pair
$\{u,v\}=\{ \myreal {\bf b }, \myimag {\bf b} \}$ and
${\bf \Delta}=x + i y$ by pair
$\{x,y\}=\{ \myreal {\bf \Delta}, \myimag {\bf \Delta} \}$. 
The basic theory implies continuity and smoothness of solution $u$ and $v$
with respect to initial conditions and
parameters $x$ and $y$. Second, observe that the Cauchy-Riemann conditions
$u_x=v_y$ and $u_y=-v_x$ are preserved by the form of the ordinary
differential equation.  Since the initial conditions and parameter
dependence are holomorphic functions of $\dm$ it follows that ${\bf
  b}(t,\dm)$ is a holomorphic function of $\dm$ at times $t$ within
the maximum extension of the solution.

Inspection shows that the differential equation is singular only at ${\bf b}=0$. For a
particular value of $\dm=\dm'$, the solution to the initial
value problem can be extended to a
maximum time $t_{mx}$ such that ${\bf b}(\dm',t_{mx}) =0$ or
to infinity. The existence of a finite $t_{mx}$ signals that a pole 
in the complex analytic function ${\bf b}(\dm,t)$ forms at
$\dm=\dm'$  and $t=t_{mx}$.
For times $t$ such that $t_0 \le
t < t_{mx}$, the solution $b(\dm, t)$ is analytic in a small
neighbourhood around the point $\dm'$. Of course, there may be poles
elsewhere in the complex $\dm$ plane.

The relationship between the original, real-valued physical problem
and the complexified system is the following. In the original problem
$\Delta$ is a real, positive quantity at $t_0$. LPT is a power
series expansion in $\Delta$ about the origin (the point $\Delta=0$).
LPT's convergence at any time $t$ can be understood by study of the
complexified system.  Consider the complex disk ${\cal D}$ centred on the origin
and defined by $|\dm| < \Delta$. At $t_0$ each point in ${\cal D}$ determines a
trajectory ${\bf b}(\dm,t)$ for the complexified system extending to
infinity or limited to finite time $t=t_{mx}(\dm)$ because of the occurrence of
a pole. The time of validity is defined as $T(\Delta)=\min_{\cal D} t_{mx}$,
i.e. the minimum $t_{mx}$ over
the disk. Since there are no poles in ${\cal D}$ at $t_0$ the time of
validity is the span of time when ${\cal D}$ remains clear of any
singularities. If a function of a complex variable is analytic
throughout an open disk centred around a given point in the complex
plane then the series expansion of the function around that point is
convergent (\citealt{brownandchurchill}).  The LPT expansion for the
original problem converges for times less than the time of
validity because the complex extension 
${\bf b}(\dm,t)$ is analytic throughout ${\cal D}$
for $t<T(\Delta)$. 
If $\Delta_1 < \Delta_2$ then, in an obvious notation, the disks are
nested ${\cal D}(\Delta_1) \subset {\cal D}(\Delta_2)$ and the times
of validity are ordered $T(\Delta_1) \ge T(\Delta_2)$.

\begin{figure}
\includegraphics[width=16cm]{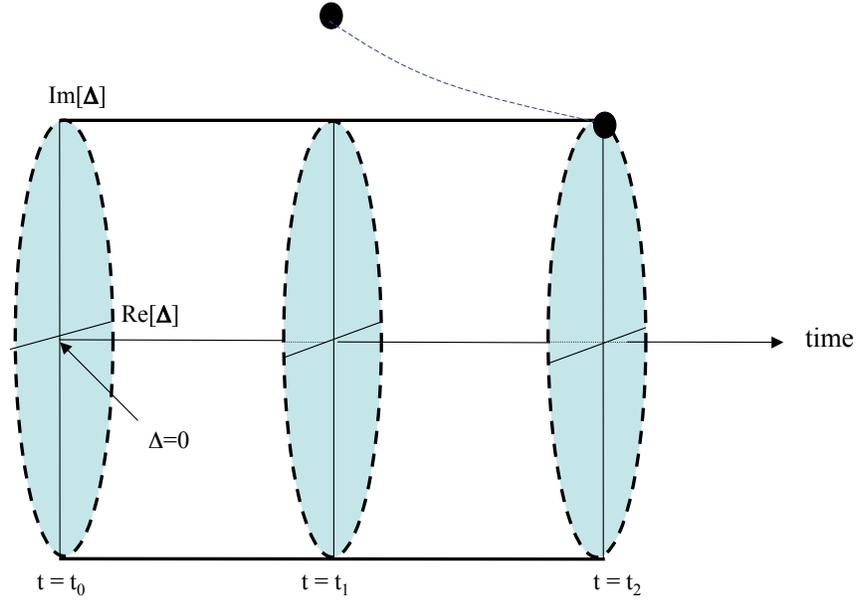}
\caption{This figure is a schematic illustration of how the time of
  validity is determined.  The initial conditions imply a specific,
  real $\Delta$ at time $t_0$. The LPT series is an expansion about
  $\Delta=0$, convergent until a pole appears at some later time
  within the disk of radius $\Delta$ (shown in cyan) in the
  complex $\dm$ plane. Typically, the pole's position forms a curve
  (blue dashed) in the three dimensional space $(\myreal[\dm], \myimag[\dm],
  t)$.  The black dots mark the pole at times $t_1$ and $t_2$. At
  $t_1$ the pole does not interfere with the convergence of the LPT
  series; at $t_2$ it does.  The time of validity may be determined by
  a pole that appears within the disk without moving through the
  boundary (not illustrated). }
\label{fig_tvalid} 
\end{figure}

This idea is shown in \figref{fig_tvalid}. No singularities are
present for the initial conditions at $t_0$; at $t_1$ a
singularity is present {\it outside} the disk but it does not prevent
the convergence of the LPT expansion with $\Delta$ equal to the disk
radius shown; at $t_2$ a singularity is present in the disk or on
its boundary and it may interfere with convergence.

A distinct but related concept is the maximum amplitude perturbation
for which the LPT expansion converges at the initial time and at all
intermediate times up to a given time.  The radius of convergence $R_\Delta(t)$ is
the maximum disk radius $\Delta$ for which $t>T(\Delta)$. Because the
disks are nested if $t_1 < t_2$ then $R_\Delta(t_1) \ge
R_\Delta(t_2)$. 

The time of validity and the radius of convergence are inverse
functions of each other. If the initial perturbation is specified,
i.e. $\Delta$ is fixed, and the question to be answered is ``how
far into the future does LPT work?'' then the time of validity gives
the answer. However, if the question is ``how big an initial
perturbation will be properly approximated by LPT over a given time
interval?'' then the radius of convergence provides the answer.

Finally, note that one can trivially extend this formalism to
deal with time intervals in the past.

\subsection{Calculating radius of convergence and time of validity}

The following recipe shows how to calculate the radius of convergence
$R_\Delta(t)$ and the time of validity $T(\Delta)$ efficiently.  Fix
$a_0$, $H_0$, $t_0$ and $\theta$; these are all real constants set by
the initial conditions. Assume that it is possible to find ${\bf
  b}(\dm, t)$ for complex $\dm$ and real $t$ by solving
\eqnref{cdiffeq}. There exist explicit expressions for ${\bf b}$ as
will be shown later.

Start with $t=t_0$ and $R_\Delta(t)=\infty$. The iteration below maps
out $R_\Delta(t)$ by making small increments in time $\delta t$.

\begin{itemize}

\item{Store old time $t_{previous}=t$, choose increment $\delta t$ and
form new time of interest $t=t_{previous}+\delta t$. }

\item{Locate all the $\dm$ which solve ${\bf b}(\dm, t)=0$. The
roots correspond to poles in the complex function. Find the
root closest to the origin and denote its distance as $|\dm_{near}|$.}

\item{The radius of convergence is $R_\Delta(t) = \min(|\dm_{near}|,R_\Delta(t_{previous}))$.}

\item{Continue.}

\end{itemize}

\begin{figure}
\includegraphics[width=14cm]{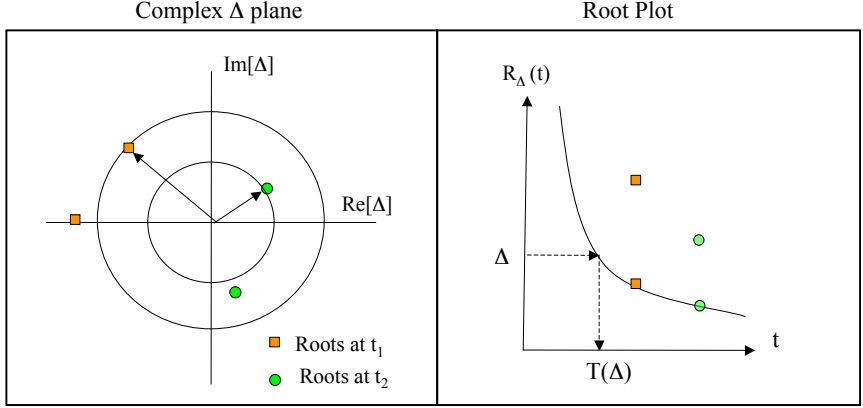}
\caption{
A schematic illustration of the radius of convergence and the time of
validity. The left panel shows the location of poles in the complex
$\dm$ plane at times $t_1$ and $t_2$, denoted by orange squares and
green dots, respectively. At a fixed time, the pole nearest the origin
determines the disk (black circle) within which a series expansion about the
origin converges. The right panel shows $|\dm|$ for $t_1$ and $t_2$. The
black line is $R_\Delta(t)$, 
the minimum $|\dm|$ calculated for a continuous range
of times (where $t_0$, the initial time, lies
far to the left). The arrows show how the time of validity is inferred
for a given $\Delta$.}
\label{fig-defn} 
\end{figure}

Since $R_\Delta$ is decreasing, the inversion to form $T(\Delta)$ is
straightforward.  \capfigref{fig-defn} shows a schematic cartoon of
the construction process.

\section{Explicit solutions}
\label{exactsoln}
The usual parametric representation provides an efficient
method to construct an explicit complex representation for ${\bf
  b}(\dm,t)$.

\subsection{Real (physical) solutions}

The original system \eqnref{rdiffeq} depends upon $a_0$, $H_0$,
$\theta$ and $\Delta$. The assumed Einstein-deSitter background has
$a_0 > 0$ and ${\dot a}_0 > 0$; as defined, the perturbation amplitude
$\Delta \ge 0$ and the relative density and velocity components
are determined by phase angle $\theta$ with $-\pi <\theta \leq \pi$.  The
quantity $(1+ \Delta \cos \theta)$ is proportional to total density
and must be non-negative.  The sign of ${\dot b}_0$ is the sign of $1+
\Delta \sin \theta$ and encodes expanding and contracting initial
conditions.  

Briefly reviewing the usual physical solution, the integrated form is
\beq 
{\dot b}^2 = H_0^2 a_0^3\left[\frac{(1+\Delta \cos \theta)}{ b} + \frac{(1+ \Delta \sin\theta)^2 -(1+\Delta \cos \theta)}{a_0}\right]. 
\label{velocityeq}
\eeq
The combination 
\beq E(\Delta, \theta) = (1+ \Delta \sin \theta)^2 -(1+\Delta \cos \theta) 
\label{energy}
\eeq 
is proportional to the total energy of the system. If $E > 0$ the
model is open and if $E<0$ it is closed and will re-collapse
eventually. \capfigref{phaseplot} shows the parabola $E=0$ which
separates open and closed regions.  For infinitesimal $\Delta$ the
line of division has slope $\tan \theta = 1/2$. Models with $\theta
\in [\theta_c^-,\theta_c^+] = [-\pi+\tan^{-1}(1/2),\tan^{-1}(1/2)]=
[-2.68,0.46]$ are closed while those outside this range are open.

\begin{figure}
\includegraphics[width=16cm]{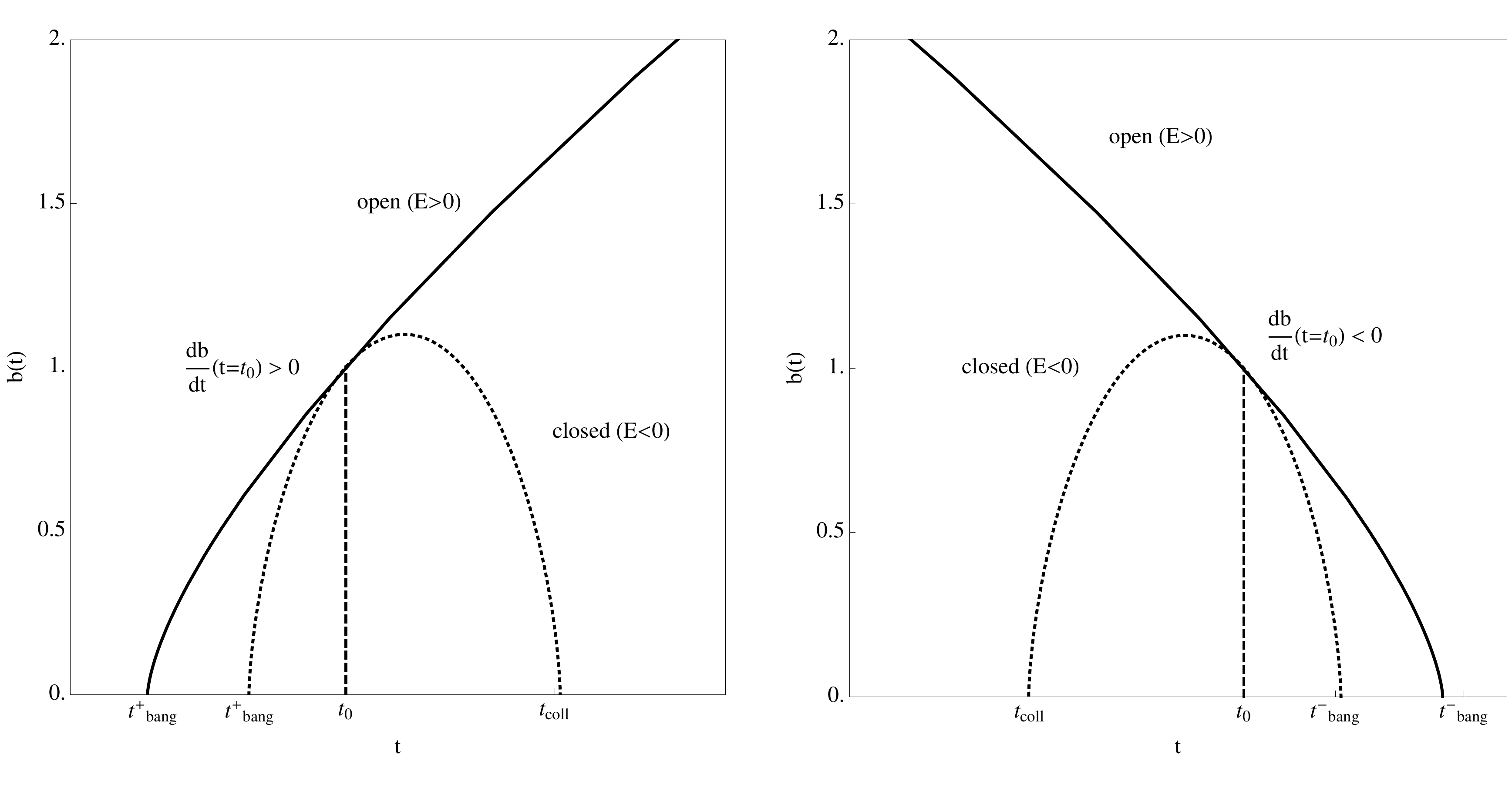}
\caption{Scale factor as a function of time.  The initial
  conditions ($b_0=a_0=1$ and varying ${\dot b}_0$) 
are given at time $t_0$ (dashed
  blue line). The left (right) panel illustrates initially expanding
  (contracting) models.  $t_{bang}^\pm$ corresponds to $\eta=0$;
  $t_{coll}$ to $\eta=2\pi$.  For expanding solutions
  $t_{age}=t_0-t_{bang}^+$ is the time interval since the initial
  singularity and $t_{coll}$ is the future singularity for closed
  models.  For contracting solutions $t_{age}=t_{bang}^- -t_0$
  is the time until the final singularity and $t_{coll}$ is the past
  singularity for closed models.}
\label{dbdtposneg} 
\end{figure}

There are four types of initial conditions (positive and negative
$E$, positive and negative ${\dot b}_0$) and four types of
solutions, shown schematically in \figref{dbdtposneg}.
The solutions have well-known parametric forms involving trigonometric
functions of angle $\eta$ or $i \eta$ (see Appendix
\ref{solutiondetails}). The convention adopted here is that the singularity
nearest the initial time $t_0$ coincides with $\eta=0$ and is
denoted $t_{bang}^+$ ($t_{bang}^{-}$) for initially expanding
(contracting) solutions (see \figref{dbdtposneg}).  The time
interval between the singularity and $t_0$ is $t_{age} = | t_0 -
t_{bang}^\pm | \ge 0$.

The parametric solution for the models can be written as 
\bea
b(\eta, \Delta, \theta) &=& \frac{a_0}{2} \frac{(1+\Delta \cos \theta)}{\left[-E(\Delta,\theta)\right]} (1 -\cos \eta)  \nonumber \\
t(\eta, \Delta, \theta) &=& t_0 \pm 
\left(
\frac{1}{2H_0} \frac{(1+ \Delta \cos \theta )}{\left[-E(\Delta, \theta)\right]^{3/2}} (\eta -\sin \eta ) - t_{age}(\Delta, \theta) \right).
 \label{paramsoln}
\eea
The plus and minus signs give the solution for initially
expanding and initially contracting models respectively. Parameter $\eta$ is
purely real for closed solutions and purely imaginary for open
solutions. The distance to the nearest singularity is
\beq  
\label{tage}
t_{age} = \int_{b=0}^{b=a_0} \frac{db}{[{\dot b}^2]^{1/2}} = 
\frac{1}{H_0} \int_{y=0}^{y=1} \frac{dy}{\left[(1+\Delta \cos \theta) y^{-1} + E(\Delta, \theta) \right]^{1/2}}.
\eeq
The second equality uses \eqnref{velocityeq} and the substitution
$y=b/a_0$.

\subsection{Complex extension}
\label{complexextension}

To extend the above parametric solution to the complex plane, one
might guess the substitution $\Delta \rightarrow \Delta e^{i\phi}$
where $-\pi <\phi \leq \pi$ in \eqnref{paramsoln} and
\eqnref{tage}.  The physical limit is $\phi=0$. However, this leads to
two problems.  First, the integral for $t_{age}$ can have multiple
extensions that agree for physical $\phi=0$
but differ elsewhere including the negative real axis.
This is tied to the fact that the operations of integration and
substitution $\Delta \rightarrow \Delta e^{i\phi}$ do not commute
because of the presence of the square root in the expression for
$t_{age}$.  A second related problem is the presence of multiple
square roots in the parametric form for $t$. These give rise to
discontinuities along branch cuts such that one parametric form need
not be valid for the entire range of $\phi$, but instead the solution
may switch between different forms. Directly extending the
parametric solution is cumbersome.

However, the original differential \eqnref{cdiffeq} is manifestly
single-valued.  The equation can be integrated forward or backward
numerically to obtain the correct solution for complex $\dm$. One can
then match the numerical solution to the above parametric forms to
select the correct branch cuts.  This procedure was implemented to
obtain the form for all $\dm$ and $\theta$.  The main result is that
the solution space for all $\theta$ and $\dm$ is completely spanned by
complex extensions of the two real parametric forms which describe
initially expanding and contracting solutions.  The expressions for
${\bf t}_{age}$ and details are given in Appendix
\ref{complexformdetails}.  

The traditional textbook treatment relating physical cosmological
models with real $\Omega > 1$ and $\Omega < 1$ typically invokes a
discrete transformation $\eta \rightarrow i \eta$ in the parametric
forms and one verifies that this exchanges closed and open solutions.
However, starting from the second order differential equation it is
straightforward to use the same type of reasoning as
above to construct an explicit analytic continuation from one physical regime
to the other.

In addition, note that the differential equation and its solution
remain unchanged under the simultaneous transformations $\dm
\rightarrow -\dm$ and $\theta \rightarrow \theta + \pi$. Every complex
solution with $-\pi <\theta \le 0$ can be mapped to a
complex solution with $0 < \theta \le \pi$ and vice-versa. For
determining the radius of convergence and the time of validity
the whole disk of radius $|\dm|$ is searched for poles so it
suffices to consider a restricted range of $\theta$ to handle
all physical initial conditions.

\subsection{Poles}

The condition ${\bf b}=0$ signals the presence of a pole.
Inspection of the parametric form shows that this condition
can occur only when $\eta=0$ or $\eta = 2\pi$. The corresponding
time
\bea
{\bf t}(\dm, \theta)  &=& \left \{
 \begin{array}{cc}
 t_0 \pm\left( \frac{\pi}{H_0} \frac{(1+ \dm \cos \theta)}{ \left[-{\bf E}(\dm)\right]^{3/2}} - {\bf t_{age}}(\dm )\right)  & (\eta =2\pi)\\
  t_0 \mp {\bf t_{age}}(\dm ) &  (\eta = 0) \\
  \end{array}
  \right.
  \label{caseA} 
\eea
is immediately inferred. Since the independent variable $t$ is real
the transcendental equation
\beq
\myimag {\bf t}(\dm,\theta)=0
\label{transcendental}
\eeq
must be solved.
It is straightforward to scan the complex $\dm$ plane and calculate
${\bf t}$ to locate solutions.  Each solution gives a root of ${\bf b}=0$
and also implies the existence of a pole at the corresponding $\dm$.
Note that relying upon the parametric solutions is a far more efficient
method for finding the poles than integrating the complex
differential equations numerically. We have verified that both
methods produce the same results.

In practice, we fix $\theta$, scan a large area of the complex $\dm$
plane, locate all purely real ${\bf t}$ and save the $\{\dm, t\}$ pairs. These
are used to create a scatter plot of $|\dm|$ as a function of time
(hereafter the ``root plot''). Generally, the location of the poles
varies smoothly with $t$ and continuous loci of roots are readily apparent.
Finding $R_\Delta$ and $T(\Delta)$ follows as indicated in \figref{fig-defn}.

\section{Results from the complex analysis}
\label{Results}

Root plots were calculated for a range of angles $0 \leq \theta \leq
\pi $.  Since the root plots depend upon $|\dm|$ they are invariant under
$\theta \rightarrow \theta - \pi$ and this coverage suffices for all
possible top-hat models.  For the results of the
full survey in $\theta$ see Appendix
\ref{fullroots}.  The theoretical radius of convergence
$R_\Delta(t)$ and time of validity $T(\Delta)$ follow directly.

This section analyses the theoretical convergence for specific open
and closed models derived from the root plots. These estimates are
compared to the time of validity inferred by numerical evaluation of
the LPT series.  The range of models with limited LPT convergence is characterised. The concept of mirror models is introduced
to elucidate a number of interconnections between open and closed convergence.
The physical interpretation of roots introduced by the
complexification of the equations but lying outside the physical
range are discussed. Finally, the special case where the background
and the perturbation have the same big bang time is analysed.

\subsection{Open models}
\label{resultA}

\begin{figure}
\includegraphics[width=8cm]{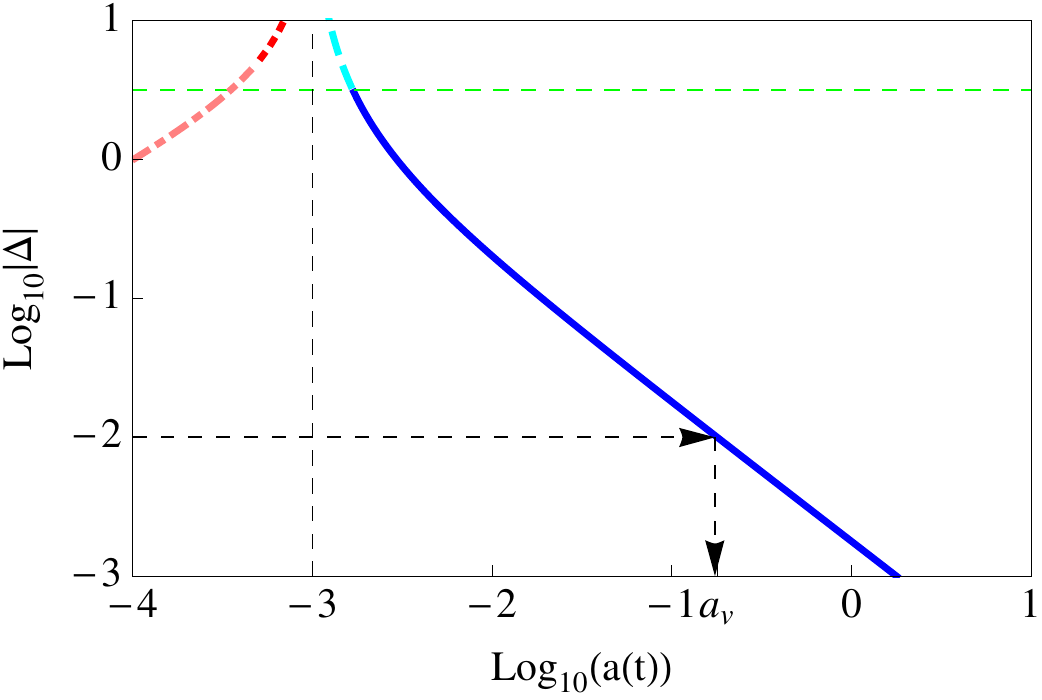}
\caption{$R_\Delta$ for $\theta = 2.82$ and $a_0=10^{-3}$ (vertical
  dashed line).  To determine the time of validity for LPT expansion
  with a given $\Delta$, move horizontally to the right of $a=a_0$
  following the dashed line with arrow and locate the first coloured
  line with ordinate equal to $\Delta$ and then move vertically down
  to read off the scale factor at the time of validity $a_v$. The
  specific case illustrated ($\Delta=10^{-2}$) matches that of the model
  with problematic convergence in \figref{openplot}. The time of
  validity is correctly predicted. The meaning of
  the colours is discussed in the text. Coloured version of the figure is available online.}
\label{openexplained} 
\end{figure}

\capfigref{openexplained} shows $R_\Delta(t)$ for
$\theta = 2.82$ and initial scale factor $a_0 = 10^{-3}$. All
$\Delta$ yield expanding open models for this $\theta$;
one choice corresponds to the model whose LPT series appeared in  
\figref{openplot} ($\Delta = 0.01$, $\theta =
2.82$, $a_0 = 10^{-3}$).  The x-axis is $\log a$ 
and is equivalent to a
measure of time. The y-axis is $\log |\Delta|$, i.e. the distance
from the origin to poles in the complex $\Delta$ plane. In principle,
future evolution may be limited by real or complex roots.  The blue solid
line and the red dotted line indicate real and complex roots of $\eta
= 2\pi$ respectively.  The cyan dashed and pink dot-dashed lines
indicate the real and complex roots of $\eta = 0$ respectively. 
Future evolution is constrained by real roots (blue and cyan) in
this example.

The time of validity is the first instance when a singularity appears
within the disk of radius $\Delta$ in the complex $\dm$ plane. For the
specific case, starting at ordinate $\Delta=10^{-2}$, one moves
horizontally to the right to intersect the blue line and then
vertically down to read off the scale factor $a_v= a[T(\Delta)] =
0.179$. The time of validity inferred from
the root plot agrees quantitatively with the numerical results in
\figref{openplot}.  

Appendix \ref{fullroots} presents a comprehensive set of results.
The time of validity is
finite for any open model. As expected, smaller amplitudes imply
longer times of validity. The poles do not correspond to
collapse singularities reached in the course of normal physical
evolution since the open models do not have any real future
singularities. A hint of an explanation is already present,
however. The green dashed line is $\delta_v=1$ (or $\Delta = 1/\sin \theta$)
at which point the root switches from $\eta= 2 \pi$
below to $0$ above. Such a switch might occur if varying the initial
velocity transposes an expanding closed model into a contracting
closed model. But it is expected to occur at $\delta_v=-1$ not $1$. The
open models are apparently sensitive to past and future singularities
in closed models with initial conditions that are transformed in a
particular manner. \S\ref{mirrormodels} explores this
interpretation in detail.

\subsection{Closed models}
\label{resultB }

\begin{figure}
\includegraphics[width=8cm]{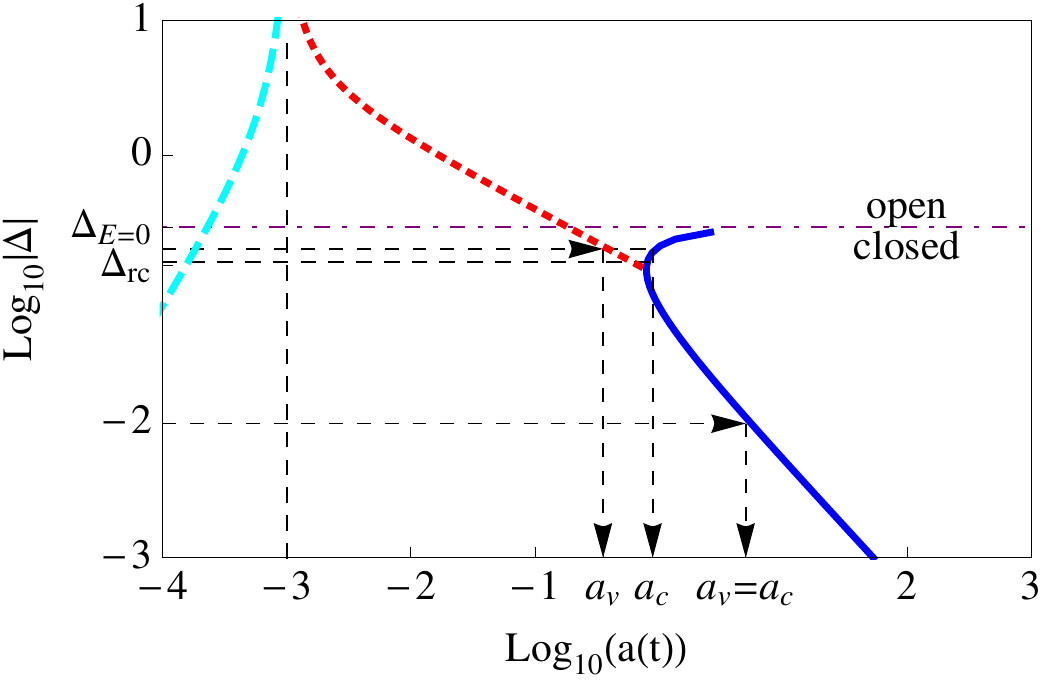}
\caption{$R_\Delta$ for $\theta = 0.44$ and $a_0=10^{-3}$.  The line
  $\Delta_E=0$ separates open and closed models. The scale factor at the
  time of validity is $a_v$. For closed models the scale factor at
  time of collapse is $a_c$.  Blue solid line and red small 
dashed line denote real and complex roots of $\eta=2 \pi$, respectively.   The
cyan dashed lines denotes the real roots of $\eta=0$.
When the first singularity 
  encountered is real, $a_v=a_c$, the time of validity is the
  future time of collapse. However, when the singularity is complex
  the time of validity is less than the actual collapse time. In the
  range $\Delta_{rc} < \Delta < \Delta_{E=0}$, there are closed models
  with $a_v<a_c$.
 }
\label{closedpredicted} 
\end{figure}

\capfigref{closedpredicted} presents $R_\Delta(t)$ for models with
$\theta = 0.44$ and $a_0=10^{-3}$. There are several new
features. Over the angular range $\theta_c^- < \theta < \theta_c^+$
the cosmology is closed for small $\Delta$ (see shaded region in
\figref{phaseplot} near $\Delta=0$). Conversely, a straight line
drawn from $\Delta=0$ within this angular range must eventually cross
the parabola $E=0$ except for the special case $\theta=0$.  Since the
velocity contribution to energy $E \propto \Delta^2$ while the density
contribution $\propto -\Delta$ it is clear that eventually $E>0$ as
$\Delta$ increases. The critical value, $\Delta_{E=0}$, is a function
of $\theta$.  Below the brown horizontal
dot-dashed line in \figref{closedpredicted} the models are closed,
above they are open (line labelled $\Delta=\Delta_{E=0}$).

The root plot has, as before, blue solid and red dotted lines denoting
the distance to real and complex $\dm$ poles, respectively, for $\eta
= 2 \pi$. The cyan dashed line denotes real roots for $\eta = 0$ and
does not restrict future evolution.

\begin{figure}
\includegraphics[width=15cm]{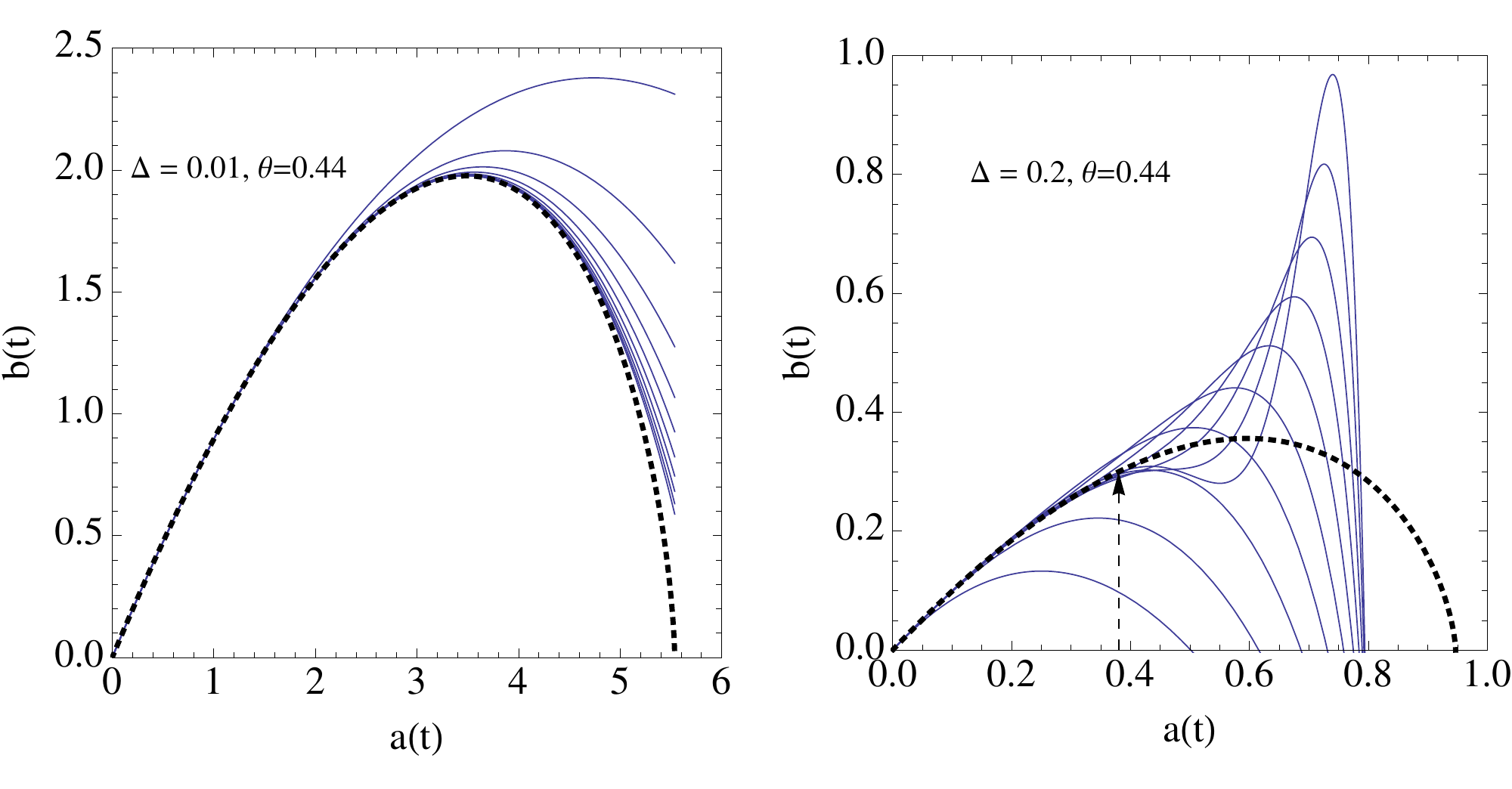}
\caption{The exact solution (black, dashed) and LPT expansions of
  successively higher order (blue) for two expanding, closed models
  with $\theta = 0.44$. The left hand panel has $\Delta=0.01$.
  LPT converges to the exact solution at all times up to the
  singularity at $a=5.5$.  The right hand panel has $\Delta=0.2$. LPT
  does not converge beyond $a=0.38$.}
 \label{closedplot}
\end{figure}

For small $\Delta$ real roots determine the time of validity. These
roots correspond exactly to the model's collapse time. In other words, the
time of validity is determined by the future singularity.  For
example, for $\Delta= 0.01$, the root plot predicts that a series
expansion should be valid until the collapse at $a = 5.5$ denoted by
``$a_v=a_c$'' on the x-axis . This prediction is confirmed 
in the left hand panel of
\figref{closedplot}. The root diagram is consistent with the
qualitative expectation that small overdensities should have long times
of validity because collapse times are long:
$\lim_{\Delta \to 0} T(\Delta) \to \infty$.

As $\Delta$ increases from very small values, i.e.
successively larger initial density perturbations,
the collapse time decreases.  Eventually
the velocity perturbation becomes important so that
at $\Delta = \Delta_{rc}$ a minimum in the collapse time is reached. For
$\Delta_{E=0} > \Delta > \Delta_{rc}$ the collapse time increases
while the model remains closed. As $\Delta \to \Delta_{E=0}$ the
collapse time becomes infinite and the model becomes critical. 
All models with $\Delta > \Delta_{E=0}$ are open.

The root diagram shows that for $\Delta > \Delta_{rc}$, the time of
validity is determined by complex not real $\dm$ for $\eta = 2\pi$. 
Closed models with $\Delta_{rc} < \Delta <
\Delta_{E=0}$ have a time of validity less than the model collapse
time.  For example, for $\Delta = 0.2$, the collapse occurs at $a=
0.94$ but convergence is limited to $a \le 0.38$. This prediction is
verified in the right panel of \figref{closedplot}.

The convergence of LPT expansions for some
closed models is limited to times well before the future singularity.
This general behaviour is observed for $\theta_c^- < \theta <
\theta_c^+$ and $\Delta_{rc} < \Delta < \Delta_{E=0}$ where both
$\Delta_{rc}$ and $\Delta_{E=0}$ are functions of $\theta$.
Appendix \ref{fullroots} provides additional details.

\subsection{Mirror models, real and complex roots}
\label{mirrormodels}

The parametrization of the perturbation in terms of $\Delta>0$ and
$-\pi < \theta \le \pi$ and the complexification of $\Delta \to \dm$
can give rise to poles anywhere in the complex $\dm$ space.  When
$R_\Delta$ is determined by a pole along the real positive axis, a
clear interpretation is possible: the future singularity of the real
physical model exerts a dominant influence on convergence. LPT
expansions for closed models with $\Delta < \Delta_{rc}$ are limited
by the future collapse of the model and are straightforward to
interpret.

The meaning of real roots for open models is less clear cut.  The
roots determining $R_\Delta$ at large $t$ are negative real and
small in magnitude.  Negative $\Delta$ lies outside the parameter
range for physical perturbations taken to be $\Delta>0$.  Nonetheless
the mapping ($\Delta,\theta$) $\to$ ($-\Delta, \theta \pm \pi)$
preserves ($\delta$, $\delta_v$) and the original equations of motion.
The poles of the models with parameters $(\Delta, \theta)$ and
$(\Delta, \theta\pm \pi)$ are negatives of each other. Let us
call these ``mirror models'' of each other. 

For infinitesimal $\Delta$ if the original model is open then the
mirror model is closed. \capfigref{phaseplot} shows that the
$\Delta_{E=0}$ line has some curvature (in fact, it is a parabola)
whereas the mirror mapping is an exact inversion through
$\Delta=0$. Small $\Delta$ points are mapped between open and closed;
large $\Delta$ points may connect open models to other open models.

\begin{figure}
\includegraphics[width=15cm]{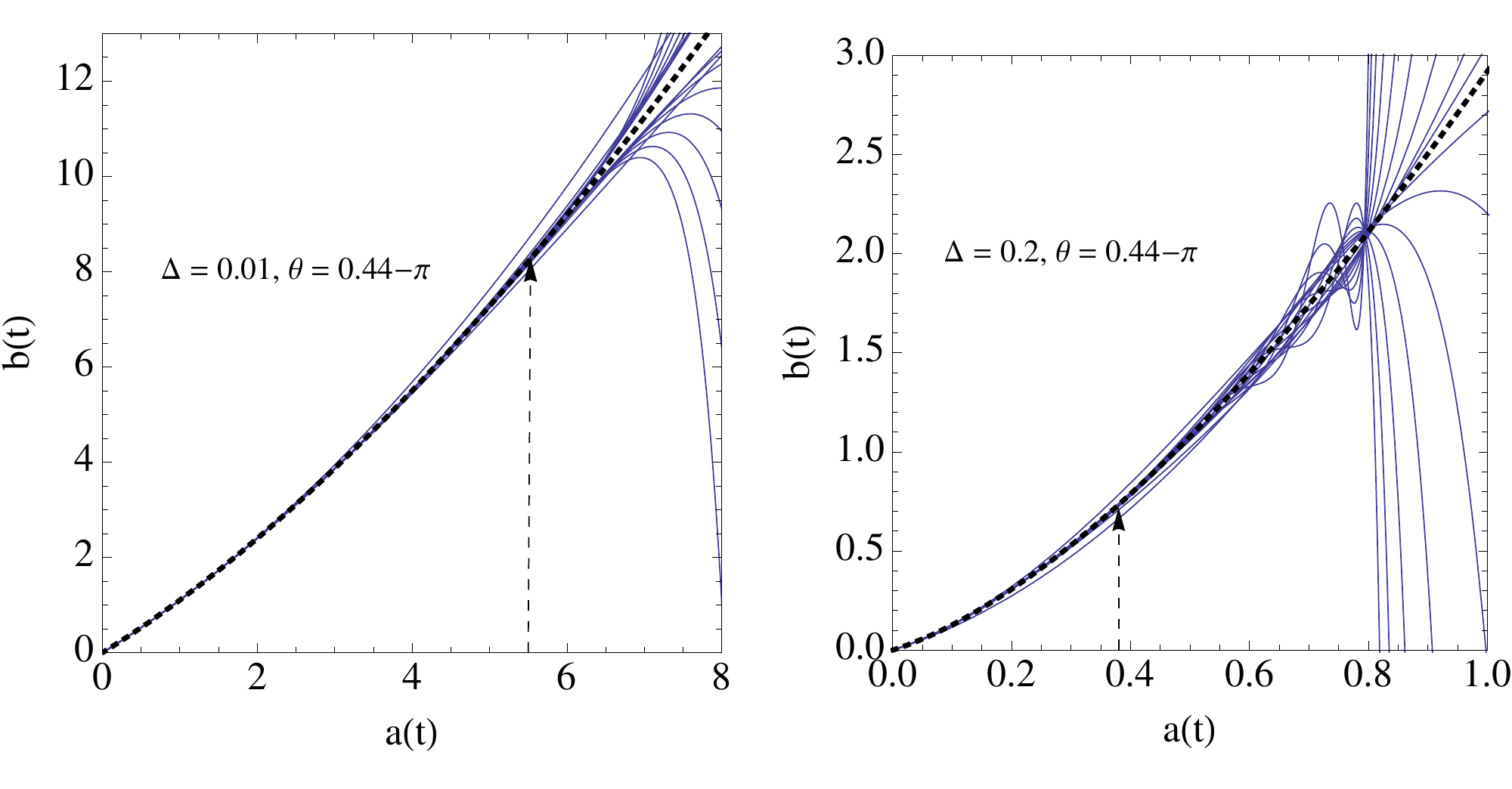}
\caption{Mirror models of the closed models of
  \figref{closedplot}.  Each graph shows the exact solution (black,
  dashed) and the LPT expansion to successively higher orders (blue) of
  one mirror model. The original model and the mirror have the
  same time of validity for the LPT expansion.}
 \label{thirdquad}
\end{figure}

If the original model is open with limiting pole which is negative
real of small magnitude then it corresponds to a future singularity
of the closed mirror model.  For example, the closed model with
parameters $(\Delta= 0.01, \theta = 0.44)$ in the left panel of
\figref{closedplot} and the open model with parameters $(\Delta= 0.01,
\theta = 0.44 -\pi)$ shown in the left panel of \figref{thirdquad}
are mirrors. The time of the validity of the open model equals the
time to collapse of its closed mirror.

The notion of mirror models explains other features of the root
diagrams. The time of validity of open models was previously discussed
using \figref{openexplained} ($\theta=2.82$).  The blue solid line
indicated real roots. Such roots are the future singularities of
closed mirror models lying in the fourth quadrant along $\theta =
2.82-\pi=-0.32$.  As $\Delta$ increases the sequence of mirror models
crosses the $\delta_v = -1$ line (the horizontal dashed line) to
become initially contracting cosmologies and, in our labelling, the
future singularity switches from $\eta=2 \pi$ to $\eta = 0$. This
explains the switch in root label from blue solid to cyan dashed
seen in \figref{openexplained}, which occurs at $\delta_v=1$
in the original model.

The symmetry of the mirroring is not limited to cases when $\dm$
is real. It applies for complex $\dm$, too. For example,
the models in the right panels of \figrefs{closedplot} and
\figrefbare{thirdquad} are mirrors of each other.  Their time of
validity is the same and determined by complex roots which are
negatives of each other. These singularities are non-physical and have
no interpretation in terms of the collapse of any model yet they
limit the LPT convergence in the same way.

\begin{figure}
\includegraphics[width=8cm]{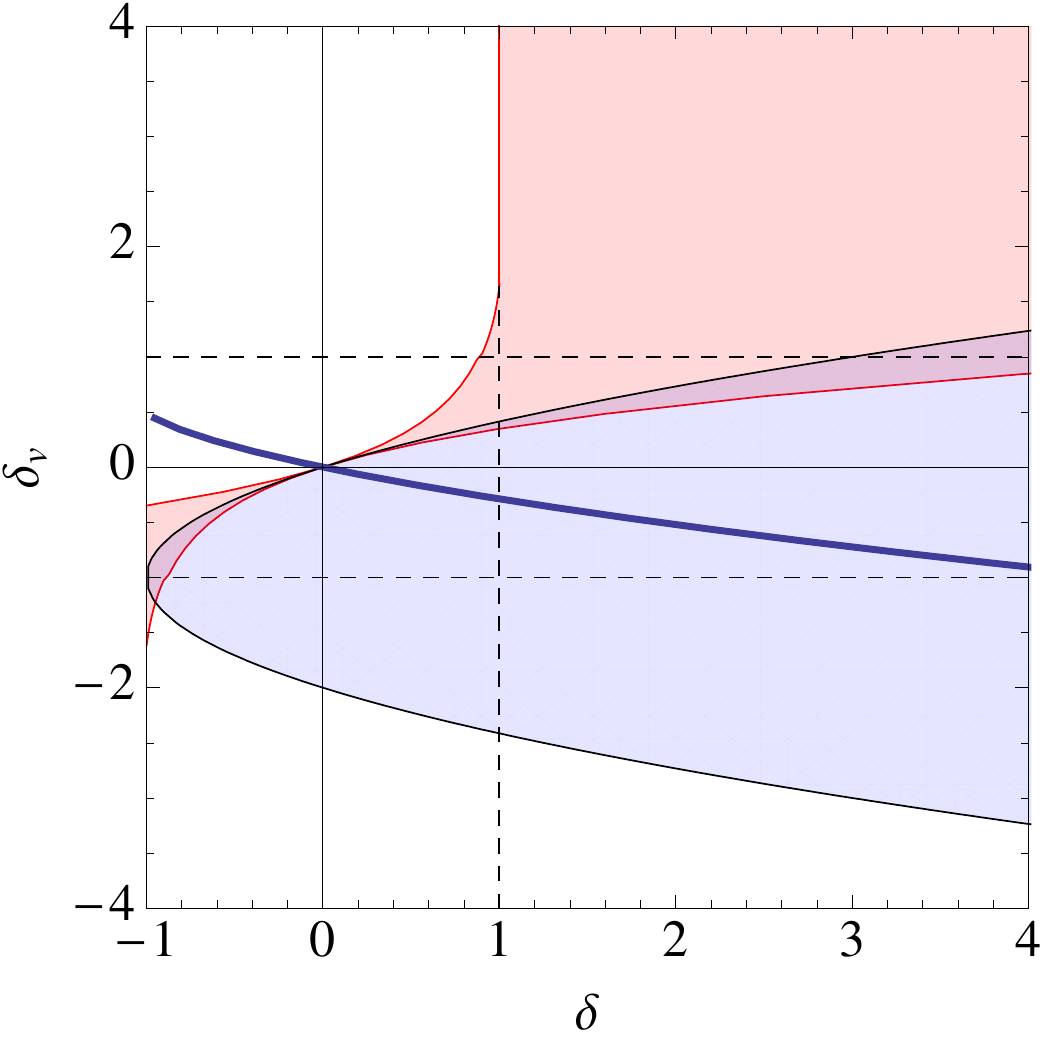}
\caption{The red shaded region denotes part of phase space where complex roots play a role. The solid blue line represents the initial conditions which correspond to the background and perturbation having the same big bang time. The black solid parabola separates the closed and open models. Coloured version online.}
\label{crootphasespace}
\end{figure}

\capfigref{crootphasespace} shows the areas of phase space where
complex roots determine the time of validity in light red. The area
within the parabola (light blue) contains closed models. Most of
the light blue region has a time of validity determined by real roots,
i.e. the time to the future singularity. The area with both light blue
and red shading encompasses closed models with the unexpected feature
that the time of validity is less than the time to collapse.

The area outside the parabola contains open models. The time of
validity of the unshaded region is determined by real roots. The
original observation of LPT's non-convergence for an underdensity
(\citealt{sahni_accuracy_1996}) is an example that
falls in this region. For small amplitude perturbations the time of
validity is simply related by mirror symmetry to the occurrence of
future singularities of closed models. The right hand plot in
\figref{thirdquad} is an example of an open model with time of validity
controlled by complex roots (red shading outside the parabola).

\begin{figure}
\includegraphics[width=16cm]{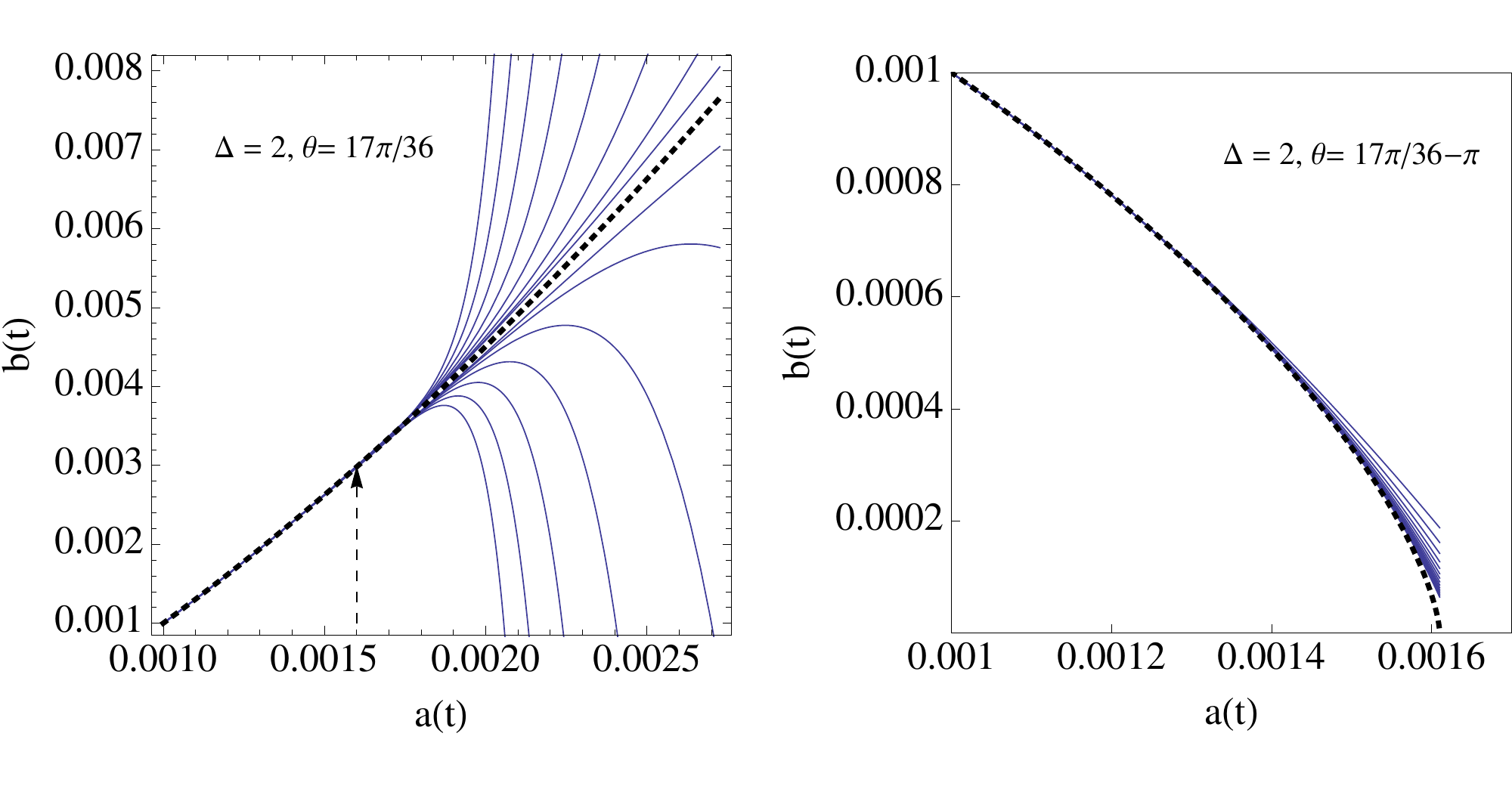}
\caption{Two open models which are mirrors of each other. Each plot
  shows the exact solution (black, dashed) and the LPT series
  expansion to successively higher orders (blue).  The left panel is
  an initially expanding, open model whose convergence is limited to
  scale factors less than $a_v = 0.0016$ (arrow). The right panel
  shows the initially contracting mirror model whose bang time at $a_v =
  0.0016$ is responsible for the limitation.}
\label{opentoopen}
\end{figure}

\begin{figure}
\includegraphics[width=10cm]{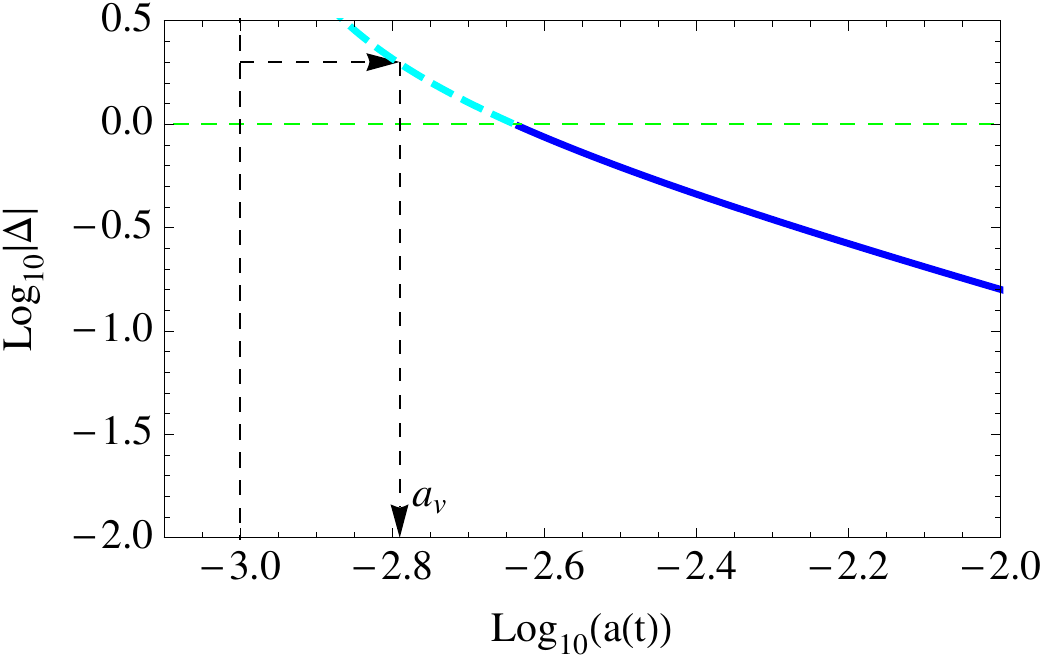}
\caption{$R_{\Delta}$ for $\theta = 17 \pi/36$ and
  $a_0 = 10^{-3}$. The blue solid and cyan dashed lines denoted real roots
with $\eta = 2 \pi$ and $\eta = 0$ respectively. For
  $\Delta=2$, the time of validity is set by the root with $\eta =0$,
  which is the bang time of the mirror
  model with $\theta = 17 \pi/ 36 -\pi$. See \figref{opentoopen} for
  the evolution of both models. }
\label{opentoopenroot}
\end{figure}

Finally, some open models (especially those with large $\Delta$) have
mirrors that are open models.  \capfigref{opentoopen} shows mirror
models ($\Delta =2$, $\theta = 17 \pi /36$) and ($\Delta =2$, $\theta
= 17 \pi/36 -\pi$).  These are initially expanding and contracting
solutions respectively.  The root plot in
\figref{opentoopenroot} predicts that the series is valid until $a_v
=0.0016$. The real root with $\eta=0$ (cyan line) sets the time
of validity and corresponds to the bang time (the future
singularity) of the initially
contracting model.

In all cases, the analysis correctly predicts the convergence of the LPT
series.

\subsection{Zeldovich and equal bang time models}
\label{discussion}

The large expanse of phase space shaded light red in 
\figref{crootphasespace} suggests that complex roots should
play a ubiquitous role in LPT applications but the situation
is somewhat more subtle.
For good physical reasons purely gravitational cosmological
calculations often start with
expanding, small amplitude, growing modes at a
finite time after the big bang. The absence of decaying modes implies that
the linearized perturbations decrease in the
past \footnote{Our analysis is restricted to the case of
initially expanding models, i.e. near $\Delta=0$. 
For initially contracting closed models, similar physical
arguments motivate a consideration
of the behaviour near the initial singularity (not the future
bang time). For initially contracting
open models the epoch of interest is $t \to -\infty$.
These models have large $\Delta$ and are not described by the
linear limit discussed in the text.}. A non-linear version of this
condition is that the perturbation amplitude is exactly zero at $t=0$.
The same condition can be formulated as
``the background and the perturbation have the same big bang time'' or
``the ages of the perturbation and the background are identical.'' The
condition is
\beq 
\frac{1}{H_0} \int_{y=0}^{y=1} \frac{dy}{\left[(1+\Delta \cos \theta) y^{-1} + E(\Delta, \theta) \right]^{1/2}} = \frac{2}{3H_0}.
\label{Zeldovich}
\eeq
This is a nonlinear relationship between the two initial
parameters $\Delta$ and $\theta$ which is shown by a thick blue line
on the phase space diagram in \figref{crootphasespace}. 
We have adopted the name ``Zeldovich'' initial conditions for the top-hat models
that satisfy the equal bang time relation. There are a variety of
definitions for Zeldovich initial conditions given in the
literature. Generally, these agree at linear order. This one has the
virtue that it is simple and easy to interpret.
Note that the blue curve does not intersect the region of phase space
where complex roots occur except, possibly, near $\dm=0$.

In the limit of small $\Delta$ \eqnref{Zeldovich} becomes 
\beq \Delta (3 \sin \theta - \cos \theta) = 0.  
\eeq 
The solutions are $\theta=\theta_{Z\pm}$ where $\theta_{Z+} = 2.82$
and $\theta_{Z-} = \pi-\theta_{Z+}=-0.32$.  The second quadrant
solution $\theta_{Z+}$ corresponds to open models
while its mirror in the
fourth quadrant $\theta_{Z-}$ to closed models. Only when $\Delta
\to 0$ can complex roots approach the loci of Zeldovich initial
conditions but they intersect only in the degenerate limit.

In the next section, we will show that points starting close to the Zeldovich curve continue to stay near it as they move through phase space. Such models have real, not complex, roots. This implies that closed systems along the curve always have a convergent series solution. Hitherto, LPT convergence has been studied only for initial conditions close to the Zeldovich curve. This is why problems have been noted only in the case of voids. The existence of the complex roots is a new finding. All of the above is based on the spherical top-hat model which has a uniform density.

As emphasized above, there are good physical motivations for adopting Zeldovich-type initial conditions. The fact that cosmological initial conditions must also be inhomogeneous (i.e. Gaussian random fluctuations) is not captured by the top-hat model. One can imagine two extreme limiting cases for how the simple picture of top-hat evolution is modified. If each point in space evolves independently as a spherical perturbation then at any given time one expects to find a distribution of points along the Zeldovich curve. As time progresses this distribution moves such that the underdense points cluster around the attracting point $(-1,0.5)$ and overdense points move towards collapse. The distribution of initial density and velocity perturbations yields a cloud of points in phase space but complex roots never play a role because nothing displaces individual points from the Zeldovich curve. Each moves at its own pace but stays near the curve. Alternatively, it is well known that tidal forces couple the collapse of nearby points. These interactions amplify the initial inhomogeneities leading to the formation of pancakes and filaments. As time progresses motions transverse to the Zeldovich curve will grow. If these deviations are sufficient they may push some points into areas with complex roots. In a subsequent paper, we will explore these issues for general inhomogenous initial conditions. 

\section{LPT re-expansion}
\label{sec_extension}

To overcome the constraints above, an
iterative stepping scheme that respects the time of validity is
developed for LPT. The initial parameters at the first step
determine the solution for some finite step size. The output at the
end of the first step determines the input parameter values for the
next step and so on.

\subsection{The Algorithm}
\label{algorithm}

Choose the background ($a_0$, $H_0$, $\Omega_0=1$, $Y_0$) and
the perturbation ($b_0=a_0$, $H_{p0}$, $\Omega_{p0}$, $X_0$)
at initial time $t_0$. The perturbed model is fully characterised by
$H_{p0}$ and $\Omega_{p0}$ or by $\delta_0=\rho_{p0}/\rho_0 - 1$ and 
$\delta_{v,0}=H_{p0}/H_0 - 1$ or by $\Delta_{0}$ and $\theta_0$.
Extra subscripts have been added to label steps.

LPT converges for times $t < T(\Delta_{0},\theta_{0})$.
Use LPT to move forward to time $t_*$ satisfying $t_0 <
t_* < T(\Delta_{0},\theta_{0})$. At $t_*$, the background and
perturbed scale factors and time derivatives are $a_*$, $b_*$, ${\dot a}_*$, and ${\dot b}_*$. The fractional density and velocity perturbations  
with respect to the background are
\bea 
\delta_*&=&(1+\delta_0)\left(\frac{a_*}{b_*}\right)^3 -1\\
\delta_{v,*}&=&\frac{{\dot b}_*/b_*}{{\dot a}_*/a_*}-1 .
\eea
Re-expand the perturbation around the background
model as follows. First, let the time and Lagrangian
coordinate for the background (inner edge of the unperturbed sphere)
be continuous: $t_1 = t_*$ and $Y_1=Y_0$. These imply
$a_1=a_*$ and ${\dot a}_1 = {\dot a}_*$, i.e. the scale factor and Hubble
constant for the background are continuous.

At the beginning of the first step we assumed $a_0=b_0$.  This is
no longer true at the end of the first step. Define a new Lagrangian
coordinate $X_1=X_0 b_*/a_*$, new scale factor $b_1 =a_*$,
and new scale factor derivative ${\dot b}_1 = {\dot  b}_* a_*/b_*$.
These definitions leave the physical edge of the
sphere and its velocity unaltered
\bea
r_{physical,*} & = & b_* X_0 = b_1 X_1 \label{Xrescale}\\
{\dot r}_{physical,*} & = & {\dot  b}_* X_0 = {\dot b}_1 X_1 .
\eea
The re-definitions relabel the fluid elements with a new set
of Lagrangian coordinates and re-scale the scale factor.
The perturbation parameters are unchanged $\delta_1 = \delta_*$ and
$\delta_{v,1}=\delta_{v,*}$ because physical quantities are
unmodified. Consequently, $\Delta_{1}=\Delta_*$ and $\theta_1=\theta_*$.

\subsection{Flow dynamics in the phase space}
\label{flow}

To examine how Lagrangian re-expansion works 
consider how the Lagrangian parameters $\Delta$ and $\theta$ would
vary if they were evaluated at successive times over the course of a
specific cosmological history. Let $\delta(t)$ and $\delta_v(t)$ be
defined via \eqnref{defdeltas} and apply the second-order equations of
motion \eqnrefs{secondorder} and \eqnrefbare{eqn:unperturbedscalefactor}
to derive the coupled first-order system
\bea
\frac{d \delta}{dt} &=& - \frac{2}{t} \delta_v (1+\delta) \\
\frac{d \delta_v}{dt} &=&  \frac{1}{3 t} \left \{ (1+\delta_v) (1- 2\delta_v) - (1+\delta) \right \}
\label{dynamics}
\eea
where all occurrences of $\delta$ and $\delta_v$ are functions of
time.  From $\delta(t)$ and $\delta_v(t)$ one infers the
parameters, $\Delta(t)$ and $\theta(t)$. These have
the following simple interpretation: a Lagrangian treatment starting
at time $t'$ has $\Delta=\Delta(t')$ and $\theta=\theta(t')$ in the
LPT series.

\begin{figure}
\includegraphics[width=16cm]{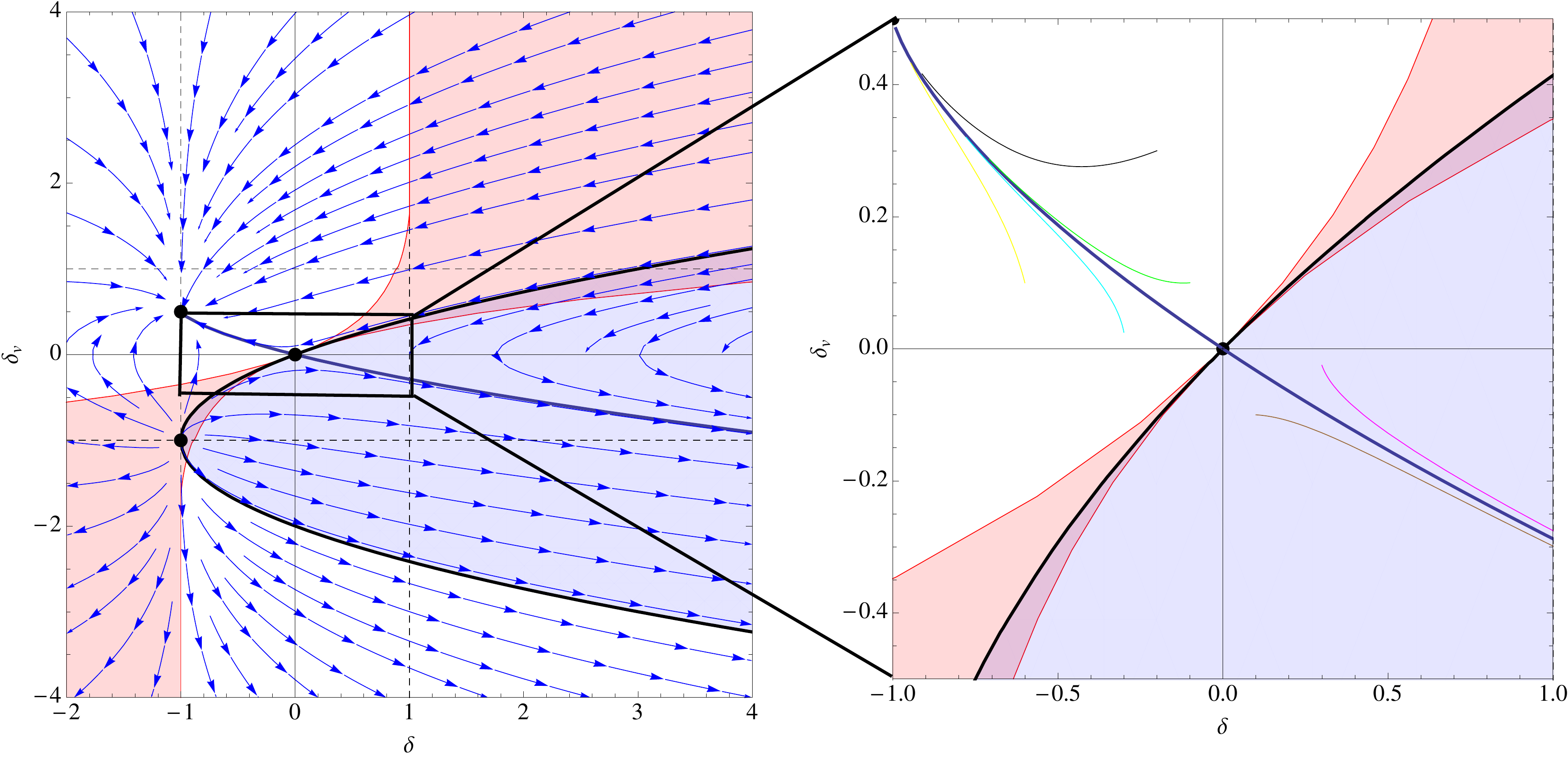}
\caption{The left panel shows streamlines of the flow described by
  \eqnref{dynamics}. The colour coding of the plot is same as
  \figref{crootphasespace}. The right panel zooms in on the area
  near the origin which is where all models are located at
  sufficiently early times. At late times, open models move away from
  the origin towards the attracting fixed point at $(\delta,
  \delta_v)= (-1,0.5)$. The attraction to the Zeldovich solution is
  shown for a set of initial conditions (yellow, cyan, green and black
  lines) that begin near but not on the critical trajectory.  Closed
  models move out to infinity along the fixed big bang time curve. Coloured version online. }
\label{phasedynamics} 
\end{figure}

Since the system is autonomous it reduces to a simple
flow in phase space.
The flow has three fixed points at $(\delta,\delta_v)=(0,0)$, the unperturbed,
background model, $(-1, -1)$, a vacuum static model,
and $(-1, 0.5)$, a vacuum expanding model.
Linearizing around $(0,0)$ shows it
is a saddle fixed point. The tangent to the $E=0$
curve at the origin is the attracting direction and the tangent to
the equal big bang curve is the repelling direction. The fixed
point at $(-1,0.5)$ is a degenerate attracting node and that at
$(-1,-1)$ is an unstable node. The flow vectors are plotted in the left
panel of \figref{phasedynamics}. The blue shaded region
indicates closed models and red shaded region indicates models where
complex roots limit the time of validity for LPT.

Note that the flow lines smoothly cover the whole phase space. The
interpretation is 
that the continuous relabelling of Lagrangian coordinates and
re-scaling of the scale factor has the potential to
overcome the convergence limitations
discussed thus far. Otherwise one might have seen
ill-defined or incomplete flows or flows that were confined
to a given region.

\subsubsection{Asymptotic limits of open and closed models}

The right panel of \figref{phasedynamics} zooms in on the area near
the origin. Initial points that correspond to open models starting
near the origin approach the Zeldovich curve and asymptotically
converge to the strong attractor at $(\delta, \delta_v) = (-1,0.5)$.

Closed models collapse and the density $\delta \rightarrow \infty$. In
the asymptotic limit, the solution to (\ref{dynamics}) is given by
$\delta \sim \delta_v^2 + K$ with integration constant $K$. 
From \figref{phasedynamics}, the flow
lines of closed models that start in the vicinity of the origin
trace a parabolic path that is parallel and essentially equivalent to the
Zeldovich curve.

The flow shows where re-expansion is needed. Closed model flow lines
that start near the origin never pass through the red shaded region
where complex roots play a role; the time of validity equals the time
to collapse and no re-expansion is needed.  However, closed models
that originate in the red region must be re-expanded. The flow suggests
that they eventually move into the blue region. So even though a closed
model may initially have an LPT series with limited convergence, re-expansion
makes it possible to move into the part of phase space
where a single step suffices to reach collapse.

\subsection{Finite steps and feasibility}

This section and the next examine the feasibility of extending a solution from recombination to today. The results will be applied to fully inhomogeneous evolution in future paper.

Let the asymptotic time of validity for an open model be expressed in
dimensionless form $\chi = \lim_{t\to\infty}H(t)
T(\Delta(t),\theta(t))$.  Here, $\Delta \to \sqrt{5/4}$ and
$\theta \to \tan^{-1}(-1/2) = 2.677$ and $T(\Delta,\theta)$ is determined by
the future time to collapse of the closed mirror
model. The result is $\chi = 2.62$ (numerical results in Appendix
\ref{fullroots}), the time of validity is proportional to the
characteristic age of the background and individual steps grow
larger and larger.

\begin{figure}
\includegraphics[width=8cm]{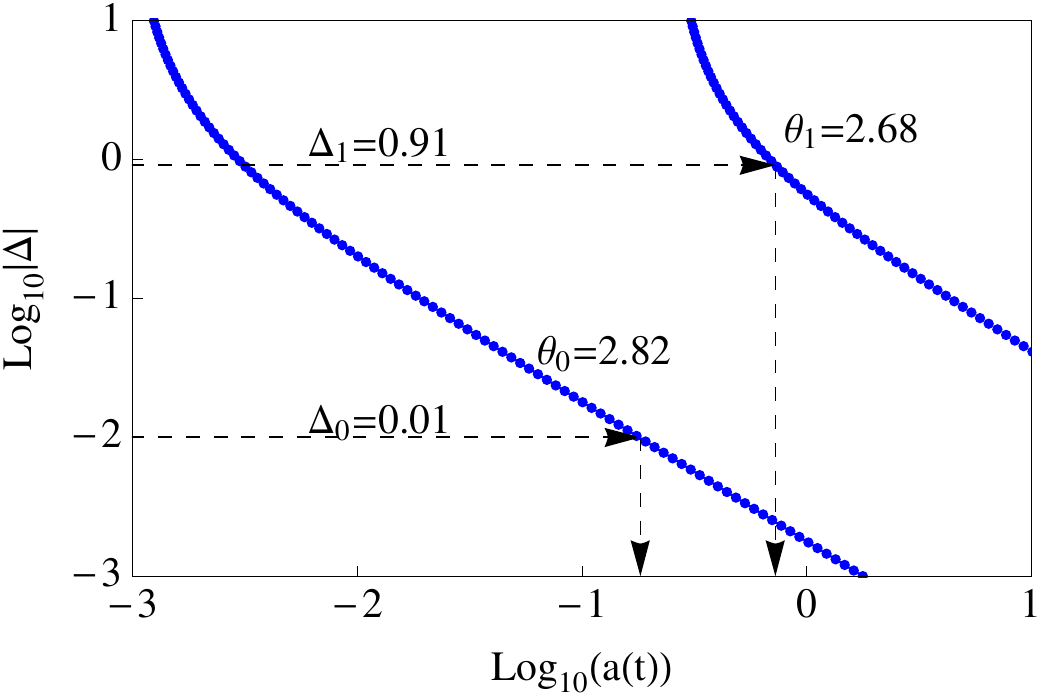}
\caption{Extending the time of validity of LPT. The first step has
$\Delta_0=10^{-2}$ and $\theta=2.82$ and implies scale factor at the
time of validity $a_v=0.179$. Incrementing by half the allowed step
gives initial conditions for the second step 
$(\Delta_1, \theta_1)=(0.91,2.68)$. Note that the
new time of validity has increased. }
\label{timestep} 
\end{figure}

An example shows that
the basic effect can be seen even before the asymptotic regime is achieved.
\capfigref{timestep} sketches the first two steps where the
assumed model parameters at the first step are $(\Delta_0, \theta_0) =
(0.01, 2.82)$. The scale factor at the
time of validity is $a
= 0.179$. A step with half the allowed increment in time is taken and the
system is re-initialised. The re-initialisation implies ($\Delta_1,
\theta_1) = (0.91,2.68)$ or
$(\delta_1, \delta_{v,1}) = (-0.82,0.4)$. Afterwards the new time of
validity is larger in this example.

\begin{figure}
\includegraphics[width=10cm]{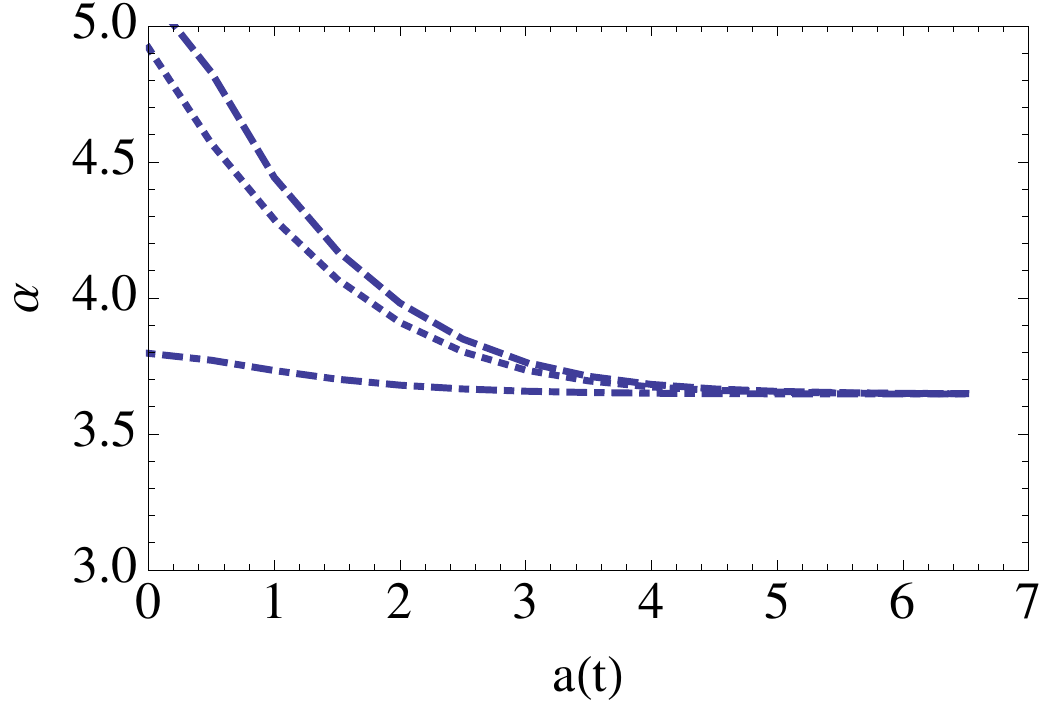}
\caption{ The ratio of successive times of validity ($\alpha$) vs. $a(t)$. The dashed, dot-dashed and dotted lines indicate three initial starting points $(0.5, 0.5), (0,1), (-0.2, 0.2)$ respectively. The ratio converges to about 3.6 and
the time of validity increases geometrically with $N$. }
\label{alpharatio} 
\end{figure}

The feasibility of the re-expansion
scheme can be examined by evaluating the ratio of the time of validity
before ($T$) and after ($T'$) a step
\beq 
\alpha = \frac{T'}{T} .
\eeq
\capfigref{alpharatio} shows $\alpha$ evaluated along the continuous flow as a
function of scale factor for three different starting initial
conditions. Since $\alpha > 3$ at all times, starting at initial time $t_i$
the time after $N$ steps is roughly $t \sim \alpha^N t_i > 3^N t_i$.

Consider, for example, the number of steps needed to extend an open
solution from recombination to today. Let $t_f$ ($t_i$) be the final
(initial) time of interest where $t_f/t_i \sim a_f/a_i \sim 10^{4.5}$.
Estimating $\alpha=3$ implies $N \sim \log_3
10^{4.5} \sim 10$ steps are needed. This numerical result for $N$ is 
an overestimate and one can do better. It is important to recall that
it based on an arbitrarily high order expansion which
achieves an exact solution. If one is limited to
calculations of finite Lagrangian order and imposes a maximum
numerical error at the end of the calculation then more than $N$ steps
may be required. At least $N$ steps are needed for series
convergence and more than $N$ steps may be needed for error control.

One can extend any open model to an arbitrary future time while
respecting the time of validity of the LPT series. The number of
steps is governed by a geometric progression.

One can also extend any closed model to the future singularity while
respecting the time of validity of the LPT series. Only a single step
is needed for a closed model when the root is real (blue shaded region
of \figref{phasedynamics}). When it is complex (the region shaded both
blue and red) the model flows first toward the node at $(0,0)$
($\Delta$ decreases) and ultimately reaches the region of real
roots. Multiple steps will generally be necessary to escape the region
of complex roots.  An approximate fit (\eqnref{timfit}) shows that
$\chi \sim T(\Delta, \theta) H(t) \propto \Delta^\beta$ for small
$\Delta$ where $\beta<-2.5$.  Both $\chi$ and the time of validity
increase as the node is approached.  Time advances at least as quickly
as a geometric progression and this is analogous to the manner in
which the open model steps towards its limit point.
However, unlike the open case, once the trajectory crosses into the
blue region (assuming it does not lie exactly on the unstable
attracting trajectory) a single final step is needed. The specific
number of steps will depend upon the starting initial conditions but
will be small because of the property of geometric progression.

\subsection{Demonstrative examples}

LPT re-expansion can solve the problematic convergence in previously
analysed open and closed models.

Open models have asymptotic values of $\Delta$ and $\theta$ and simple
evolution.  The first section below includes numerical results that
provide a practical demonstration of the success of LPT re-expansion
in this case. Convergence as Lagrangian order increases and/or time
step size decreases is observed qualitatively.

Closed models have a somewhat more complex behaviour (before and after
turnaround). The second section provides both a qualitative and
quantitative discussion of convergence. The scaling of the leading
order error and the time step control which are derived are of general
applicability.

\subsubsection{Open model}

\begin{figure}
\includegraphics[width=16cm]{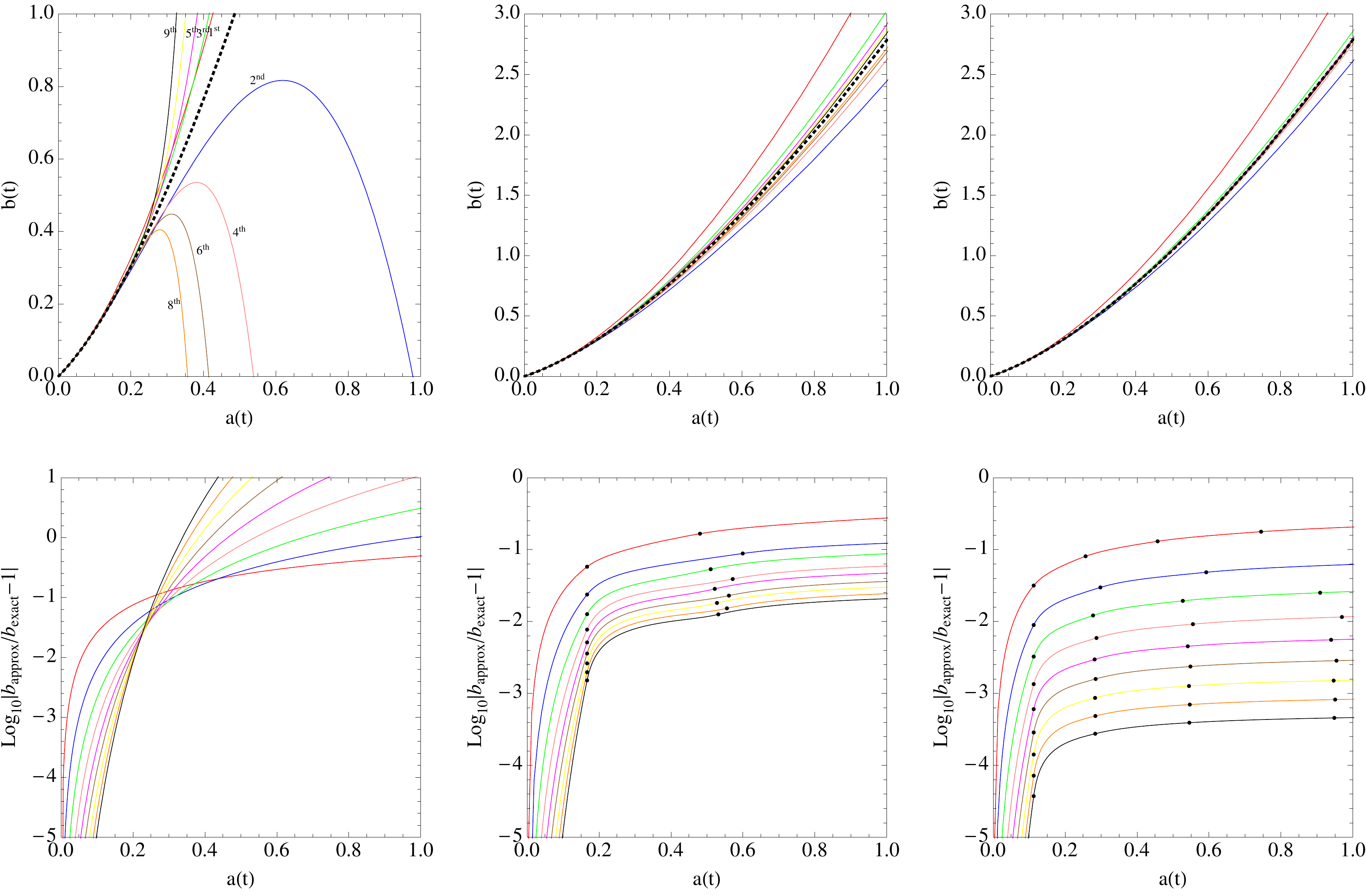}
\caption{LPT re-expansion of an open model with $\Delta_{0} = 0.01$
  and $\theta_0= 2.82$.  The top three figures show the scale factor
  for the same initial conditions calculated with one step (left),
  three steps (middle) and five steps (right). The black dots indicate
  the position of the time steps. In the middle and right panels, the
  solution was advanced $9/10$ and $1/2$ the allowed time of validity,
  respectively. The bottom figures show the errors for all LPT
  approximations to $b(t)$ including the unphysical negative ones.  The
  order of the LPT expansion are colour-coded according the top left
  figure. The single step expansion does not respect the time of
  validity whereas both the three and six step examples do. The
  original expansion does not converge over the full time range
  whereas the re-expansions do. Coloured version online.}
\label{extension}
\end{figure}

\capfigref{extension} investigates the effect of time step and
order on the evolution of the open model introduced in
\figref{openplot} ($\Delta = 0.01$, $\theta= 2.82$, $a_0=10^{-3}$). 
The series convergence breaks down at $a= 0.179$.  The left panel
shows an attempt to take a single step to $a=1$ using successively
higher LPT series orders. As expected, higher order terms do not
improve the accuracy of the description because the time of validity
is violated. The middle panel employs three steps to reach $a=1$, each
respecting the time of validity.  Now the LPT series with higher order
improves the accuracy just as one desires.  The right panel employs
six steps to reach $a=1$, each respecting the time of validity. Again,
higher order improves the description. Note that
more frequent re-expansion, i.e. smaller steps in time,
improve the errors at fixed LPT order.

\subsubsection{Closed model}

\begin{figure}
\includegraphics[width=16cm]{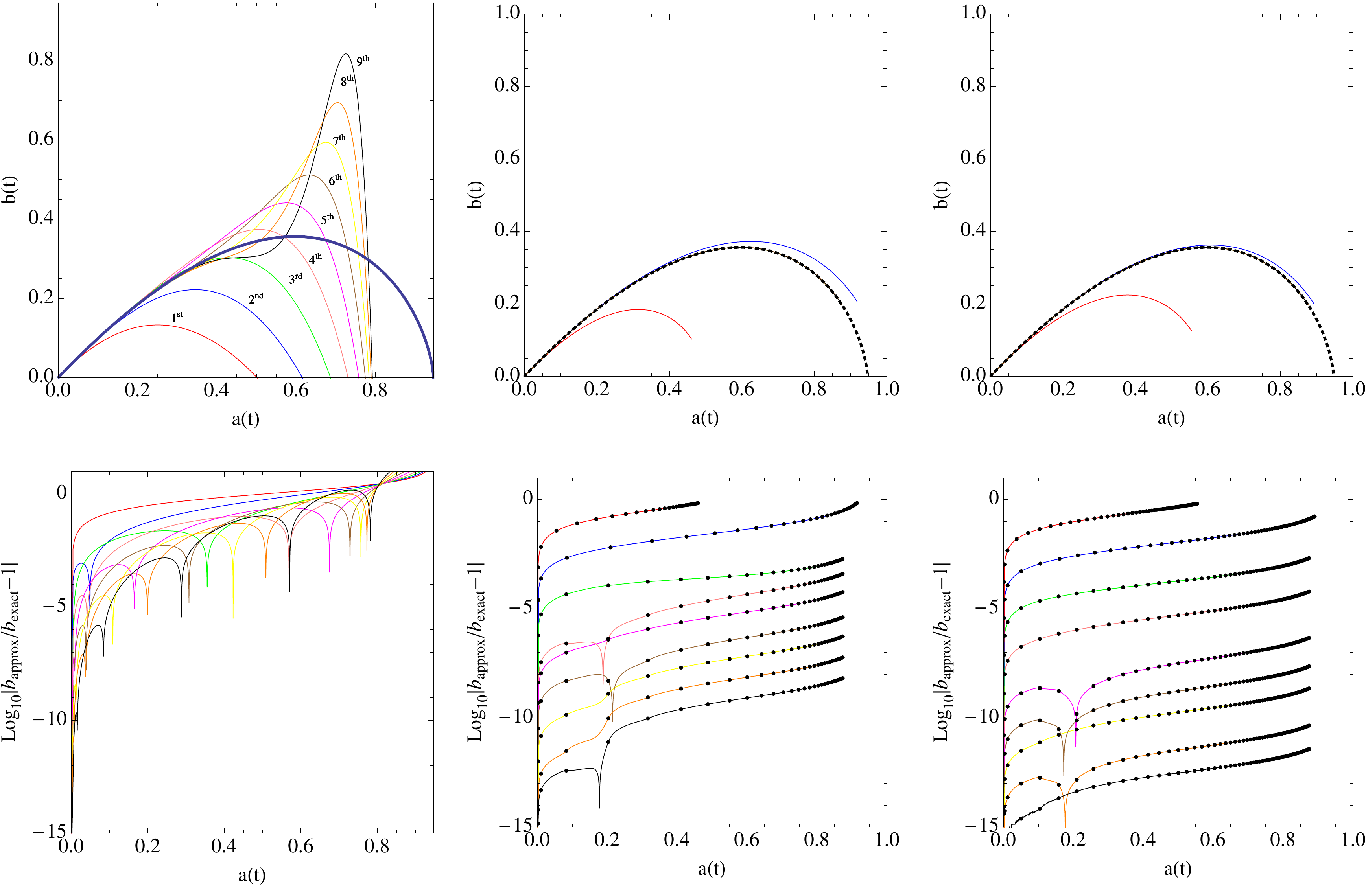}
\caption{LPT re-expansion of a closed solution with
  $\Delta = 0.2$, $\theta = 0.44$.  The top three figure show the
  scale factor calculated with a single step (left) and multiple steps
  with $\epsilon =0.5$ (middle) and $\epsilon =0.2$ (right) (refer to
  text for definition of $\epsilon$). The bottom figures show the
  errors for all LPT approximations to $b(t)$ including the unphysical
  negative ones. The order of the expansion is colour-coded as in the
  top left figure. The single step expansion does not respect the time
  of validity whereas both the other cases do.  The black dots
  indicate the position of the time steps. The original expansion does
  not converge over the full time range whereas the re-expansions do. Coloured version online. }
\label{closedext}
\end{figure}

\capfigref{closedext} investigates the closed model introduced in
\figref{closedplot} ($\Delta = 0.2, \theta = 0.44, a_0 =
10^{-3}$). The time of validity is determined by a complex root. The
first panel shows that the series begins to diverge at $a=0.38$ well
before the collapse singularity is reached at $a=0.94$.

A single time step less than the time of validity is guaranteed to
converge as the order of the Lagrangian expansion increases. LPT
re-expansion utilises a set of such time steps each of which is
likewise guaranteed to converge.  However, since a calculation of
infinite order is never achieved in practice, it is worth characterising
how convergence depends upon two calculational choices one has at hand, the time
step and the order of the Lagrangian expansion.

A single small step beginning at $t=t_0$ and ending at $t_f$ has leading order error for the $m$-th order Lagrangian approximation \footnote{Typically, the numerical coefficient is of order unity and varies with $m$ as well
  as the particular value of $\theta$.  For the purposes of a
  discussion of the scaling of the error term, we assume the numerical
  coefficients to be constant as $m$ and $\theta$ vary.}
$\propto (t_f/t_0 - 1)^{m+2} \Delta^{m+1}$, where $\Delta$ is the value at the initial time.  If the same small interval is covered in $N$ smaller steps, the error after $N$ steps scales as $N^{-m} (t_f/t_0 -1)^{m+2} \Delta^{m+1}$ (see Appendix \ref{errorchar} for details). If the step size increases in a geometric sequence such that $\delta t/t$ is a constant for each intermediate step, then $t_f = t_0 (1+ \delta t/t)^{N} $ and the error after $N$ steps scales as $N (t_f/t_0 -1) (\delta t/t)^{m+1}\Delta^{m+1} $. This leads to the interpretation that the error per intermediate step scales as $(\delta t/t)^{m+1} \Delta^{m+1}$. Define $\epsilon = (\delta t/t)\Delta$. The leading order error scales as 
$\epsilon^{m+1}$ which is numerically small if $\epsilon < 1$.  The
sum of all the missing higher order terms is finite if $\delta t <T$,
i.e respects the time of validity. 

In a practical application, the initial and final times are not close. 
A reasonable time step criterion is to choose $\epsilon <1$ fixed throughout the evolution and to infer $\delta t$ for a given $\Delta$. Other choices are possible but
$\delta t$ must always be less than
the time of validity. If $\epsilon$ is held fixed throughout the
evolution, then the net error after $N$ steps for the $m$-th order
approximation $\propto \epsilon^{m+1} N$.

The number of steps required to go from the initial to the final time can be estimated. As a special case assume that $\Delta$ is constant.  The time step
criterion implies that the number of steps to move from the initial time $t=t_0$ to
the final time $t_f$ for given $\epsilon$ is $N = \log (t_f/t_0)/\log
(1+(\epsilon/\Delta))$. For limited total
intervals ($t_f-t_0<<t_0$) and small steps ($\epsilon/\Delta<<1$) the
exact answer reduces to $N \sim (t_f-t_0) \Delta/\epsilon =
(t_f-t_0)/\delta t$. Here $\delta t = \epsilon t\Delta$ does not
grow appreciably over the interval so the estimate for $N$ is
a maximum. In this limit, the net error $\propto \epsilon^m \Delta$. The leading order error for the m-th order Lagrangian scheme decreases at least as quickly as $\epsilon^{m}$.

In more general situations the value of $\Delta$ varies. Once the
closed model turns around $\Delta$ increases without bound.  For fixed
$\epsilon$ the step size $\delta t$ decreases monotonically to zero as
$t \to t_{coll}$ where $t_{coll}$ is the time of the future
singularity.  At any order it would take infinitely many steps to
follow the solution up until collapse.  Consider the problem of
tracking the solution up to a large, finite value of $\Delta =
\Delta_f$. This moment corresponds to a fixed
time $t_f \lta t_{coll}$ in the exact solution. The number of steps $N <
N_{max} \sim t_f/\delta t_f$ where $\delta t_f$ is the step size for
the system near $\Delta_f$; $\delta t_f \propto \epsilon/\Delta_f$.
The leading order error
after $N$ steps at the $m$-th Lagrangian order $\propto
\epsilon^{m+1} N < \epsilon^{m+1} N_{max} \sim \epsilon^m \Delta_f$.
This method of step control forces the leading order error at fixed
time $t_f < t_{coll}$ to decrease as the Lagrangian order $m$
increases and/or the control parameter $\epsilon$ decreases.

The second and third panels in \figref{closedext} show the runs with
$\epsilon = 0.5$ and $\epsilon = 0.2$ respectively. The Lagrangian
orders are colour-coded; dots show time steps determined by the above
criterion. At each order the solution was terminated when the
numerically determined $\Delta > 100$ so as to avoid the infinite step
regime. This required 32 steps for the $\epsilon = 0.5$ run and 77
steps for the $\epsilon =0.2$ run.  As the red solid lines
illustrates, the first order solution turns around before all other
solutions. This explains why its step size begins to shrink near the
midpoint of the graph. By contrast, all the step sizes for higher
order solutions are very similar up to that point.

The numerical errors may be analysed from two points of view.
\begin{enumerate}
\item A comparison of different coloured lines (different Lagrangian
  orders) in a single panel shows that error decreases as $m$
  increases. This is true in a quantitative as well as qualitative
  sense. For example, in the second panel at $a=0.64$ a plot of the
  log of the absolute error is approximately linear in $m$, as
  expected.

\item A comparison of the same coloured lines in the middle and right
  panels shows that smaller $\epsilon$ implies better accuracy. Again,
  this is true in a quantitative as well as qualitative sense. For
  example, the observed ratio of errors at $a=0.64$ for the $9$-th
  order calculations is $5\times 10^{-4}$. To evolve up to this time
  with $\epsilon =0.5$ (middle panel) takes 10 steps; with $\epsilon
  =0.2$ (right panel) it takes 22 steps. The expected ratio of errors
  is $(0.2/0.5)^{9+1} (22/10) \sim 2 \times 10^{-4}$, the same order
  of magnitude as the observed ratio.

\end{enumerate}
These comparisons lead to the important conclusion that the leading
order error for LPT re-expansion varies with Lagrangian order and time
step as theoretically expected.

It is clear that considerable benefit accrues not only from implementing
higher order Lagrangian schemes but also by limiting time step
size (which must always be less than the time of validity). For simple examples like the top-hat it is feasible to work to
very high Lagrangian order but this is not likely to be true in the
context of more complicated, inhomogeneous problems.  On the other
hand, marching forward by many small time steps using LPT re-expansion
is generally feasible. In the example above the initial perturbation
is $\Delta = 0.2$ whereas a practical calculation starting at
recombination would start with $\Delta \sim 10^{-5}$. For the same
$\epsilon$ the practical application requires more steps for the phase
before turnaround but the net increase is only a modest logarithmic
factor. In fact, most of the steps in the example were taken after
turnaround and the total number varies with the depth of the
collapse. This will continue to be true for the practical calculation.
The choice of step size and order for such applications will be the
subject of a forthcoming paper.

\section{Conclusion}
\label{conclusion}

We have investigated the time of validity of Lagrangian perturbation
theory for spherical top-hat cosmologies with general initial
conditions.  Using techniques from complex analysis we showed that the
time of validity is always limited for open models. We also discovered
a class of closed models whose time of validity is less than their
time to collapse. We introduced the concept of the mirror model and
derived a symmetry principle for the time of validity of mirror
models.  For small initial perturbations the time of validity of LPT
series expansion of an open model corresponds to the collapse time of 
a closed mirror model. 

A qualitative analogy is useful.  A single LPT series expansion is
similar to a single step in a finite difference approximation for
advancing a hyperbolic partial differential equation like the wave
equation. The time of validity of the LPT expansion is analogous to
the Courant condition which guarantees stability. In LPT the
constraint is an acceleration-related time-scale; in the wave equation
it is a sound-crossing time-scale.

We developed the method of LPT re-expansion which overcomes the
limitations intrinsic to a single expansion. We demonstrated how to
iteratively re-expand the solution so as to link convergent series
expressions that extend from initial to final times. The time of
validity of the expansions set the minimum number of re-expansion
steps ($\sim 10$) necessary for cosmological simulations starting at
recombination and proceeding to the present epoch. Finite as opposed
to infinite order Lagrangian expansions required extra steps to
achieve given
error bounds.  We characterised how the leading order numerical error
for a solution generated by LPT re-expansion varied with the choice of
Lagrangian order and of time step size. We provided a recipe for time
step control for LPT re-expansion based on these results.

Our long-term goal and motivation for this study is to develop a
numerical implementation of LPT re-expansion for fully inhomogeneous
cosmological simulation. Top-hats with Zeldovich initial conditions have
special properties with respect to LPT convergence. We found that all
underdense models must be treated by re-expansion while none of the
overdense ones need be. However, during the course of an inhomogeneous
simulation the density and irrotational velocity perturbations
(with respect to a homogeneous background cosmology) at an
arbitrary point will generally not fall on the top-hat's Zeldovich
curve. Hence, the convergence of LPT in inhomogeneous applications
must be guided by the analysis of more general models. Top-hats with
arbitrary initial conditions are the simplest possibility and constitute
the main focus in this paper. The limitations on LPT convergence
which we have elucidated in this generic case are considerably more
complicated  than in the top-hat with Zeldovich initial conditions.
Our plan is to use the generic time of validity criterion to
determine the time-stepping for inhomogeneous evolution.
This should allow us to develop high-precision simulations with
well-defined control of errors. The practical impact of a refined
treatment of LPT convergence is not yet clear.

The convergence issues we have dealt with should not be confused with
the breakdown when orbit crossing takes place and the Jacobian of the
transformation from Lagrangian to physical
coordinates becomes singular. At that time
the flow becomes multi-streamed and much of the
simplicity and advantage of the Lagrangian approach vanishes.  The
aim of the current work is to make sure it is possible to reach the
epoch of multi-streamed flow
but offers nothing new on how to proceed beyond it. In fact, it may
be necessary to include an effective pressure term in the equations to
account for the velocity dispersion induced by orbit crossing
(\citealt{adler_lagrangian_1999}; \citealt*{buchert_extendingscope_1999}) or
to adopt alternative approximations for the basic dynamics (such as
the adhesion approximation; see \citealt{sahni_approximation_1995} for a
review and references therein) to make progress.

\section*{acknowledgments}
S. N. thanks Varun Sahni for discussions on convergence of Lagrangian
theory at the IUCAA CMB-LSS summer school, Paul Grabowski, Sergei
Dyda and Justin Vines for useful conversations and Saul Teukolsky for feedback on the manuscript. The authors would like to thank Thomas Buchert for useful comments on the paper. This material is
based upon work supported by the NSF under Grant No. AST-0406635 and
by NASA under Grant No. NNG-05GF79G.

\clearpage
\bibliographystyle{mn2e.bst}
\bibliography{mybibtex,mybibtex2}

\appendix

\section{Formal set-up of the spherical top-hat}
\label{formalsetup}

We intend to study an inhomogeneous universe. It contains a single,
compensated spherical perturbation evolving in a background
cosmology. To describe two spatially distinct pieces of the
inhomogeneous universe (the background and the central perturbation)
we invoke the language of homogeneous cosmology.

\subsection{Description of the background}

The origin of the coordinate system is the centre of the sphere.
The background system at the initial time $t_0$ is set by the
physical size of the inner edge $r_{b,0}$, the velocity ${\dot
  r_{b,0}}$ and density parameter $\Omega_0$. The Lagrangian coordinate
system is extended linearly throughout space once the Lagrangian
coordinate of the inner edge is fixed.  Let the Lagrangian coordinate
of the inner edge be 
\beq 
Y = \frac{r_{b,0}}{a_0} .
\eeq 
Either choose the initial background scale factor
$a_0$ and determine the coordinate system or, alternatively,
fix $Y$ and infer the background scale factor. 
In either case, the scale
factor embodies the gauge freedom associated with the radial
coordinate system.

The future evolution of the inner edge of the background is given by 
$r_{b}(t) = a(t) Y$. 
The velocity at the initial time satisfies 
${\dot r}_{b,0} = \dot{a}_0 Y$. 
The density at any later time is
\beq 
\rho_b(t) = \frac{\rho_{b0}a_0^3}{a^3},  
\eeq
and the Hubble parameter for the background is 
\beq 
H_0 = \frac{{\dot r}_{b,0}}{r_{b,0}} = \frac{{\dot a}_0}{a_0}. 
\eeq 
The evolution of the scale factor is
\beq 
\frac{{\ddot a}}{a} = - \frac{4 \pi G \rho_{b0} a_0^3}{a^3} = - \frac{1}{2} \frac{H_0^2 a_0^3 \Omega_0}{a^3} 
\eeq
The quantities, $r_{b,0}$, ${\dot r}_{b,0}$, $\Omega_0$ and $t_0$ along
with the choice of the coordinate system, completely specify the
background universe.

\subsection{Description of the innermost perturbation}

The perturbation can be described by four
physical quantities: the physical position $r_{p,0}$ and velocity
${\dot r}_{p,0}$ of the edge (or the ratio $H_{0p} = {\dot
  r}_{p,0}/r_{p,0}$), the density parameter $\Omega_{p0}$ at the
initial time $t_0$. The Lagrangian coordinate system for the
perturbation is
\beq 
X = \frac{r_{p,0}}{b(t_0)}. 
\eeq
It can be linearly extended
throughout space. 

Like $a_0$, $b(t_0)$ embodies the gauge freedom associated with the
choice of the coordinate system.  Without loss of generality, one can
pick this gauge to satisfy 
\beq 
b(t_0) = a_0 .
\eeq  
Note that the Lagrangian coordinate systems for the background
and perturbation are different.

Let $\rho_0$ and $\rho_{p,0}$ denote the densities of the background and perturbation respectively. 
Define the perturbation parameters
\bea
\delta & = & \frac{\rho_{p0}}{\rho_{b0}} -1 \\
\delta_v & = & \frac{H_{0p}}{H_0} -1 
\eea
giving
\beq 
\Omega_{0p} = \frac{(1+\delta)}{(1+\delta_v)^2 } .
\eeq

\subsection{Inhomogeneous model}

\begin{figure}
\includegraphics[width=13cm]{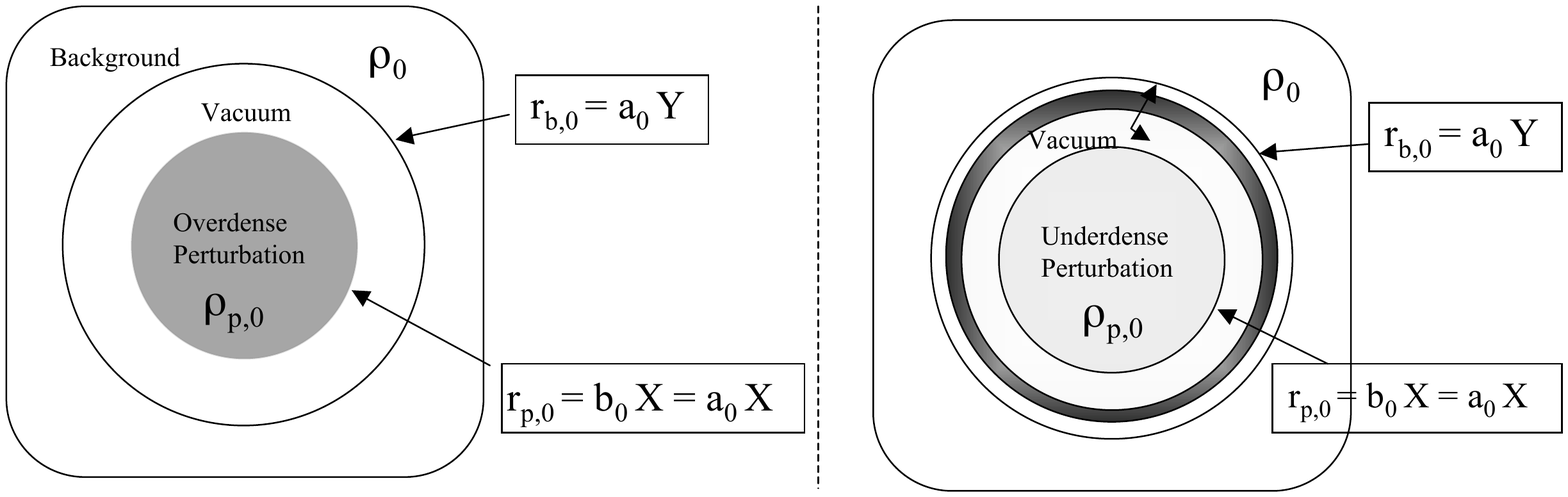}
\caption{A cartoon showing the physical set-up of the problem. }
\label{setup}
\end{figure}

\capfigref{setup} shows how an overdense and underdense innermost
sphere may be embedded with compensation in a homogeneous background
universe.
The assumption that the background cosmology evolves like a
homogeneous model, fully described in terms of its Hubble constant and
density, imposes consistency conditions. At the initial instant the ``inner
edge'' of the unperturbed background distribution is at physical
distance $r_{b,0}$ from the centre of the sphere. 
The region with $r > r_{b,0}$ will evolve like an unperturbed homogeneous cosmology 
as long as
\begin{enumerate}
\item the mass within equals the mass that an unperturbed sphere
would contain;
\item matter motions within the perturbed region do
not overtake the inner edge of the homogeneous region.
\end{enumerate}
These conditions which are obvious in the Newtonian context have general
relativistic analogues (\citealt{landaulifschitz}).

Next, consider the innermost perturbed spherical region. At the
initial time let $r_{p,0}$ be the ``outer edge'' of this region.
The physical properties and evolution of the innermost region are
fully described in terms of its Hubble constant and density as long as
its outer edge does not overtake matter in surrounding
shells. While this is obvious in a Newtonian context there
exists a relativistic analogue (\citealt{tolman_effect_1934}; \citealt{landaulifschitz}).

The inhomogeneous model is incomplete without specification of the transition
region between the innermost sphere and the background.  For the background to
evolve in an unperturbed fashion the mass within $r_{b,0}$ must be
exactly $4 \pi \rho_0 r_{b,0}^3/3$. There are many ways to satisfy
this requirement.  For example, when $\delta > 0$ a simple choice is
to place an empty (vacuum) shell for $r_{p,0} < r < r_{b,0}$ so
that $(\rho_{p0}/\rho_0) = (r_{b,0}/r_{p,0})^3 =
(Y/X)^3$. The evolution of each matter-filled region
proceeds independently as long as the trajectories of the inner and
outer edges do not cross.  When $\delta < 0$, a more complicated
transition is required. For example, one choice is to nest
sphere, empty shell and dense shell (see \figref{setup})
 so that the mass within $r_{b,0}$ matches that of the unperturbed background. 
 In this case $\rho_{p0} r_{p,0}^3 = f \rho_0 r_{b,0}^3$ for some $f<1$ (the remaining
fraction $1-f$ is placed in the dense shell). Varying the specifics of
the compensation region while keeping the properties of the sphere
fixed leaves $\delta$ and $\delta_v$, as defined above, invariant.

For fixed $\delta$ and $\delta_v$ the solution $b(t)$ is independent
of the details of the transition. Nonetheless, variation in $f$,
$r_{b,0}/r_{p,0}$ and $Y/X$ all go hand-in-hand. Hence, the extent of
time that the sphere's evolution may be treated as independent of the
matter-filled outer regions also varies. A basic premise of this paper
is that it is meaningful to determine the limitations arising from the
convergence of the LPT series independently of limitations associated
with crossing of separate matter-filled regions. For a given a
$\delta$ and $\delta_v$ this separation can be achieved for
specific constructions by choosing the radius
and (hence velocity) of the inner sphere and the energy of the
compensating region appropriately.

\subsection{Number of degrees of freedom for the innermost sphere}

If the innermost sphere corresponds to an overdensity then the
compensating region can be a vacuum as shown in \figref{setup}.
Having picked the co-ordinate system, having selected equal initial times for
the background and perturbation (not equal bang times but equal times
at which we
give the background and perturbation values), and required the
correct amount of mass, only two degrees of
freedom remain: $\delta$ and $\delta_v$.

To reiterate, the background and the perturbation can have different
big bang times. Setting them equal would imply a relationship between
$\delta$ and $\delta_v$ and leave a single free parameter.

If the innermost sphere corresponds to an underdensity then the
compensating region is not vacuum but a spherical shell.  In this
case, in addition to $\delta$ and $\delta_v$, one must specify $f$ or,
equivalently, $r_{p,0}$. But the solution for $b(t)$ is independent of
the size of the innermost sphere so, again, only two degrees of
freedom remain.

\subsection{Preventing shell crossing}

There are two sorts of limitations for the solution of
$b(t)$. One is the calculation-dependent limitation arising from the
convergence properties of the Lagrangian series expansion. It involves
the scale factors only. The other is a
physical limitation arising from collisions of the innermost region
with surrounding non-vacuum regions (either the background or a compensating
shell). We show that it is possible to delay the epoch of collisions
indefinitely without altering the evolution of the innermost region.

Fix $H_0$, $H_{p,0}$, $\rho_0$ and $\rho_{p,0}$.  This implies that
the expansion parameters in LPT, $\delta$ and $\delta_v$, and the time
of validity of the LPT solution are all fixed. Consider the case of an
overdensity surrounded by vacuum. To stave off the collision of the
outer edge of the innermost region with inner edge of the homogeneous
background hold $r_{b,0}$ fixed and reduce $r_{p,0}$. The velocity
$\dot r_{p,0} = H_{0p} r_{p,0}$ becomes arbitrarily small.  The time
for the edge to reach any fixed physical distance increases without
bound. Shell crossings may be put off indefinitely. However, we have
altered the mass within the innermost edge of the background so
we add back a thin, dense shell just inside $r_{b,0}$ and set it on
a critical trajectory outward. This accomplishes our goal.

The case of the underdensity surrounded by a compensating shell is
identical. First, we must make sure that the compensating shell does
not overrun the homogeneous model.  Choose the shell to be thin, fix
its initial physical distance from the centre and adjust is velocity
(based on how the interior mass changes) to give a critical solution.
The two power laws, one for the compensating shell and one for the
innermost boundary of the homogeneous model, cannot cross in the
future.  Second, as above, note that reducing $r_{p,0}$ reduces the
outward velocity of the edge so that it
takes more time to reach the initial position of the
compensating shell. The time can be made arbitrarily long.

The limitations in LPT convergence are completely distinct from those
associated with physical collisions in inhomogeneous model.

\section{Series expansions for a function of two variables}
\label{example} 

In this section we elucidate by example some qualitative features of
the expansion of $b(t,\Delta)$, the central quantity in the Lagrangian
treatment of the top-hat. We assume a very simple form denoted
$f(t,\Delta)$ and look at convergence with respect to expansions in
$t$ and $\Delta$. Let
\beq 
f(t,\Delta) = t^{2/3} \left(\frac{1}{t}+\Delta \right)^{1/3} .
\eeq 
The series expansion of this function around $\Delta=0$ at fixed $t$ is
\beq f \sim t^{1/3} + t^{4/3} \Delta - 
 \frac{1}{9} t^{7/3} \Delta^2 + \frac{5}{81} t^{10/3} \Delta^3 - 
 \frac{10}{243}  t^{13/3}\Delta^4 + \frac{22}{729} t^{16/3} \Delta^5 + {\mathcal O}(\Delta^6) \eeq
which is supposed to mimic the Lagrangian expansion in $\Delta$.
One can also expand the function as a series in $t$ around $t=t_i$
\bea
f & \sim & \sqrt[3]{\Delta+\frac{1}{t_i}} t_i^{2/3} + \frac{(2 \Delta
  t_i+1) (t-t_i)}{3 \left(\Delta+\frac{1}{t_i}\right)^{2/3}
   t_i^{4/3}}  +  \frac{\left(-\Delta^2 t_i^2-\Delta t_i-1\right)
   (t-t_i)^2}{9 \left(\Delta+\frac{1}{t_i}\right)^{2/3}
   t_i^{7/3} (\Delta t_i+1)}  \nonumber \\
   & &  + \frac{\left(4 \Delta^3 t_i^3+6 \Delta^2
   t_i^2  +  12 \Delta t_i + 5\right) (t-t_i)^3}{81
   \left(\Delta+\frac{1}{t_i}\right)^{2/3} t_i^{10/3} (\Delta
   t_i +1)^2} +\mathcal{ O}\left((t-t_i)^4\right) . \eea
Both expansions involve the complex power $z^{1/3}$. 
There are two branch cuts which
extend to $z=0$ so at $\Delta=-1/t$ the function is not analytic.
Additionally, the expansion in $t$ is not analytic at $t=0$.

\begin{figure}
\includegraphics[width=15cm]{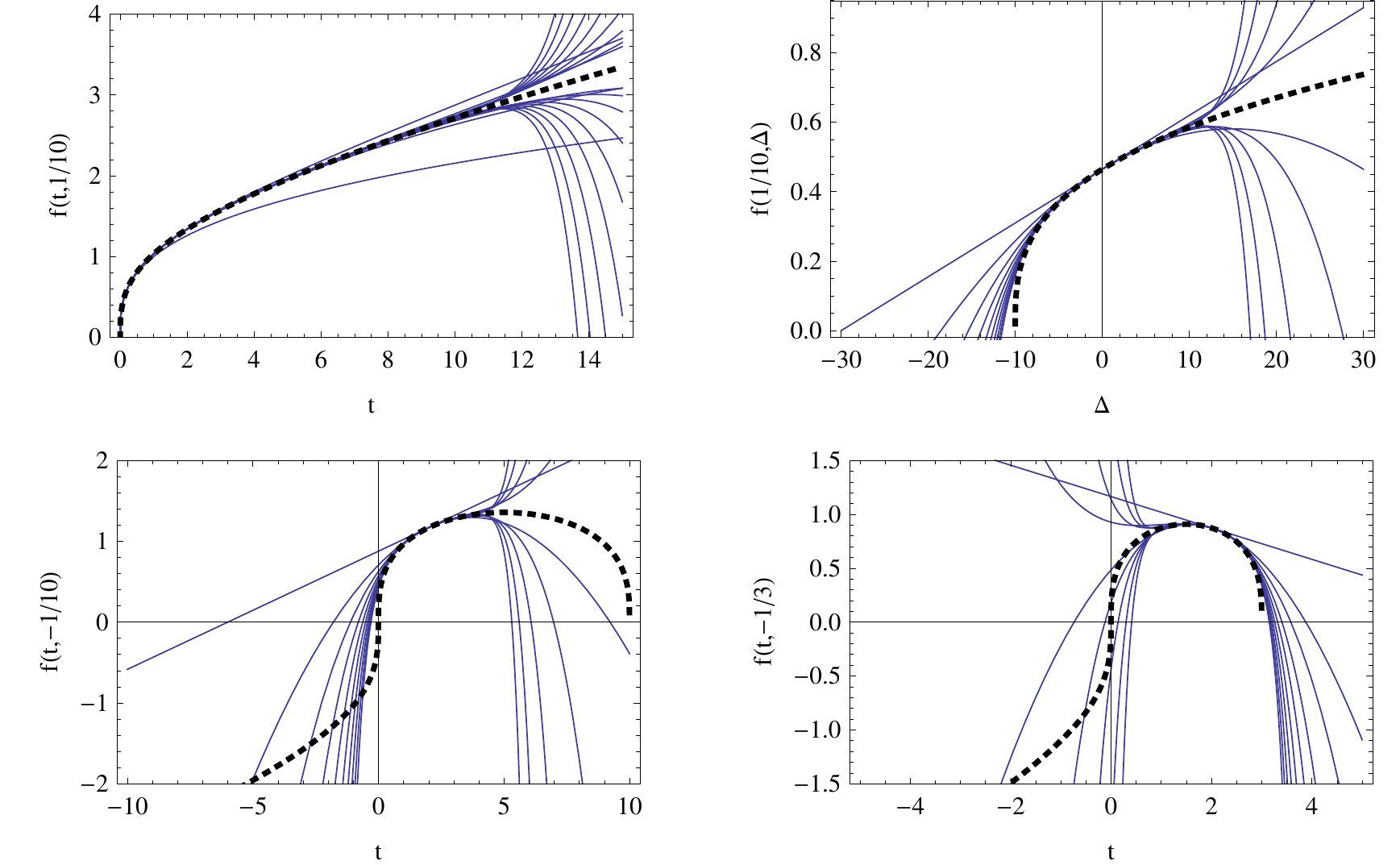}
\caption{Series expansions in $t$ and $\Delta$ for an illustrative
  function $f(t, \Delta)$ (see text). The black dotted line indicates
  the exact function $f$ and the blue solid lines indicate successive
  approximations. The top left and right panels are series expansions
  in $\Delta$ around $\Delta=0$ plotted as a function of $t$ (for
  $\Delta = 1/10$) and function of $\Delta$ (for $t=1/10$)
  respectively. The bottom left and right panels are series expansions
  in the $t$ around $t=2$ plotted as functions of $t$ for
  $\Delta=-1/10$ and $\Delta = -1/3$ respectively.  }
\label{examplegraph}
\end{figure}

The efficacy of various expansions are illustrated in
\figref{examplegraph}. In all the plots the black dotted line
indicates the exact function.  The top left panel shows successively
higher order series approximations in $\Delta$ as a function of $t$
for the specific case $\Delta=1/10$. The question here is whether the
pole at a given time lies with a disk of radius $1/10$? The location
of the pole is $\Delta=-1/t$ so the answer is ``yes'' when $t >
10$. This pole interferes with the convergence of the series expansion
for $\Delta=1/10$.  The figure demonstrates the (future) time of validity
is $t<10$.

The top right panel shows the series in $\Delta$ at a fixed
$t=1/10$. The question here is how big a perturbation will converge
at $t=1/10$? Since the location of the pole is $\Delta=-1/t$ the
radius of convergence at the indicated time is $10$. Perturbations with
$|\Delta| > 10$ are not expected to converge and
the figure shows that this is indeed the case.

The bottom left panel shows the series in $t$ expanded around $t_i=2$
for fixed $\Delta=1/10$. The poles are at $t=-10$ and $t=0$ in the
complex $t$ plane.  The expected radius of convergence is
$\min(|2-0|,|2-(-10)|) = 2$ or $t_i-2 < t < t_i+2$.  As seen in the
plot, the series converges only in the expected range $(0,4)$

The bottom right panel shows the series in $t$ expanded around $t_i=2$
for $\Delta=-1/3$. The poles are at $t = 3$ and $t=0$ in the complex
$t$ plane.  The expected radius of convergence is $\min(|2-0|,|2-3|) =
1$ or $t_i-1<t<t_i+1$.  As seen in the plot, the series converges only
in the expected range $(1,3)$.
 
\section{Parametric Solution}
\label{solutiondetails}

The background model has scale factor $a_0$ and Hubble constant $H_0 =
{\dot a}_0/a_0$. The model, perturbed in density and velocity,
is parameterized by $\Delta$ and $\theta$ and has scale factor
$b(t)$.  For the choice of
coordinate system given in the text the second order equation for
$b$ is
\beq 
\frac{{\ddot b}}{b} = -\frac{1}{2}  \frac{H_0^2 a_0^3(1+\Delta \cos \theta)}{b^3} \label{secondorder2} 
\eeq
with the initial conditions that at $t=t_0$, $b(t_0) = a_0$, ${\dot
  b}(t_0) = {\dot a}_0 (1+\Delta \sin \theta)$.  The scale factor
$a_0$ and the velocity of the background ${\dot a}_0$ at the initial
time $t_0$ are positive. The parametrization of
${\dot b}(t_0)$ allows either positive or negative values
where $\Delta$ is non-negative and $-\pi
<\theta \leq \pi$.  The quantity $(1+ \Delta \cos \theta)$,
proportional to total density, is non negative.

This equation once integrated is 
\beq {\dot b}^2 = H_0^2 a_0^3\left[\frac{(1+\Delta \cos \theta)}{ b} + \frac{(1+ \Delta \sin\theta)^2 -(1+\Delta \cos \theta)}{a_0}\right]. 
\label{velocityeq2}
\eeq
The combination 
\beq E(\Delta, \theta) = (1+ \Delta \sin \theta)^2 -(1+\Delta \cos \theta) 
\label{energy2}
\eeq
is proportional to the total energy and determines the fate of the
system.  If $E(\Delta, \theta) > 0$, the model is open and if
$E(\Delta, \theta)<0$, the model is closed and will re-collapse
eventually. Four cases (positive and negative $E$, positive and negative
${\dot b}_0$) are shown in \figref{dbdtposneg}.

\subsection{Initially Expanding Solutions}

The expanding case with ${\dot b}_0 >0$ for open models ($E>0$) has solution
\bea  b(\eta, \Delta, \theta) &=& \frac{a_0}{2} \frac{(1+\Delta \cos \theta)}{E(\Delta, \theta)} (\cosh \eta -1) \\
t(\eta, \Delta, \theta) &=&  \frac{1}{2H_0} \frac{(1+\Delta \cos \theta )}{E(\Delta, \theta)^{3/2}} (\sinh \eta - \eta )+ t_{bang}^+(\Delta ,\theta)
\eea
and the singularity $b=0$ occurs at $\eta=0$. 
For closed models ($E<0$) the solution is  
\bea  b(\eta, \Delta, \theta) &=& \frac{a_0}{2} \frac{(1+\Delta \cos \theta)}{|E(\Delta, \theta)|} (1 - \cos \eta ) \\
t(\eta, \Delta, \theta) &=&  \frac{1}{2H_0} \frac{(1+\Delta \cos \theta )}{|E(\Delta, \theta)|^{3/2}} (\eta - \sin \eta)+ t_{bang}^+(\Delta ,\theta). 
\eea
For closed models, the convention adopted sets $\eta=0$ at
the singularity nearest in time to $t_0$.
For both models, the time at $\eta=0$ is
denoted $t_{bang}^+$. For closed models the time at $\eta = 2\pi$
is denoted $t_{coll}^+$.

At the initial time the solutions (both open and closed)
satisfy $b(t_0) = a_0$, ${\dot b}(t_0) = {\dot a}_0 (1+ \Delta \sin
\theta)$ and $t=t_0$. The condition
$b(t_0) = a_0$ sets the value of the parameter at the initial
time $\eta_0$.
The velocity condition is then manifestly satisfied
from the form of \eqnref{velocityeq2}. The condition $t=t_0$ at 
$\eta = \eta_0$ sets the value of the bang
time
\beq  t_{bang}^+ = t_0 -
\left\{
\begin{array}{cc}
 \frac{1}{2H_0} \frac{(1+\Delta \cos \theta )}{|E(\Delta, \theta)|^{3/2}} (\eta_0 - \sin \eta_0) & E < 0 \\
\frac{1}{2H_0} \frac{(1+\Delta \cos \theta )}{E(\Delta, \theta)^{3/2}} (\sinh \eta_0 - \eta_0 )& E > 0.
\end{array}
\right.
\label{tbangpos1}
\eeq
The bang time for the model can also be written as 
\beq t_{bang}^+ = t_0 - \int_{b=0}^{b=a_0} \frac{db}{({\dot b}^2)^{(1/2)}},
\label{tbangpos2}
 \eeq
where ${\dot b}^2$ is given by \eqnref{velocityeq2} with the sign for the
square root positive. The age of the model since its birth is
\beq 
t_{age}(\Delta, \theta) = 
\int_{b=0}^{b=a_0} \frac{db}{({\dot b}^2)^{(1/2)}} = 
\int_{\eta=0}^{\eta =\eta_0} \frac{db/d\eta\cdot d\eta}{({\dot b}^2(\eta))^{(1/2)}} .
\eeq 
Inserting the appropriate parametric solution, 
one can verify that the bang times
obtained from (\ref{tbangpos1}) and (\ref{tbangpos2}) are identical.
Generally $t_{bang}^+\neq 0$.

The velocity at the initial time is 
\beq {\dot b}_0 = {\dot a}_0 |E|^{1/2} 
\left\{
\begin{array}{cc}
\frac{\sin \eta_0}{1-\cos\eta_0} & E < 0 \\
\frac{\sinh \eta_0}{\cosh \eta_0 - 1} & E > 0.
\end{array}
\right.
\eeq
First, ${\dot b}_0>0$ implies $\eta_0>0$.
Second, if the age of the model
increases, $\eta$ increases. For the open solution if $\eta$
varies from $0$ to $\infty$ time increases from $t_{bang}^+$ to
$\infty$. For a single cycle of the closed
solutions, $\eta$ increases from $0$ to $2 \pi$ and time increases
from $t_{bang}^+$ to $t_{coll}^+$.

In summary, the parametric solutions solve \eqnref{secondorder2} and
\eqnref{velocityeq2} for the specified initial conditions. As a final
useful step, rewrite \eqnref{tbangpos2} by defining $y=b/a_0$
\beq
t_{bang}^+ =  t_0 -\frac{1}{H_0} \int_{y=0}^{y=1} \frac{dy}{\left[(1+\Delta \cos \theta) y^{-1} + E(\Delta,\theta) \right]^{1/2}}
\eeq
which follows from \eqnref{velocityeq2} and uses the same
positive square root convention.

\subsection{Initially Contracting Solutions}

Next, consider the case ${\dot b}_0 <0$. 
The parametric solution for $E>0$ is 
\bea 
b(\eta, \Delta, \theta) &=& \frac{a_0}{2} \frac{(1+\Delta \cos \theta)}{E(\Delta, \theta)} (\cosh \eta -1) \\
t(\eta, \Delta, \theta) &=&  \frac{1}{2H_0} \frac{(1+\Delta \cos \theta )}{E(\Delta, \theta)^{3/2}} (-\sinh \eta + \eta )+ t_{bang}^-(\Delta ,\theta) 
\eea
and for $E<0$ is 
\bea   b(\eta, \Delta, \theta) &=& \frac{a_0}{2} \frac{(1+\Delta \cos \theta)}{|E(\Delta, \theta)|} (1 - \cos \eta ) \\
t(\eta, \Delta, \theta) &=&  \frac{1}{2H_0} \frac{(1+\Delta \cos \theta )}{|E(\Delta, \theta)|^{3/2}} (-\eta + \sin \eta)+ t_{bang}^-(\Delta ,\theta). 
\eea
Again, for closed models, the convention adopted is that the
singularity nearest to $t_0$ corresponds to $\eta=0$.  The time at
$\eta= 0$ is $t_{bang}^-$ and the collapse time for closed models is
$t_{coll}^-$.

The parametric form of the solutions satisfies
\eqnref{secondorder2} and \eqnref{velocityeq2}. Just as in the
previous case, the initial conditions set $\eta_0$ and
$t_{bang}^-$. Since the singularity at $\eta= 0$ lies to the future of
$t_0$,
\beq 
t_{bang}^- = t_0 +  \int_{b=0}^{b=a_0} \frac{db}{(\dot b^2)^{(1/2)}}. 
\eeq
where, ${\dot b}^2$ is given by \eqnref{velocityeq2}. The sign of
the square root is chosen to be positive and the integral is a
positive quantity which is added to $t_0$.  For closed models the singularity
at $\eta=2 \pi$ lies to the past of $t_0$ at $t_{coll}^-$. In this case
(see \figref{dbdtposneg}) the labelling implies $t_{coll}^- < t_0 < t_{bang}^-$.
Although this might seem backwards, it facilitates combining
the open and closed models into one complex
function as was done in the positive ${\dot b}_0$ case. The
initial velocity is
\beq
{\dot b}_0 = {\dot a}_0 |E|^{1/2} 
\left\{
\begin{array}{cc}
\frac{\sinh \eta_0}{1 - \cosh \eta_0} & E > 0 \\
\frac{\sin \eta_0}{\cos\eta_0-1} & E < 0
\end{array}
\right.
\eeq
The initial velocity ${\dot b}_0<0$ implies $\eta_0 > 0$. For the age
of the model to increase, $\eta$ must decrease. Conversely, if $\eta$
increases, the time in the open model decreases from $t_0$ to $-\infty$
and the time in the closed model decreases from $t_0$ to $t_{coll}^-$.

A table summarising the properties of the physical solutions
with $\ceta = |\ceta| \zeta = \eta \zeta$ follows.

\begin{center}
\begin{tabular}{ |c| c| c|c|c| c|} 
\hline
     & Closed & Open    & If $\eta$ increases & If $t$ increases & $t_{bang} - t_0$  \\
      \hline
$  {\dot b}_0 >0$ & $\zeta = 1$ & $ \zeta = i$ & $t$ increases from $t_{bang}^+$ & $\eta$ increases to $\infty$ or $2 \pi$&$ <0$  \\
\hline
  ${\dot b}_0 <0 $ &  $ \zeta =1 $  & $ \zeta = i $ & $t$ decreases from $t_{bang}^-$ & $\eta$ decreases to $0$ & $>0$ \\
\hline    
\end{tabular}
\end{center}

\subsection{Analytic Extension of the exact solution in parametric form}
\label{complexformdetails}

The differential \eqnref{cdiffeq} was solved numerically over the
range $0\leq \theta \leq \pi$, $0< \delta <100$ and $-\pi <\phi \leq
\pi$ where $\dm = \Delta e^{i\phi}$.
For each value of $(\Delta, \phi, \theta)$,
the numerical solution matched one of the two possible parametric
forms.

Omitting the explicit functional dependence on
$\dm$ and $\theta$ the following abbreviations are useful
\bea 
{\bf j} & = & (1+\dm \cos \theta) \\
{\bf h} & = & \frac{(1+\dm \sin \theta)^2}{\bf j} \\
{\bf E} & = & ({\bf h}-1){\bf j} .
\eea
The two possible parametric forms that agree with the numerical solution
are
 \bea
{\bf b}(\eta) &=& \frac{a_0}{2} \frac{{\bf j}}{\left[-{\bf E}\right]} (1 -\cos \eta) \\
{\bf t}(\eta) &=& t_0 \pm 
\left(
\frac{1}{2H_0} \frac{\bf j}{\left[-{\bf E}\right]^{3/2}} (\eta -\sin \eta ) - {\bf  t_{age}} \right)
 \label{paramsoln2}
\eea
where
\beq
\label{tage2}
{\bf t_{age}} = \frac{1}{H_0}  \left(
\sqrt{{\bf j}} \sqrt{{\bf h}}
 - \frac{{\bf j}}{\left[{\bf E}\right]^{3/2}}\sinh^{-1}\sqrt{\frac{\bf E}{\bf j}} \right) . 
\eeq
The branch cut lies along the negative real axis for all fractional
powers and from $-i \infty$ to $-i$ and $+i$ to $i \infty$ for
the inverse sinh function.

\begin{figure}
\includegraphics[width=10cm]{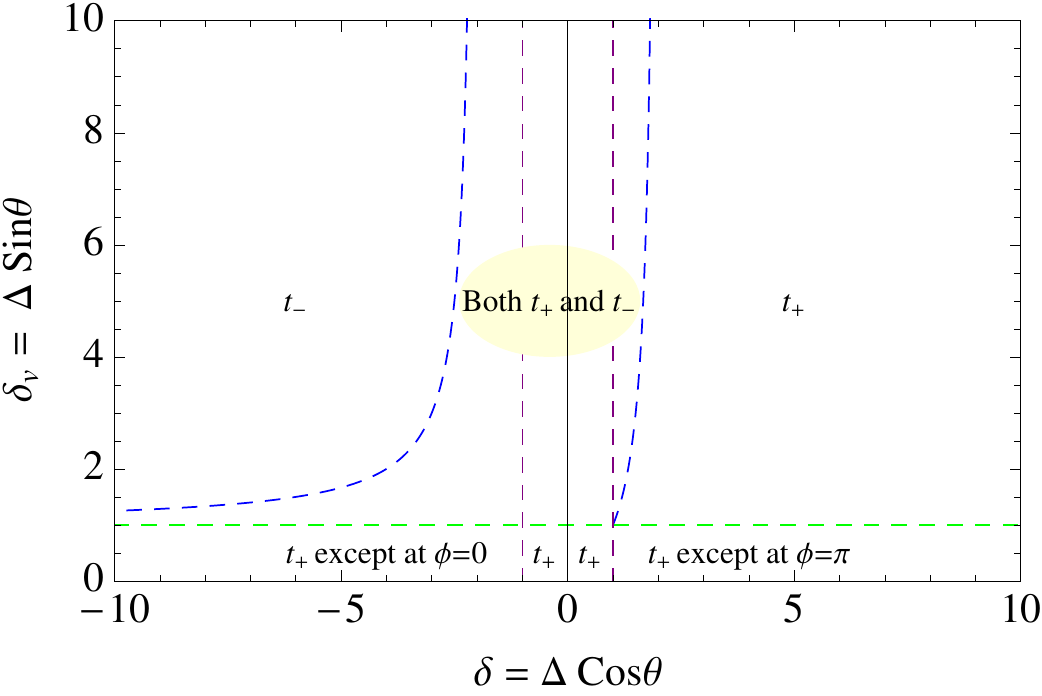}
\caption{This figure describes one aspect of the analytic extension
  of the exact solution. For a given real $\Delta$, the complex
  extension $\Delta \rightarrow \Delta e^{i\phi}$ obeys
  \eqnref{prescription} with two possible forms ${\bf t}_+$ and ${\bf
    t}_-$. The choice depends on $\phi$, $\Delta$, $\theta$. For some
  ($\Delta$, $\theta$) a single form is sufficient for all $\phi$; for
  other values both forms are needed. This figure illustrates how the
  upper half plane is partitioned based on this property. Coloured version online. }
\label{phidivision}
\end{figure}

The prescription for the correct form is for the choice of
the $\pm$ sign in ${\bf t}$ \eqnref{paramsoln2} and denoted ${\bf t}_+$
and ${\bf t}_-$.
The correct form depends upon $\theta$, $\phi$, 
$\arg[{\bf h}]$ (the arg is defined to be between $-\pi$ and
$\pi$) and the (real) value  $j={\bf j}$ when $\phi=0$ or $\pi$.
The \figref{phidivision} shows the upper half plane for the perturbation
partitioned into areas where the complex extension of the solution has
one of two forms. The lower half plane has the same structure inverted
through the origin. The horizontal red dashed
line denotes $\Delta \sin \theta = 1$ and the vertical red dashed
lines denote $\Delta \cos \theta = \pm1$. In some areas a single form
applies as marked but in the central area both occur. The detailed
prescription is
\beq
{\bf t} = 
\left\{
\begin{array}{lr}
0  \leq \theta \leq \pi/4 &
\left\{
\begin{array}{lr}
 \phi = \pi ,|\Delta| \sin \theta < 1 \mbox{  and  }  j  < 0 & {\bf t_-} \\ \\
\mbox {otherwise} & {\bf t}_+ \\ \\
\end{array}
\right.\\ \\
\pi/4  < \theta \leq \pi &
\left\{
\begin{array}{lr}
0< \phi < \pi \mbox{  and  }  \arg{\bf h} > 0 & {\bf t}_+  \\ \\
-\pi < \phi < 0 \mbox{  and  }    \arg{\bf h} < 0 & {\bf t}_+ \\\\
\phi=0 \mbox {  and  }
\left\{
\begin{array}{c}
\cos \theta > 0 \\
\mbox{ or} \\ 
\cos \theta < 0 \mbox{  and  }  j >0 
 \end{array} 
\right.
& {\bf t}_+ \\ \\
\phi= \pi \mbox {  and  }
\left\{
\begin{array}{c}
\cos \theta < 0 \\
\mbox{ or} \\ 
\cos \theta > 0 \mbox{  and  }  j  < 0 
 \end{array} 
\right.
& {\bf t}_- \\ \\
\mbox{otherwise} & {\bf t}_- 
 \end{array}
\right.\\ \\
\end{array}
\right.
\label{prescription}
\eeq

\section{Numerical solutions}
\label{fullroots}
\subsection{Algorithm}

The initial conditions are parameterized by $\Delta > 0$ and $-\pi <
\theta \leq \pi$. The transformation $\theta \rightarrow \pi \pm
\theta$ and $\Delta \rightarrow -\Delta$ leaves the solution
unchanged. At any time the roots for $\theta$ and $\pi \pm \theta$ are
negatives of each other. The root plot only depends upon the absolute
value of the root so the plots for $\theta$ and $\pi \pm \theta$ are
identical.  It is sufficient to consider the upper half plane.

For a given $\theta$ the algorithm to map out $R_{\Delta}(t)$ is the following:
Vary $\Delta$ from $0$ to an arbitrarily
large value ($\sim 100$) in small increments. For each $\Delta$ select
$\dm=\Delta e^{i \phi}$ by varying the phase angles $\phi$ over the
range $0$ to $2 \pi$. For each $\dm$ evaluate ${\bf t}(\eta)$ at
$\eta=0$ and $\eta=2 \pi$ calculated according to
\eqnref{prescription}. Finally, hunt for solutions that set the imaginary
part of ${\bf t}$ to $0$. This last step involves one-dimensional
root-finding in $\phi$ at fixed $\Delta$. A solution leads to a specific
pair $(t,\dm)$ that is a pole in the function ${\bf b}(\dm,t)$.

Roots with $t>t_0$ limit future evolution; those with $t<t_0$ limit
backwards evolution. Both sets are shown in the results. Roots are
classified based on whether they are real or complex. For closed
models the real roots can represent a singularity that is nearby
($\eta=0$) or far away ($\eta=2 \pi$) from $t_0$. This classification
at the initial time
is independent of whether the singularity is in the past or future and
is independent of whether the model is expanding or contracting.
For open models the real roots are always considered
nearby ($\eta=0$).

In what follows the numerical answers are first described in
qualitative terms. In the next section simple analytic estimates
for the time of validity are developed.

\begin{figure}
\includegraphics[width=16cm]{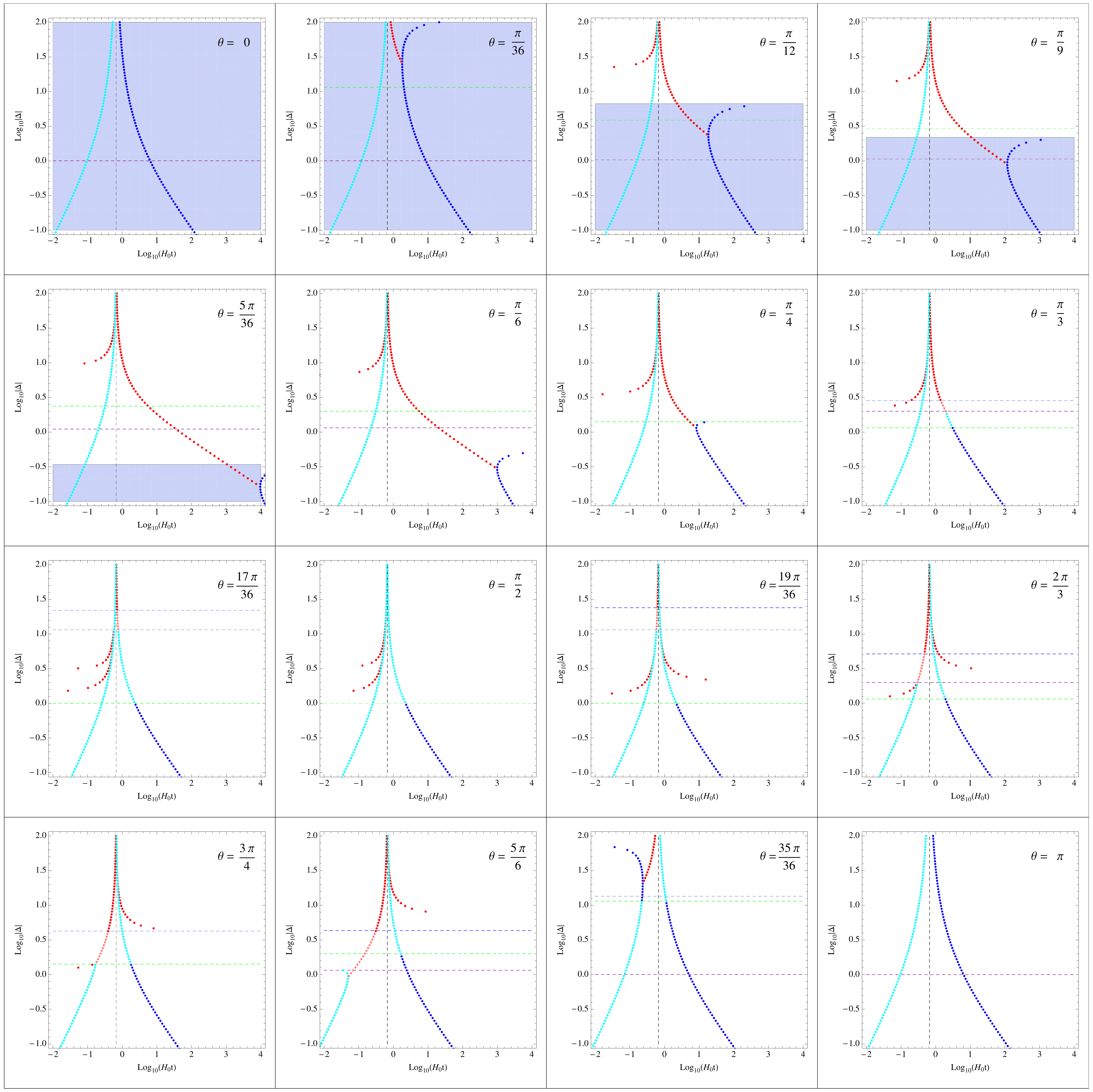}
\caption{Root plots for $\theta$ in the range $0\leq \theta \leq
  \pi$. In each plot the abscissa is $\log_{10} H_0 t$ and the
  ordinate is the logarithm of the magnitude of the root. The vertical
  black dashed line marks the initial time. The shaded area
  corresponds to closed models. The blue and red points show real and
  complex roots with $\eta = 2\pi$, respectively. The cyan and
  pink show real and complex roots with $\eta =0$, respectively. The
  green and purple dashed lines are $|\delta_v| =1$ ($\Delta = |\sin
  \theta|^{-1}$) and $|\delta|=1$ ($\Delta = |\cos \theta|^{-1}$),
  respectively. The blue dashed line indicates the switch between two
  forms and a single form of the parametric solution at $\Delta=
  |2\sec \theta - \csc \theta|$. Coloured version online. }
\label{allroots}
\end{figure}

\capfigref{allroots} shows the root plots on a log-log scale. Sixteen
panels, each with a particular value of $\theta$ listed at the top,
are displayed. The x-axis is $\log_{10} H_0 t$ and the y-axis
is the log of the distance of the singularity from the origin in the complex
$\dm$ plane. The initial time, $H_0 t_0 = 2/3$, is marked by the
vertical black dashed line.

For each $\theta$, the shaded region indicates the range of $\Delta$
that gives rise to closed models. \capfigref{phaseplot} shows that
closed models occur only for $\theta < \theta_c^+= 0.463$ in the upper
half plane so only some of the root plots have shading and then
only at smaller $\Delta$.

The colour coding of the dots indicates four types of roots: real and
complex roots where $\eta = 2\pi$ are in blue and red, respectively;
real and complex roots with $\eta = 0$ are in cyan and pink,
respectively.  The radius of convergence at the initial time $t_0$ is
infinite, i.e. the Lagrangian series is exact at the initial time by
construction. At times very close to the initial time the root loci
lie off the plot. Only the roots to the right of $H_0 t_0$ are relevant
for forward evolution and, conversely, only those to the left are
relevant for backwards evolution. The discussion is focused on the
case of forward evolution but it is straightforward to consider
the restrictions on marching backwards in time.

The phase of the root (of smallest magnitude) appears in
\figref{imaginary}. When closed models have real roots they are
positive; when open models have real roots they are negative.
However, some open and closed models also possess complex roots.  The
set of models with complex roots (of smallest magnitude)
is evident from the shading in
\figref{crootphasespace}. The phase of each root of smallest
magnitude in \figref{allroots} is indicated by the colour
shading in \figref{imaginary}.

\begin{figure}
\includegraphics[width=16cm]{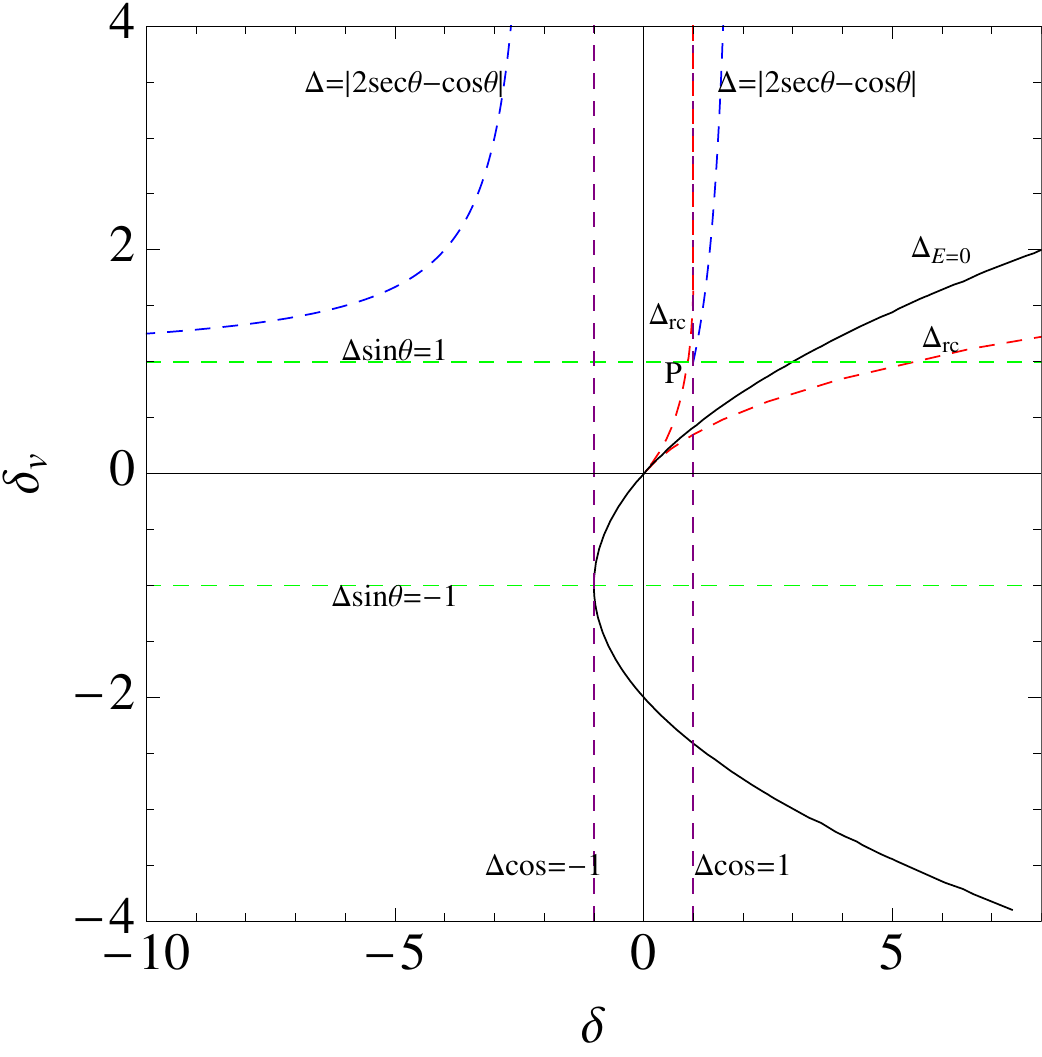}
\caption{
Several conditions determine the nature of the roots in
phase space. The most significant are schematically
illustrated here. The green horizontal lines are 
$|\delta_v|=\Delta |\sin \theta| = 1$; purple vertical lines
are $|\delta| = \Delta |\cos \theta| = 1$; the black curved
line is the $E=0$ critical solution. The red lines $\Delta_{rc}$
mark where real roots associated with closed models (or
closed mirror models) transform to complex roots. The blue
dashed lines mark the division between one and two complex
forms (see also \figref{phidivision}).
Physical models lie to the right of $\delta = -1$.
Expanding models lie above  $\delta_v=-1$.
The intersection $\delta=\delta_v=1$ occurs 
at $\theta = \pi /4$. The point P near
$\theta = 0.84$ is the meeting of $\delta_v=1$ and $\Delta_{rc}$. Coloured version online. }
\label{rootanalysis}
\end{figure}

There are horizontal dashed lines with colours green, blue and purple
in \figrefs{allroots} and \figrefbare{imaginary} indicating $|\delta_v|=1$,
$|\delta| = 1$ and the transition between one and two complex forms,
respectively. For each $\theta$ the lines mark the
implied, special value of $\Delta$.
These dashed lines also appear with
the same colour coding in \figref{rootanalysis}.


The roots in \figref{allroots} will be analysed in the range $0<\theta
\leq \pi/4$, $\pi/4 <\theta \leq \pi/2$, $\pi/2 <\theta \leq \pi$.

\subsubsection{$0\leq \theta \leq \pi/4$} 

The top left panel in \figref{allroots} has $\theta=0$; the blue dots
indicate real roots with $\eta=2 \pi$; the blue shading indicates a
closed model; the phase is positive (top left panel in
\figref{imaginary}). Only a single branch is evident. In sum, each
root is the collapse time of a closed, pure density perturbation.  For
an expanding model, $\eta=2 \pi$ implies that the root is the future
singularity. That $\theta=0$ is a special case can be seen by
consulting \figref{rootanalysis}: a ray starting at $\Delta=0$ with
$\theta=0$ never intersects any of the other lines of the diagram. In
general, each time a ray crosses one of the lines there is qualitative
change in the properties of the roots.

For $0<\theta\le \pi/4$ a great deal more complexity is evident in
\figref{allroots}. First, consider a ray emanating with small angle
$0< \theta < \theta_c^+$ in \figref{rootanalysis} ($\tan \theta_c^+
=1/2$ is the slope of the $E=0$ line at the origin).  Eventually such
a line will cross the black line which is the $E=0$ critical solution
labelled $\Delta_{E=0}$.  For small $\Delta$ the models are closed; for
larger $\Delta$ they are open. In \figref{allroots} this distinction
corresponds to the the blue shading (closed models) at small $\Delta$
versus the unshaded (open) models at large $\Delta$. 

Within the shaded region note that two branches of real roots are
present beyond a given time; at large $t$ (asymptotically) the lower
branch is $\Delta \to 0$ and the upper branch is $\Delta \to
\Delta_{E=0}$.  The lower branch sets the time of validity for small
$\Delta$. Each root is the collapse time of a closed model which
has both density and velocity perturbations at the initial time.

As $\Delta$ increases the time of validity inferred from the lower
branch decreases.  At the critical
point $\Delta = \Delta_{rc}$, the two real branches merge and
connect to a branch of complex roots (intersection of red and blue
points).  For $\Delta > \Delta_{rc}$, the complex roots determine the
time of validity even though the upper branch provides a real root. The
complex roots do not have a direct physical interpretation in terms
of future singularities of physical models. On \figref{rootanalysis}
the ray emanating from the origin at shallow angle crosses the
red dashed line labelled $\Delta_{rc}$ at this critical point.

Physically, when $\Delta$ exceeds $\Delta_{rc}$, the velocity
perturbation dominates the density perturbation in the sense that the
collapse time begins to increase. The real root corresponds to the
future singularity of the model. As $\Delta$ increases
further, the solution eventually becomes critical (infinite collapse
time).  The particular value where this occurs is
$\Delta_{E=0}$ and it corresponds on \figref{rootanalysis} to the ray
crossing the labelled black line.  Within the entire range
$\Delta_{rc}< \Delta < \Delta_{E=0}$ the complex root determines the
time of validity. So, even though any model in this range is closed
and possesses a real future singularity, the time of validity is
determined by the complex root. This gives the sliver on
\figref{crootphasespace} which is the overlap of light red and blue
shadings.

Both $\Delta_{rc}$ and $\Delta_{E=0}$ decrease as $\theta \to
\theta_c^+$ as is evident from \figref{rootanalysis} and both vanish
at $\theta_c^+$. On \figref{allroots} the real roots completely
disappear and only the complex roots are present, i.e. the two real
branches have been pushed out to infinite times.  The panel with
$\theta = 5 \pi/36 = 0.436$ is numerically closest to the critical
case $\theta_c^+ = 0.464$ and the real branches are just barely
visible at the right hand edge.

For the rest of the upper half plane $\theta_c^+<\theta \le \pi$ the
ray no longer intersects any closed models. 

For $ \theta_c^+ <  \theta < \pi/4$ the real roots reappear and move
back to the left in \figref{allroots} (see panel with $\theta=\pi/6$).
Now, however, the roots are negative (see \figref{imaginary}).
This is a manifestation of mirror symmetry which
relates the negative real roots of an open model to the positive real
roots of a closed model. At large $t$ the two
branches have $\Delta \to 0$ and $\Delta \to \Delta_{E=0}$ and are
completely analogous to the real branches just discussed 
for closed models. The
separation between the two real branches increases as $\theta$ increases
and the solution loci shifts upwards in $\Delta$. And just as before the
two branches join and meet a complex branch. The second red dashed line
$\Delta_{rc}$ in \figref{rootanalysis} shows the real to complex
transition for the roots for the open models. 
 
This behaviour might be expect to continue for $\pi/4 < \theta < \pi$ 
but there is an additional complication: the analytic
extension involves two forms. As the ray sweeps counterclockwise in
\figref{phidivision} it crosses $\delta_v=1$ (horizontal dashed line
and the curved
blue line. These are also schematically
illustrated in \figref{rootanalysis}.

\subsubsection{$\pi/4 < \theta < \pi/2$}

All physical models are in this range are open. Real roots
have a straightforward interpretation in terms of the mirror
models. Although some of the analysis described for $\theta <\pi/4$
continues to apply several additional
complications ensue. To understand them it is useful refer to the
phase space picture shown in \figref{rootanalysis}. As $\theta$
increases, the point where $\Delta_{rc}$ meets $\delta_v=1$ is
labelled P.

For a fixed $\theta$ consider increasing $\Delta$ from small
values near the origin to $\infty$. The order in which this ray
intersects the green ($\delta_v=1$), purple ($\delta=1$), red
($\Delta_{rc}$) and blue (one or two complex forms) curves will
correlate with the change in roots.

The roots are negative real for small $\Delta$. They correspond to the
collapse time of a closed mirror model. Increase $\Delta$ and
ignore $\Delta_{rc}$.  When the $\delta_v=1$ line is crossed,
the sign of the closed mirror model's velocity switches from expanding
to contracting.  This just means that the labelling of the
future singularity switches from
further away ($\eta=2 \pi$) to nearer ($\eta = 0$). Now recall
$\Delta<\Delta_{rc}$ implies real roots and, by definition, $\delta_v=\Delta \sin
\theta$.  Hence, $\Delta_{rc}(\theta) > 1/\sin \theta$ implies that
the label switch occurs just as outlined. On \figref{rootanalysis} rays
counterclockwise of point P belong to this case. This is responsible for
the switch from blue (real $\eta=2 \pi$) to cyan (real $\eta=0$) roots
at the green line in \figref{allroots} for $\theta=\pi/3$ and $17
\pi/36$.

Conversely, if $\Delta_{rc}(\theta) < 1/\sin \theta$ the roots are
already complex and the label switch occurs between the corresponding
complex roots. There are no pictured examples in \figref{allroots}.

In the previous section with the $0<\theta \leq \pi/4$, the physical
interpretation of $\Delta_{rc}$ (as $\Delta$ increases) was that the
velocity contribution to the perturbation became dominant in the
original model if the model was closed or in the mirror closed model
if the original model was open. In latter case the mirror
models were initially expanding. Now, the same idea continues
to apply in the regime $\pi/4 <\theta <0.84$. Here the transition from
real to complex roots occurs before the $\delta_v=1$ line is crossed.
The significance of $\Delta_{rc}$ is that it marks the
increasing importance of velocity
perturbations in the closed expanding mirror models.

However, for $0.84 < \theta <\pi/2$ as $\Delta$ increases the open
model crosses $\delta_v=1$, the mirror model swaps from $\eta =2\pi$
to $0$ and the roots (real) corresponds to the real future singularity
of a closed, contracting model.  As $\Delta$ increases further, first
the mirror model becomes critical and then an open model contracting
to a future singularity. While the magnitude of $\Delta$ grows larger
than a critical value the velocity perturbation dominates the mirror
model dynamics. When $\Delta>\Delta_{rc}$ the roots switch from real
to complex. At this point the contracting mirror model can be open,
closed or critical.

Note in \figref{rootanalysis} that $\Delta_{rc}$ asymptotes to the
vertical purple line $\delta=1$.  The corresponding mirror model hits
the line $\delta = -1$ in the third quadrant.  This is the limiting
vacuum solution. Although there are no physical models beyond
the analytic extension continues and the roots
change from real to complex. All open models with $\theta \lta \pi/2$
see a transition to complex roots as the mirror
approaches the vacuum solution.

Finally \figref{rootanalysis} shows as a blue curve the point at which
there is a switch in complex form of the analytic extension. Here,
the complex roots switch from $\eta =0$
to $\eta = 2\pi$. The roots remain complex and since there is
no physical interpretation and it is irrelevant whether they belong to
$\eta =0$ or $\eta = 2\pi$.

In \figref{allroots} the panel with $\theta=\pi/3$ and $17 \pi/36$
show these transitions: the blue to cyan transition at the green
dashed line is the mirror model switch from expanding to contracting;
the cyan to pink transition at the purple dashed line is the mirror
model moving through $\delta=-1$; the pink to red transition is
the switch from two to one complex roots and $\eta=0$ to $\eta=2 \pi$.

\subsubsection{$\theta = \pi/2$}

At $\theta = \pi/2$, only real roots of $\eta =0$ are present for
large $\Delta$. This is a special case in that a ray only intersects
one special line $\delta_v=1$ in the upper half plane.

\subsubsection{$\pi/2 <\theta \leq \pi$}

All models in this range also correspond to open models. Like the
previous cases, small $\Delta$ have
real, negative roots with $\eta = 2\pi$. The mirror models in this case
lie in the fourth quadrant. The crossover of real roots from $\eta =0$
to $\eta = 2\pi$ occurs at $\delta_v=1$, however,
unlike in the earlier case, the line $\delta = -1$ is
never approached by the mirror models in the fourth quadrant. As a result,
there is no switch from real to complex roots and all models have real
negative roots. The $\eta = 2\pi$ roots for small $\Delta$ are
collapse times of initially expanding closed mirror models and the
$\eta = 0$ are future singularities of initially contracting closed and open
mirror models for intermediate and large values of $\Delta$
respectively.

\begin{figure}
\includegraphics[width=16cm]{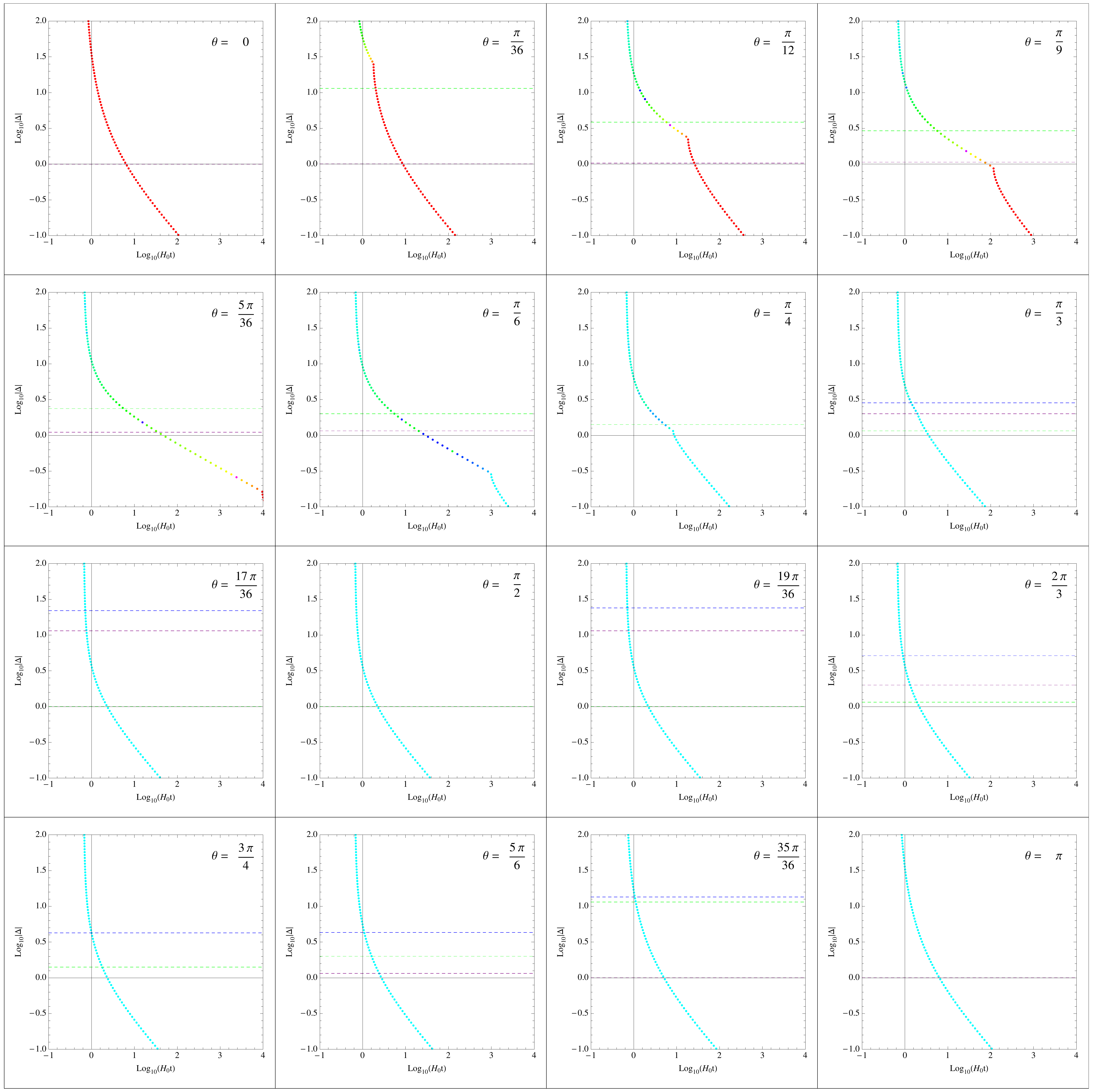}
\includegraphics[width=5cm]{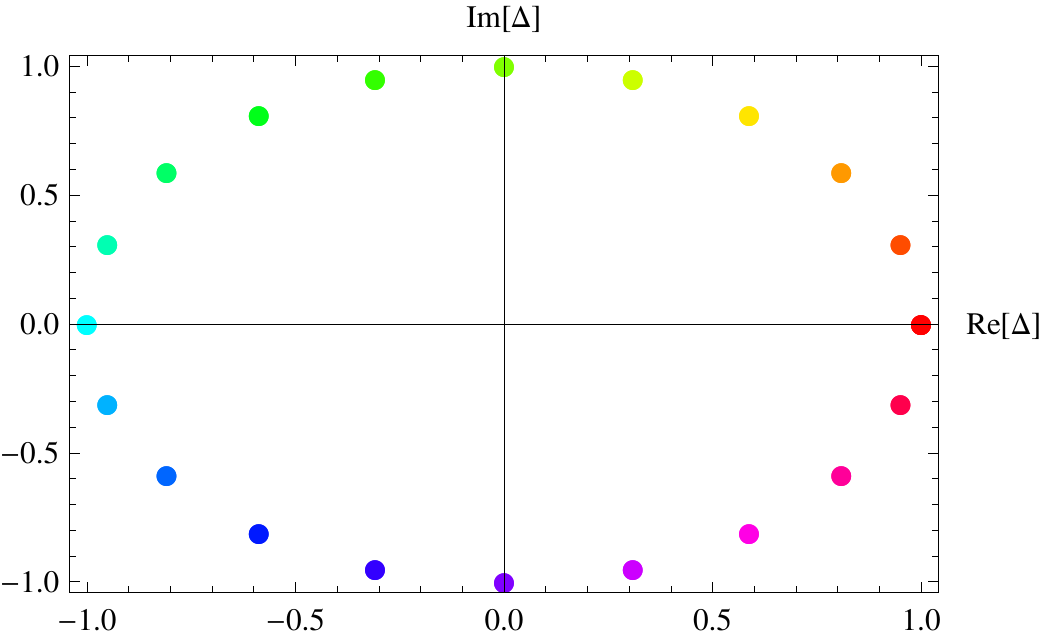}
\caption{Roots with $\eta = 2\pi$ plotted in the complex $\Delta$
  plane for $0<\theta \leq \pi$. These values of $\theta$ correspond
  to those in \figref{allroots}. The colour codes the complex phase of
  the roots ($\dm = \Delta e^{i\phi}$).  The real positive ($\phi=0$)
  and negative ($\phi=\pi$) roots are shown in red and cyan
  respectively. The complex roots can have any colour other than these
  two and the bottom figure provides the coding. By comparison with
  \figref{allroots} one sees that all open models with real roots are
  cyan (negative); likewise all closed models with real roots are red
  (positive). Note, however, that there exist complex roots for both
  open and closed models. Coloured version online. }
\label{imaginary}
\end{figure}

\subsection{Numerical Results}
\label{numericalfit}

Here we present numerical formulas that give the time of validity for
any initial $\Delta$ and $\theta$. Real roots occur for small $\Delta$
when $0 < \theta < \pi/2$; and they occur for all $\Delta$ when $\pi/2
\leq \theta \leq \pi$ or $\theta =0$. Real roots correspond to past or
future singularities of physical models and are known exactly.

\capfigref{allroots} shows that complex roots occur $0<\theta<\pi/2$.
In the range $\pi/4<\theta<\pi/2$ \figref{imaginary} shows that the
phase of the complex roots is very close to $\pi$. We can approximate
these roots as real, negative roots. Conversely, \figref{imaginary}
also shows that in the range $0 < \theta \leq \pi/4$ the phase is not
close to $0$ or $\pi$. These roots are complex only when $\Delta >
\Delta_{rc}$. First, we fit $\Delta_{rc}$ by
\beq 
\Delta_{rc,app}(\theta) =
\left| 0.41 \csc^2\theta (\cos\theta-2\sin \theta) + 3.57
(\cos\theta-2\sin \theta)(\sin \theta)^{4.39} \right| .
\label{deltaapprox}
\eeq
We cannot approximate the time of validity with the results for
physical cases but it turns out that the numerically derived time of validity
is insensitive to $\theta$ in the range
$0 < \theta < \pi/4$ and may be fit
\beq 
H_0 t_{app}(\Delta)= \frac{2}{3} + \left \{ 
\begin{array}{cc}
\frac{14.125}{\Delta^{2.5}} & 0 < \Delta \leq 1 \\
\frac{1.514}{\Delta^{2.82}} & 1 <\Delta \leq 2 \\
\frac{1.778}{\Delta^{3.13}} & 2 <\Delta \leq 5 \\
\frac{63095}{\Delta^{9.6}} & 5 <\Delta \leq 10 \\
\frac{2\times 10^6}{\Delta^{8.5}}  & \Delta > 10 .
\label{timfit}
\end{array}
\right.
\eeq
Using these quantities, the table below gives an approximation to the time of validity,
$T_{app}$, for all values of $\theta$ and $\Delta$. 
The times for collapse and the bang times are equivalent
to \eqnref{paramsoln2} and reproduced here for convenience:
\bea
t_{coll}(\Delta, \theta) &=& 
t_0 + \frac{1}{2H_0} \frac{(1+ \Delta  \cos \theta )}{\left[-E(\Delta, \theta)\right]^{3/2}} (2 \pi ) - t_{age}(\Delta, \theta)\\
t_{bang}^- (\Delta, \theta) &=& 
t_0 +  t_{age}(\Delta, \theta)
\eea
where, 
\bea  
t_{age} (\Delta, \theta) &=& 
\frac{1}{H_0} 
\sqrt{(1+\Delta \cos \theta)} \sqrt{\frac{(1+\Delta\sin \theta)^2}{(1+\Delta  \cos \theta)}} 
 -\frac{1}{H_0} \frac{(1+\Delta  \cos \theta)}{\left[E(\Delta , \theta)\right]^{3/2}}\sinh^{-1}\sqrt{\frac{E(\Delta , \theta)}{(1+\Delta  \cos \theta)}}, \\ 
 E(\Delta, \phi, \theta) &=& (1+ \Delta  \sin \theta)^2- (1+ \Delta \cos\theta) 
\eea

\begin{center}
\begin{table}
\caption{Approximation to time of validity, $T_{app}(\Delta, \theta)$,
  for $0\leq \theta \leq \pi$. Note that $\Delta_{rc,app}$ is
an approximation to $\Delta_{rc}$ in \eqnref{deltaapprox}.}
\begin{tabular}{|c|cc|c|}
\hline
\multicolumn{3}{|c|}{Parameter range}& $T_{app}$\\
\hline 
                    & $0 < \Delta <\Delta_{rc,app}$ & $E(\Delta, \theta) < 0$ 
                    & $t_{coll}(\Delta, \theta)$ \\
$0<\theta <\pi/4$   & $0 < \Delta <\Delta_{rc,app}$ & $ E(\Delta, \theta) > 0$  
                    & $t_{coll}(-\Delta, \theta)$ \\ 
                    & $\Delta >\Delta_{rc,app}$ & 
                    & $t_{app}(\Delta, \theta) $ \\
\hline
                    & $0 < \Delta < \frac{1}{|\sin \theta|}$ &
                    & $t_{coll}(-\Delta, \theta)$ \\
$\pi/4 <\theta \leq \pi/2 $  
                    & $\frac{1}{|\sin \theta|} \leq \Delta \leq 
                           \left|\frac{2 \sin \theta -\cos \theta}
                           {\sin \theta \cos \theta} \right|$ &
                    & $\myreal\left[t_{bang}^-(-\Delta, \theta)\right] $\\
                    & $\Delta > \left|\frac{2 \sin \theta -\cos \theta}
                           {\sin \theta \cos \theta} \right|$ &
                    & $\myreal\left[t_{coll}(-\Delta, \theta)\right] $ \\
\hline
                    & $ 0 < \Delta \leq \frac{1}{|\sin \theta|} $ &
                    & $t_{coll}(-\Delta, \theta)$\\	 		
$\pi/2 \leq \theta \leq \pi$ & $   \Delta >\frac{1}{|\sin \theta|} $ &
                    & $t_{bang}^-(-\Delta, \theta)$\\
\hline
\end{tabular}
\end{table}
\label{tabletvalid}
\end{center}

The error in the fit is estimated as 
\beq 
{\cal E} = \frac{T - T_{app}}{T} . 
\eeq
If ${\cal E} > 0$ then the approximation is conservative in this
sense: the approximate time of validity is less than the true value.
Conversely, if ${\cal E} <0$, then the approximation overestimates
the time of validity. Using the above fits the worst case is ${\cal E}
\simeq -0.02$.  We always use a time step $\delta t$ which satisfies
$\delta t < 0.98 T_{app}$ so that the inaccuracy in the approximation
is irrelevant.

\section{Error characterisation of the Lagrangian series}
\label{errorchar}

We want to characterise the errors associated with calculating the
solution at time $t_f$ given some fixed initial conditions at time
$t_0$. Errors arise because any real calculation involves truncating
the Lagrangian expansion. We want to compare the errors that result
from different choices of truncation order and of the number of
re-expansion steps assuming all series expansions are convergent
(i.e. all respect the time of validity). Let $N_m$ represent the final
physical coordinate generated with a $m$-th order calculation using
$N$ steps.  Ultimately, we seek to characterise differences like
$N_m - {N'}_{m'}$. The quantity $1_\infty$ is the exact answer.

\subsubsection{Single step}
The Lagrangian series solution for a single step has the form 

\beq
1_\infty = a(t) \left( 1 + \sum_{i=1}^\infty \frac{b^{(i)}(t)}{a(t)}\Delta_0^i \right ) X_0, 
\eeq
where each $b^{(i)}$ satisfies 
\beq
{\ddot b}^{(i)} - \frac{H_0^2 a_0^3 b^{(i)}}{a^3} = S^{(i)}
\eeq
and initial conditions are specified at $t=t_0$. 
The initial conditions at each order and the forms for the first few $S^{(i)}$ are given in the text. 

If $t_f -t_0 = \delta t << t_0$, then the solutions can be expanded in the small parameter $\delta t/t_0$. The solutions are
\bea
b^{(1)}(t)/a(t) &\sim& c^{(1)} \frac{\delta t}{t_0} ,\\
b^{(i)}(t)/a(t) & \sim & c^{(i)} \left ( \frac{\delta t}{t_0}\right)^{i+1} \mbox{   for   } i\geq 2. 
\eea
The coefficients $c^{(i)}$ depend on the angle $\theta$ and have a weak dependence on the Lagrangian order. The difference between the exact answer and the $m$-th order approximation for a single step is 
\beq 
 1_\infty -1_m  =\left ( \sum_{i= m+1}^\infty c^{(i)} \left( \frac{\delta t}{t_0}\right )^{i+1} \Delta_0^{i} \right) X_0.
\eeq 
As long as $t_f$ is within the time of validity of LPT, by definition, the LPT series  converges and from the equation above, the leading order error scales as 
$ \sim (\delta t/t_0)^{m+2} \Delta^{m+1}$ (order terms first by powers of
$\Delta$ and then by powers of $\delta t/t_0$).

\subsubsection{Multiple steps} 
In general for a practical application one is limited to working at a finite Lagrangian order. 
In such cases, it becomes necessary to ask if convergence can be achieved by working at a finite Lagrangian order with increasing number of steps. 

First, we outline the calculation.  The initial data is subscripted by
``0''. For example, let the initial perturbed scale factor be
$b_0=b(t_0)$, the initial background scale factor $a_0$, the initial
density contrast $\delta_0$ and the initial velocity perturbation
$\delta_{v0}$.  The Lagrangian expansion parameter $\Delta_0$ and
angle $\theta_0$ follow from the relations $\delta_0 = \Delta_0 \cos
\theta_0$ and $\delta_{v0}=\Delta_0 \sin \theta_0$. The physical
coordinate is $r_0 = b_0 X_0$; for given $r_0$ the initial Lagrangian
coordinate $X_0$ is fixed by choosing $b_0$ to be equal to $a_0$.

Consider taking $N$ steps from initial to final time with an $m$-th
order Lagrangian expansion. Assume that the final time is within the
time of validity of the Lagrangian expansion. For definiteness, let the $j$-th time be
$t_j = t_0 \beta^{j/N}$ where $\beta = t_f/t_0$ (so $t_N$ is
just the final time $t_f$).  This geometric sequence of increasing
steps is well-suited for an expanding background with a small growing
perturbation. The scaling of differences like
$N_m - 1_\infty$, $(N+1)_m-N_m$ and $N_{m+1}-N_m$ with $N$ and $m$ are
all of interest. We expect the same scaling of these differences with
$N$ and $m$ for any uniformly refined set of time steps.

A finite order Lagrangian expansion accurate to order $m$ is
a truncated representation of the full Lagrangian solution
\beq
b(t) = \sum_{i=0}^{m}  b^{(i)}(t) \Delta_0^i .
\eeq
At the beginning of the first step the scale factor at $t_0$ is
advanced to $t_1$ and written as $b(t_0 \to t_1;
\Delta_0,\theta_0)$. Note the explicit dependence on the
perturbation parameters at $t_0$.  Abbreviate the scale factor and its
derivative for the truncated expression as $b$ and ${\dot b}$. 
The background scale factor at time $t_1$ is $a_1$. At the end of
the first step
the Lagrangian coordinate $X_1$ and the new $b_1$ are inferred as
described in the main body of the text by re-scaling quantities
calculated at $t_1$. The new $b_1$ is {\it not}
$b$. The net result for the full step $t_0 \to t_1$ is
\bea
X_1 & = & X_0 \frac{b}{a_1} \\
b_1 & = & a_1 \\
{\dot b}_1 & = & \frac{{\dot b}}{b} a_1 \\
\delta_1 & = & (1 + \delta_0)\left(\frac{a_1}{b}\right)^3 - 1 \\
\delta_{v1} & = & \frac{a_1 {\dot b}}{{\dot a}_1 b } - 1 .
\eea
The newly defined quantities subscripted by ``1'' will be used to initiate the next step. 
The updated perturbations imply new Lagrangian expansion parameter and angle according to
\bea
\Delta_1 & = & \sqrt{\delta_1^2 + \delta_{v1}^2} \\
\cos \theta_1 & = & \frac{\delta_1}{\Delta_1} \\
\sin \theta_1 & = & \frac{\delta_{v1}}{\Delta_1} .
\eea
The new physical position is $r_1 = b_1 X_1 = b X_0$.
In a numerical calculation the truncated $b$ is exact to floating point precision but contains an error because of the omitted orders; in a
symbolic calculation $b$ is known to order $\Delta_0^m$.

The next step from $t_1 \to t_2$ involves a similar
update with $b=b(t_1 \to t_2; \Delta_1,\theta_1)$
\bea
X_2 & = & X_1 \frac{b}{a_2} \\
b_2 & = & a_2 \\
{\dot b}_2 & = & \frac{{\dot b}}{b} a_2 \\
\delta_2 & = & (1 + \delta_1)\left(\frac{a_2}{b}\right)^3 - 1 \\
\delta_{v2} & = & \frac{a_2 {\dot b}}{{\dot a}_2 b } - 1 .
\eea
The new physical position is $r_2=b_2 X_2$.
This iterative scheme repeats for a total of $N$ steps. 
It ultimately yields an approximation to the position at the final
time denoted $N_m = b_N X_N$.

A difference like $(N+1)_m-N_m$ may be calculated numerically for
various $N$ and $m$ and the scaling fitted and inferred. In addition,
one can approach the problem symbolically. To write $N_m$ we need to
expand the final result in powers of $\Delta_0$.  Note, for example,
that $\Delta_1$ and $\theta_1$ are known as expansions in $\Delta_0$
with coefficients that depend upon $\theta_0$.  Perturbation-related
quantities are re-written systematically in terms of initial
quantities.  For example, $b(t_1 \to t_2; \Delta_1,\theta_1)$ may be
expanded in powers of $\Delta_1$ with coefficients depending upon
$\theta_1$. Next, all occurrences of $\Delta_1$ and $\theta_1$ are
replaced by expansions in powers of $\Delta_0$ and coefficients
depending upon $\theta_0$.  All terms up to and including $\Delta_0^m$
are retained in the final result. This procedure is systematically
repeated until all quantities are expressed in terms of initial data.
Finally, the difference $(N+1)_m - N_m$ is calculated symbolically.
Similar strategies allow construction of all the differences of
interest.

To make analytic progress assume that $f_t=\beta-1=(t_f-t_0)/t_0 << 1$
is a small parameter. In a difference like $(N+1)_m-N_m$ many ``lower
order'' terms will coincide.  Consider an ordering of terms by the
powers of $\Delta_0$ (first) and by powers of $f_t$ (second).  Define the
leading-order difference to be the first non-vanishing term
proportional to $\Delta_0^p f_t^q$ for smallest $p$ and then smallest
$q$. It is straightforward to apply this ordering to simplify the
differences like $(N+1)_m - N_m$. The leading
order differences satisfy the following simple equalities
\bea
\label{eq:errscale}
\nonumber | 1_\infty - N_m | & \sim & |g_{N,m}| \\
| N_{m+1} - N_m | & \sim & |g_{N,m}| \\
\nonumber  |(N+1)_m - N_m | & \sim & |g_{N+1,m}-g_{N,m}| 
\eea
where
\bea
\nonumber g_{N,m} & = & K_{N,m} \cos \theta \sin^m \theta \Delta^{m+1} f_t^{m+2} \\
\nonumber K_{N,m} & = & \frac{1}{9} 
             \left( \frac{-2}{3} \right)^m 
             \left( \frac{ N - \frac{m}{2+m} }{N^{m+1}} \right) .
\eea
These differences can be compared with the numerical differences
for which no expansion in $f_t$ is carried out. 

We verified the analytical scaling by the following numerical experiment. The parameters of the problem at the starting time $t_0 $ are $\Delta_0 = 1/2$, $\theta_0 = -\pi/4$. The final time of interest $t_f$ is close to the initial time so that $(t_f -t_0)/t_0 = 1/4$. The $m$-th order Lagrangian approximation is evaluated at this fixed final time with successively increasing number of steps. Values of $m$ from 1 to 4 and values of $N$ from 10 to 50 were considered. 
For geometric time steps
$(t_{i+1}-t_i)/t_i = \beta^{1/N}-1$ is
independent of $i$ and denoted $\delta t/t$ below.

\begin{figure}
\includegraphics[width=16cm]{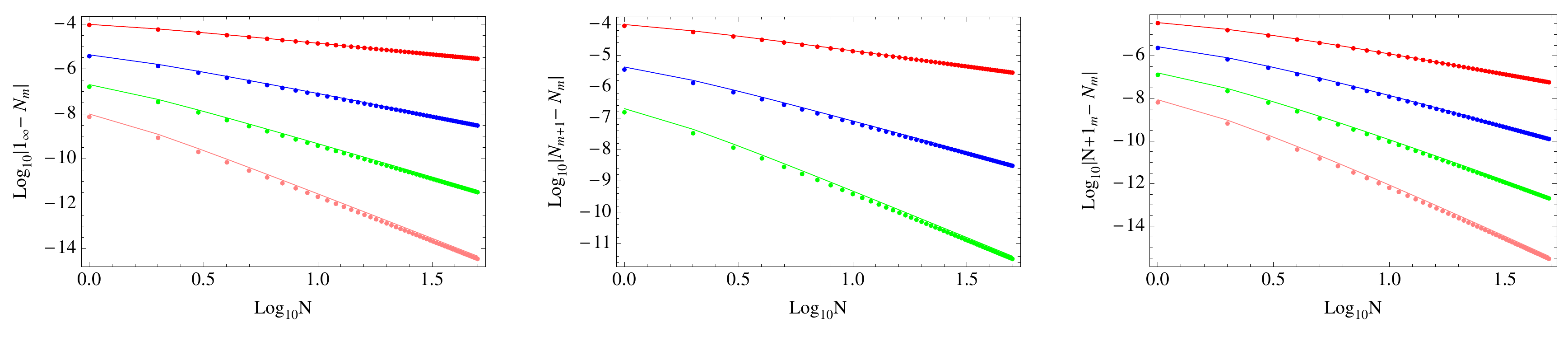}
\caption{The three panels show the log of the errors $|1_\infty -N_m|$, $|N_{m+1}-N_m|$ and $|(N+1)_m -N_m|$ vs. $N$. The final time $t_f$ is the same for all these comparisons. The dots correspond to the data generated by the numerical experiment and lines correspond to the analytical formulas given in \eqnref{eq:errscale}. The lines from top to bottom correspond to $m=1,2,3,4$ respectively for the first and third panels and to $m=1,2,3$ for the second panel. It is clear that for a fixed $m$, increasing the number of steps improves convergence. Conversely, for a fixed $N$, increasing the Lagrangian order $m$ improves convergence.  }
\label{errorsorder}
\end{figure}

The results are plotted in \figref{errorsorder}. The points indicate the numerical data points and the solid lines indicate the analytical functions defined in \eqnref{eq:errscale}. The numerical calculation was done with a high enough precision that even small errors of the order of $10^{-14}$ are not contaminated by floating point errors. 
The agreement between the numerical experiment and the symbolic differences is very good. 

Thus, the scaling of the errors implies that for a small total time step, any finite order Lagrangian scheme will yield convergent results upon taking multiple steps. Conversely, for a fixed number of steps, a higher order Lagrangian calculation will give better results. 

It is useful to express the scaling in terms of the individual small step size $\delta t/t$. Under the 
assumptions that $(t_f -t_0)/t_0$ is small, $(t_f -t_0)/t_0 \sim N \delta t/t$.  The scaling  $|1_\infty 
- N_m|  \sim N^{-m} \Delta^{m+1} ((t_f-t_0)/t_0)^{m+2}$ can be re-written as 
$|1_\infty - N_m|  \sim N ((t_f-t_0)/t_0) \cdot \Delta^{m+1} (\delta t/t)^{m+1}$, which can be interpreted as an error of $((t_f-t_0)/t_0) \Delta^{m+1} (\delta t/t)^{m+1}$ per step. 
In the text, the quantity $\epsilon = \Delta \delta t/t$ is kept
constant. For fixed initial and final times, the error scales
$\propto N \epsilon^{m+1}$. If $\Delta$ does not change appreciably
then the error is $\propto \Delta \epsilon^m$. Convergence is attained when
$\epsilon \to 0$.

\label{lastpage}

\end{document}